\documentclass[aps, prb, reprint]{revtex4-1}

\usepackage{natbib}
\usepackage{amsmath}
\usepackage{amsfonts}
\usepackage[utf8]{inputenc}
\usepackage{xcolor}
\usepackage{bbm}
\usepackage{graphicx}
\usepackage{float}
\usepackage{hhline}
\usepackage{booktabs}
\usepackage{etoolbox}
\usepackage{bbm}
\usepackage{tabularx}
\usepackage{url}
\usepackage{subcaption}
\restylefloat{table}

\usepackage{tikz}
\usetikzlibrary{shapes,arrows}

\usepackage{algorithm}
\usepackage{algpseudocode}

\AtBeginDocument{%
  \heavyrulewidth=.08em
  \lightrulewidth=.05em
  \cmidrulewidth=.03em
  \belowrulesep=.65ex
  \belowbottomsep=0pt
  \aboverulesep=.4ex
  \abovetopsep=0pt
  \cmidrulesep=\doublerulesep
  \cmidrulekern=.5em
  \defaultaddspace=.5em
}

\usepackage{graphicx}
\usepackage{amssymb}
\usepackage{subcaption}
\usepackage{booktabs}
\usepackage{hyperref}
\usepackage{cleveref}
\usepackage{xcolor}
\usepackage{caption}
\usepackage{soul}

\newcommand{\op}{\hat{\mathcal{O}}}
\newcommand{\psit}{\psi^\star}
\newcommand{\Psit}{\Psi^\star}

\newcommand{\Na}{N_\alpha}
\newcommand{\Nb}{N_\beta}
\newcommand{\braket}[1]{\langle #1 \rangle}

\newcommand{\overlap}{\mathbf{S}}
\newcommand{\opoverlap}{\hat{\mathbf{O}}}
\newcommand{\hoverlap}{\hat{\mathbf{H}}}
\newcommand{\opoverlapopt}{\hat{\mathbf{O}}^\star}

\newcommand{\psimat}{\mathbf{\Psi}}
\newcommand{\psimatopt}{\mathbf{\Psi}^{\star}}

\newcommand{\oppsimat}{\hat{\mathcal{O}}\psimat}
\newcommand{\hpsimat}{\hat{H}\psimat}
\newcommand{\hpsimatopt}{\hat{H}\psimatopt}

\usepackage{times}

\begin{document} 

\title{Accurate Computation of Quantum Excited States with Neural Networks} 
\author{David Pfau$^{1,2}$}
\email{pfau@google.com}
\author{Simon Axelrod$^{1,3,4}$}
\author{Halvard Sutterud$^2$}
\author{Ingrid von Glehn$^1$}
\author{James S. Spencer$^1$}
\affiliation{$^1$Google DeepMind, S2, 8 Handyside Street, London N1C 4DJ}
\affiliation{$^2$Department of Physics, Imperial College London, South Kensington Campus, London SW7 2AZ}
\affiliation{$^3$Department of Chemistry and Chemical Biology, Harvard University, 12 Oxford St., Cambridge, MA 01238}
\affiliation{$^4$Department of Materials Science and Engineering, Massachusetts Institute of Technology, 182 Memorial Dr., Cambridge, MA 01239}
\date{May 17, 2024}

\begin{abstract}
We present a variational Monte Carlo algorithm for estimating the lowest excited states of a quantum system which is a natural generalization of the estimation of ground states. The method has no free parameters and requires no explicit orthogonalization of the different states, instead transforming the problem of finding excited states of a given system into that of finding the ground state of an expanded system. Expected values of arbitrary observables can be calculated, including off-diagonal expectations between different states such as the transition dipole moment. Although the method is entirely general, it works particularly well in conjunction with recent work on using neural networks as variational Ans{\"a}tze for many-electron systems, and we show that by combining this method with the FermiNet and Psiformer Ans{\"a}tze we can accurately recover vertical excitation energies and oscillator strengths on a range of molecules. Our method is the first deep learning approach to achieve accurate vertical excitation energies, including challenging double excitations, on benzene-scale molecules. Beyond the chemistry examples here, we expect this technique will be of great interest for applications to atomic, nuclear and condensed matter physics.
\end{abstract}

\maketitle

\section{Introduction}
The computation of excited state properties of quantum systems is a fundamental challenge in chemistry and physics. Understanding electronic excitations is critical for predicting fluorescence in quantum dots\cite{brus1984electron}, molecular conformational changes in the presence of light \cite{polli2010conical}, and photocatalytic activity \cite{prier2013visible}. In condensed matter physics, excitations determine the optical band gap of semiconductors, which is critical for predicting the behavior of solar cells, photosensors, LEDs and lasers. Excited states are also relevant to understanding nuclear phenomena like metastable isomers\cite{chiara2018isomer}. Despite the importance of excited states for quantum phenomena, a full computational account of excited states remains challenging. Excited states are much more challenging to compute than ground states -- inexpensive methods like time-dependent density functional theory (TD-DFT) often give qualitatively incorrect results, while even gold standard methods like multireference configuration interaction (MRCI) can have large error bars on higher excited states. Techniques from machine learning can be applied to this problem, but most applications must be trained on expensive {\em ab-initio} calculations \cite{westermayr2020machine}. Far less work has focused on how insights from machine learning can improve the {\em ab-initio} calculations themselves.

Recent work using neural networks as a wavefunction Ansatz has demonstrated the ability to reach impressive levels of accuracy on ground state calculations with variational quantum Monte Carlo (VMC) \cite{carleo2017solving, hermann2023ab}, even exceeding coupled cluster (CCSD(T)) accuracy on some bond-stretching systems. VMC \cite{foulkes2001quantum, carlson2015quantum} is conceptually simple -- it works by finding an explicit functional form for a wavefunction which minimizes a variational bound -- and scales as $\mathcal{O}(N^3)-\mathcal{O}(N^4)$ with system size, which is favorable for a wavefunction method. While neural network Ans{\"a}tze are state of the art for variational optimization of ground states, they have not yet matched the accuracy of other methods for excited states calculations as they scale to larger systems, despite initial forays \cite{choo2018symmetries, entwistle2023electronic}. This could be due to the Ansatz not being sufficiently expressive, but it could also be due to the choice of variational principle. When used to optimize ground states, there are only two variational principles for QMC -- energy minimization and variance minimization. Innovations in ground state VMC primarily focus on the choice of trial wavefunction \cite{kwon1993effects, bajdich2008pfaffian}, or optimization method used to achieve the variational bound \cite{sorella1998green, toulouse2007optimization}, but the choice of objective to optimize is well-established. The same cannot be said for variational optimization of excited states.

Approaches for computing excited states by VMC either aim to find a single excited state (so-called state-{\em targeting} methods), or aim to find all of the lowest-lying exciting states (called either state-{\em averaging} or state-{\em specific} depending on whether the states overlap). Methods for finding multiple states either minimize the energy of each state sequentially or minimize the (possibly weighted) total energy of all states while maintaining orthogonality. Among state-targeting methods, there are methods which target specific energy ranges\cite{zhao2019variational, otis2023promising}, specific symmetries of the system\cite{choo2018symmetries}, or a specific ordering of the roots\cite{zimmerman2009excited}. For state-averaging and state-specific approaches, the different states must be kept from collapsing onto one another, which can be achieved by including a penalty term in the variational bound\cite{choo2018symmetries, pathak2021excited, entwistle2023electronic, wheeler2023ensemble}, or by explicitly constructing orthogonal Ans{\"a}tze by solving a generalized eigenvalue problem, sometimes repeatedly re-orthogonalizing during optimization\cite{schautz2004optimized, cordova2007troubleshooting, filippi2009absorption, cuzzocrea2020variational, dash2021tailoring}.

All of these approaches have drawbacks and limitations that makes it difficult or impossible to use them with recently-developed Ans{\"a}tze based on deep neural networks \cite{carleo2017solving, luo2019backflow, pfau2020ab, hermann2020deep} on general systems. Targeting specific symmetries or energy ranges requires prior knowledge about the states of interest which may not be available, and state-targeting by variance minimization can lose track of the desired state \cite{cuzzocrea2020variational}. Root-targeting methods are prone to root-flipping, whether they are used for QMC or other computational paradigms\cite{dorando2007targeted, lewin2008computation}. Some methods require solving a generalized eigenvalue problem from stochastic estimates of the Hamiltonian and overlap matrices, which introduces statistical bias into the gradient estimates \cite{filippi2009absorption, zimmerman2009excited}. While this is usually avoided by accumulating matrix elements for long enough that this bias is small, this is not possible in the deep learning paradigm, where parameters are optimized by a large number of small, noisy steps\cite{bottou2007tradeoffs}. Explicitly orthogonalizng Ans{\"a}tze by solving a generalized eigenvalue equation for linear coefficients is usually only possible when the Ansatz is a linear combination of basis set functions\cite{ceperley1988calculation, nightingale2001optimization}, possibly including a shared Jastrow factor \cite{schautz2004optimized}, which rules out neural networks. Other methods do not maintain true orthogonality, but only keep linearized approximations to the wavefunctions orthogonal \cite{zimmerman2009excited}. Some penalty methods which have been used with neural network Ans{\"a}tze have problems with biased gradients \cite{entwistle2023electronic}, and even with unbiased gradients, convergence is only guaranteed if the strength of the penalty term is set above a certain critical value which is not known {\em a priori}. When optimizing multiple states simultaneously with unbiased penalty methods, a weighting factor must be chosen for each state, otherwise the critical penalty threshold diverges \cite{wheeler2023ensemble}. Heuristics such as variance matching may be required to achieve good numerical results for all approaches. Despite almost four decades of work on QMC methods for excited states \cite{carlson1984variational, ceperley1988calculation}, no single variational principle has emerged which has no free parameters, has convergence guarantees when optimizing with noisy Monte Carlo estimates, and is applicable to all possible Ans{\"a}tze and all excited states, regardless of symmetry.

Here we present a new variational principle for computing the lowest excited states of a quantum system by Monte Carlo which does not suffer from any of the limitations described above. Like many state-specific and state-averaged approaches, our method minimizes the 
5
 total energy over states, but we make a particular choice of sampling distribution which does not require the states to be orthogonal. This choice of sampling distribution is equivalent to reformulating the problem of finding $K$ excited states of an $N$ particle system into the problem of finding the ground state of a $K$-fermion system where each fermion is equivalent to $N$ particles in the original system. Instead of orthogonalizing the states, the local energy is promoted from a scalar to a matrix, which gives unbiased estimates of a matrix whose eigenvalues are the energies of orthogonal states. Because wavefunction optimization can be done by stochastic gradient descent from unbiased noisy estimates of the total energy, the procedure is guaranteed to converge to a local minimum of the total energy over states\cite{robbins1951stochastic}. Due to the many desirable mathematical properties summarized above, we refer to our proposed approach as {\em natural excited states} for VMC (NES-VMC).

\section{Natural Excited States}

We aim to find the $K$ lowest eigenfunctions $\psi_1,\ldots,\psi_K$ of a Hamiltonian $\hat{H}$. To find the ground state, we can take samples $\mathbf{x}\sim \psi^2(\mathbf{x})$ of particle positions and compute unbiased estimates of the energy $\mathbb{E}_{\mathbf{x} \sim \psi^2}\left[\psi^{-1}(\mathbf{x}) \hat{H} \psi(\mathbf{x}) \right]$ as well as gradients of the energy, and then optimize the functional form of $\psi$. The scalar $E_L(\mathbf{x}) \triangleq \psi^{-1}(\mathbf{x}) \hat{H} \psi(\mathbf{x})$ inside the expectation is the {\em local energy}. This is the conventional energy minimization principle for VMC.

To generalize this to excited states, consider the function:

\begin{equation}
    \Psi(\mathbf{x}^1, \ldots, \mathbf{x}^K) \triangleq
    \mathrm{det}\begin{pmatrix}
    \psi_1(\mathbf{x}^1) & \ldots & \psi_K(\mathbf{x}^1) \\
    \vdots & & \vdots \\
    \psi_1(\mathbf{x}^K) & \ldots & \psi_K(\mathbf{x}^K)
    \end{pmatrix}
\end{equation}
where $\mathbf{x}^1,\ldots,\mathbf{x}^K$ are $K$ different sets of particle states. We call $\psi_1,\ldots,\psi_K$ the {\em single-state Ans{\"a}tze} and $\Psi$ the {\em total Ansatz}. The total Ansatz resembles a Slater determinant, except that single-particle orbitals are substituted with many-particle single-state Ans{\"a}tze.

We can compute Monte Carlo estimates of the total energy over {\em all} states by taking samples $\mathbf{x}^1,\ldots,\mathbf{x}^K \sim \Psi^2(\mathbf{x}^1,\ldots,\mathbf{x}^K)$ and generalizing the local energy from a scalar to a matrix:

\begin{widetext}
\begin{equation}
    \mathbf{E}_\Psi \triangleq \mathbb{E}_{\mathbf{x}^1,\ldots,\mathbf{x}^K \sim \Psi^2}\left[
    \begin{pmatrix}
    \psi_1(\mathbf{x}^1) & \ldots & \psi_K(\mathbf{x}^1) \\
    \vdots & & \vdots \\
    \psi_1(\mathbf{x}^K) & \ldots & \psi_K(\mathbf{x}^K)
    \end{pmatrix}^{-1}
    \begin{pmatrix}
    \hat{H}\psi_1(\mathbf{x}^1) & \ldots & \hat{H}\psi_K(\mathbf{x}^1) \\
    \vdots & & \vdots \\
    \hat{H}\psi_1(\mathbf{x}^K) & \ldots & \hat{H}\psi_K(\mathbf{x}^K)
    \end{pmatrix}
    \right]
    \label{eqn:main_text_expected_local_energy_matrix}
\end{equation}
\end{widetext}
The trace of this matrix is an unbiased estimate for the total energy over states, and unbiased gradients can be estimated as well. The determinant in the definition of $\Psi$ guarantees that the states will not collapse onto one another during minimization, even though nothing constrains the states to be orthogonal. Minimizing $\mathbf{Tr}[\mathbf{E}_\Psi]$ is mathematically equivalent to ground state VMC for an extended system that is $K$ times larger than the ground state of $\hat{H}$.

If the true minimum of the total energy could be found, it would give a set of single-state Ans{\"a}tze that are a linear combination of the lowest states of $\hat{H}$. To recover the energies of {\em individual} states, we simply diagonalize the matrix $\mathbf{E}_\Psi$ at the minimum. For other observables $\hat{O}$, a matrix similar to $\mathbf{E}_\Psi$ can be accumulated with $\hat{O}$ substituted for $\hat{H}$, and then be transformed into the same basis as the eigenvectors of $\mathbf{E}_\Psi$. We call this method {\em natural excited states} for VMC. A derivation of NES-VMC, its equivalence to ground state VMC in an extended system, and more details on the computation of observables, are given in the Methods section.

While NES-VMC is fully general and can be applied to any quantum Hamiltonian, our experimental validation is focused on electronic structure in atoms and molecules, due to the abundant experimental and computational literature to compare against. For all experiments, we are solving the electronic Schr{\"o}dinger equation in the Born-Oppenheimer approximation and in atomic units:

\begin{align}
\hat{H} = &-\frac{1}{2}\sum_i \nabla^2_i + \sum_{i > j} \frac{1}{|\mathbf{r}_i-\mathbf{r}_j|} \nonumber \\
		&- \sum_{i I} \frac{Z_I}{|\mathbf{r}_i - \mathbf{R}_I|} 	+ \sum_{I > J} \frac{Z_I Z_J}{|\mathbf{R}_I-\mathbf{R}_J|}
\end{align}
where the indices $i$ and $j$ are over electrons and $I$ and $J$ are over atomic nuclei with fixed locations.

To try to disentangle the effect that the choice of Ansatz has on performance, we investigated two different neural network architectures: the FermiNet\cite{pfau2020ab} and the Psiformer\cite{von2023self}. While the Psiformer is more accurate on large systems, it is also slower, and for ground state calculations up to approximately 15 electrons, no appreciable difference in accuracy between the two has been found.

\section{Atomic Spectra}
\label{sec:atomic_spectra}

\begin{figure*}
    \centering
    \includegraphics[width=\textwidth]{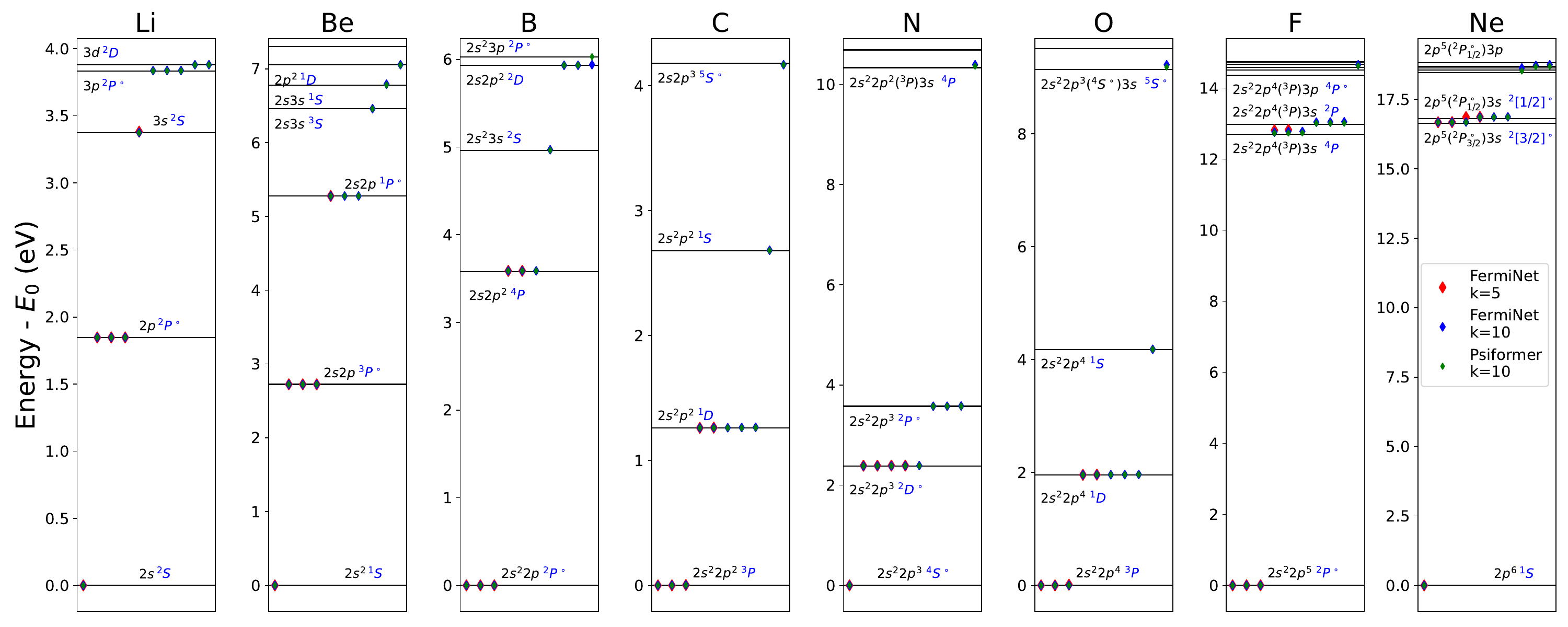}
    \caption{\textbf{Excited state energies for first row atoms from lithium to neon.} Results from natural excited state VMC applied to the FermiNet (10 states, blue, 5 states, red) are shown on top of experimental results\cite{sansonetti2005handbook}. Spectral lines which match computed states are labeled with electron configurations and atomic term symbols (except for the highest levels of F and Ne, where term symbols are omitted for clarity). For all but the largest systems and highest excited states, there is excellent agreement with experiment. The discrepancy between 5 and 10 excited states is minimal except for the highest excited states of F and Ne, where computing more states increases the accuracy of a given state. Complete numerical results are given in Table~\ref{tab:atomic_spectra}.}
    \label{fig:atomic_spectra}
\end{figure*}

To check the correctness of our method, we investigate the excited states of first-row atoms. While we do not aim to reach the accuracy of spectroscopic measurements, we can have high confidence in the accuracy of experimental data, and do not need to worry about effects such as adiabatic relaxation and zero-point vibrational energy which affect molecular measurements. All experimental data was taken from the energy level tables in the NIST Handbook of Basic Atomic Spectroscopic Data\cite{sansonetti2005handbook}. Because we are working with the nonrelativistic Schr{\"o}dinger equation, we are not able to compute fine or hyperfine structure. To remove the fine structure, experimental energy levels with different total angular momenta are averaged together weighted by the degeneracy $m_J = 2J+1$ and treated as a single level. The hyperfine structure is too small to be of concern here. To investigate the effect of the choice of Ansatz as well as the choice of number of states $k$ to compute, we ran calculations with the FermiNet with both 5 and 10 states, as well as the Psiformer with 10 states. Results are given in Fig.~\ref{fig:atomic_spectra}, with numerical results in Table~\ref{tab:atomic_spectra}.

For all atoms, NES-VMC gives results closely matching experiment on most states. From lithium up to oxygen, the error relative to experiment is far less than 1 mHa (27.2 meV) for all but the highest excited state, and is often less than 0.1 mHa. On lithium, all Ans{\"a}tze correctly converge to the $^2S$ and $^2P^\circ$ states, which are missed by the use of the PauliNet Ansatz in combination with a penalty method\cite{entwistle2023electronic}. The method struggles in some cases to get the highest energy state correct, but this seems to be improved by computing more states -- for instance, the error in the $^4P$ states of fluorine is cut in half by increasing the number of states from 5 to 10. In rare cases, the highest state seems to converge to the incorrect state, such as boron with the Psiformer, which seems to converge to the $^2P^\circ$ state rather than the last $^2D$ state. Fluorine and neon both have relatively large errors on the order of 1-2 mHa for low-lying states, but going from the FermiNet to the Psiformer Ansatz reduces this error in all cases. The largest errors are in the highest states of fluorine and neon, on the order of 10 mHa. In this case we suspect the difficulty is due to the large number of different states with similar electron configurations and energies, and hope that by computing even more states or by using even more expressive Ans{\"a}tze, the effects of individual states can be disentangled. The excellent performance on low-lying states gives us confidence that NES-VMC is mathematically sound.

\section{Oscillator Strengths}
\label{sec:oscillator_strengths}

\begin{figure*}
    \centering
    \hspace*{-1.5cm}
    \includegraphics[width=1.2\textwidth]{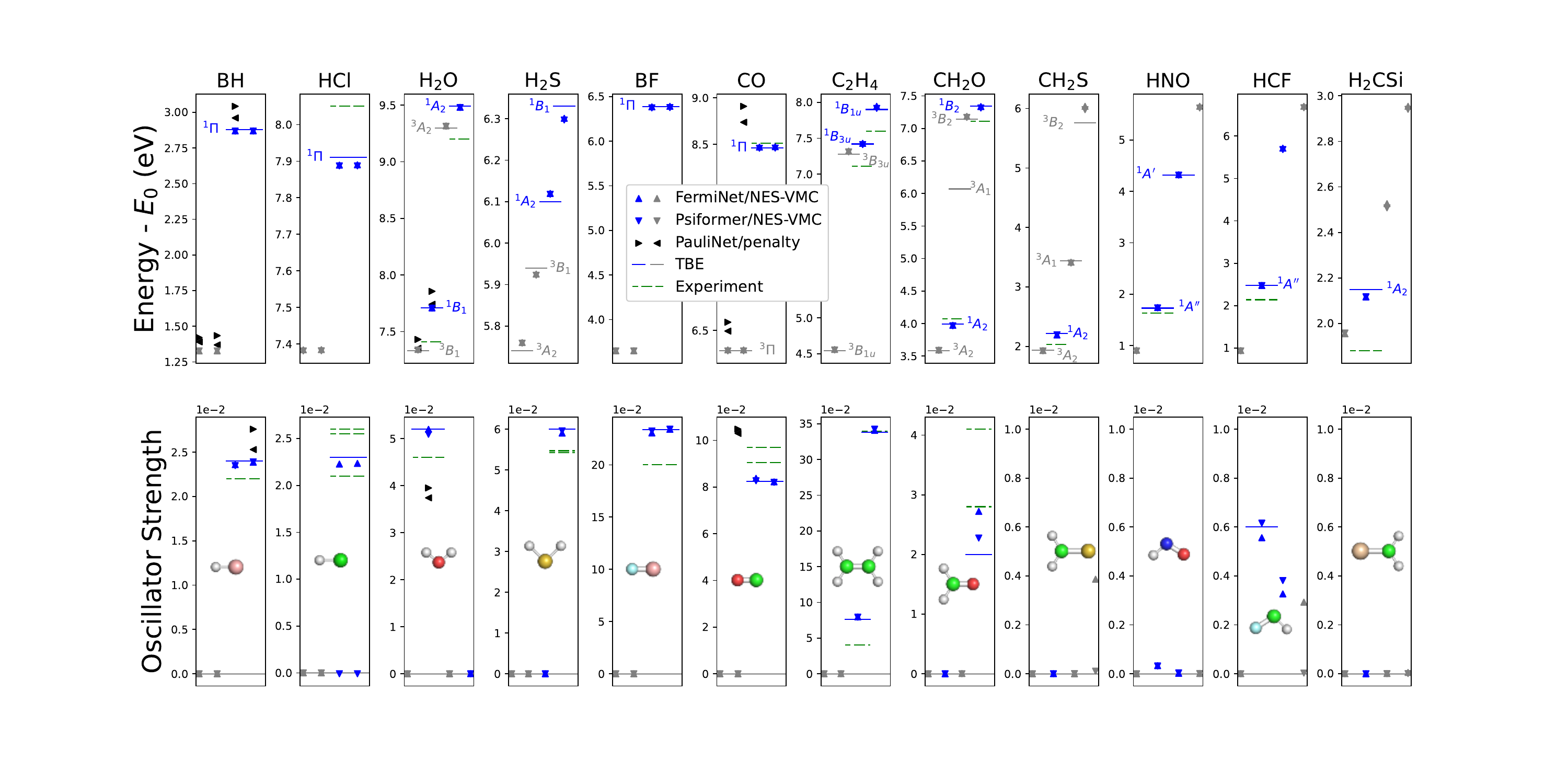}
    \caption{\textbf{Vertical excitation energies and oscillator strengths for small molecules}. Singlet states are in blue and triplet states are in gray. NES-VMC results are indicated by markers while theoretical best estimates from Chrayteh {\em et al.} \cite{chrayteh2020mountaineering} or directly from QUEST \cite{veril2021questdb} are given by the lines. When no data from QUEST is available, no TBE is given. Experimental results from Chrayteh {\em et al.} \cite{chrayteh2020mountaineering} and references thererin are given by the dashed lines in green. Where available, energies and oscillator strengths from Entwistle {\em et al.} \cite{entwistle2023electronic} are provided by the black triangles for comparison, with (pointing left) and without (pointing right) variance matching. In almost all cases, our results on both energies and oscillator strengths agree closely with theoretical best estimates. Complete numerical results are given in Table~\ref{tab:oscillator_strengths}.}
    \label{fig:oscillator_strengths}
\end{figure*}

We are interested in the performance of NES-VMC on more complicated molecular systems, as well as observable quantities other than the energy. The QUEST database\cite{loos2018mountaineering, loos2019reference, chrayteh2020mountaineering, loos2020mountaineering, loos2020mountaineeringb, loos2021reference, loos2021mountaineering, loos2022mountaineering, veril2021questdb} is an excellent source of benchmark vertical excited states calculations on molecules of various sizes, with consistent geometries and basis set extrapolations. Of particular interest is the subset of QUEST for which oscillator strengths have been computed \cite{chrayteh2020mountaineering}, as oscillator strengths are both experimentally observable and the calculations are known to be highly sensitive to the choice of basis set \cite{crossley1984fifteen}.

Oscillator strengths are a measure of the probability of transition between different states occurring as a result of photon emission or absorption. The transition dipole moment between two states gives a measure of how that transition will interact with light:

\begin{equation}
    \mathbf{d}_{ij} = \left\langle \psi^\dagger_i \sum_k q_k \mathbf{r}_k \psi_j \right\rangle
    \label{eqn:tdm}
\end{equation}
where the sum over $k$ is taken over all particles in the system with charge $q_k$ and position $\mathbf{r}_k$. For electrons, $q_k = -e$. The transition dipole moments are vector-valued quantities which include a complex phase, and are not directly observable. The oscillator strength of a particular transition is a dimensionless positive scalar that can be computed from the transition dipole moment and measured experimentally:

\begin{equation}
    f_{ij} = \frac{2}{3}\frac{m}{\hbar^2}\left(E_i - E_j \right) |\mathbf{d}_{ij}|^2
    \label{eqn:osc_strength}
\end{equation}
Computational details are discussed in more detail in Sec.~\ref{sec:appendix_oscillator}.

We applied NES-VMC to all of the small molecules investigated in Chrayteh {\em et al.}\cite{chrayteh2020mountaineering}, computing the 5 lowest energy states and the oscillator strengths of all transitions with both the FermiNet and Psiformer. Results are presented in Fig.~\ref{fig:oscillator_strengths} and Table~\ref{tab:oscillator_strengths}. Wherever possible, we take results from QUEST\cite{chrayteh2020mountaineering, veril2021questdb} to be theoretical best estimates (TBEs) for comparison, though for many of the states we converged to, especially triplets, no results exist in QUEST. For molecules with heavier atoms (HCl, H$_2$S, H$_2$CSi), we found that using pseudopotentials for the heaviest atoms significantly improved the accuracy of the results, likely because the total energy scale was reduced by ignoring core electrons. Where applicable, we also include a comparison against the VMC penalty method with the PauliNet Ansatz of Entwistle {\em et al.}\cite{entwistle2023electronic}. We omit N$_2$ because the lowest-lying excited states are all triplets. For all diatomic systems, the $^1\Pi$ state is doubly-degenerate, and so the baseline oscillator strengths are divided by two to match the computed results.

In almost all cases, both the vertical excitation energies and the oscillator strengths are in excellent agreement with the TBE. The vertical excitation energies are almost all within chemical accuracy (1.6 mHa or 43 meV) of the TBE while the oscillators strengths usually diverge from the TBE by at most an amount on the order of 0.001, comparable to the uncertainty in the calculations. This is much more accurate than the oscillator strengths presented in Entwistle {\em et al.}, even when corrections using variance matching are applied. We note that we do not use variance matching for any of the NES-VMC calculations.

There are a few cases where NES-VMC behaves oddly. While the FermiNet and Psiformer find nearly identical vertical excitation energies for the $^1\Pi$ state of HCl, and the FermiNet accurately predicts the oscillator strength, the Psiformer mistakenly finds this to be a dark state. On formaldehyde (CH$_2$O), both the FermiNet and Psiformer fail to find the $^3A_1$ state at all, and the oscillator strength for the $^1B_2$ state diverges from the TBE by a significant margin, although the Psiformer halves that margin relative to the FermiNet. Vertical excitation energies for systems with heavier atoms, such as H$_2$S, and the highest state of thioformaldehyde (CH$_2$S), are not quite as accurate as other results. For nitroxyl (HNO) and fluoromethylene (HCF), the network initialization must be carefully chosen to ensure convergence to the correct states (see Sec.~\ref{sec:pretraining}). What is clear is that NES-VMC works well in the large majority of cases, and leads to state-of-the-art results for neural network Ans{\"a}tze.

\section{Carbon Dimer}

\begin{figure*}
     \centering
    \includegraphics[width=0.8\textwidth]{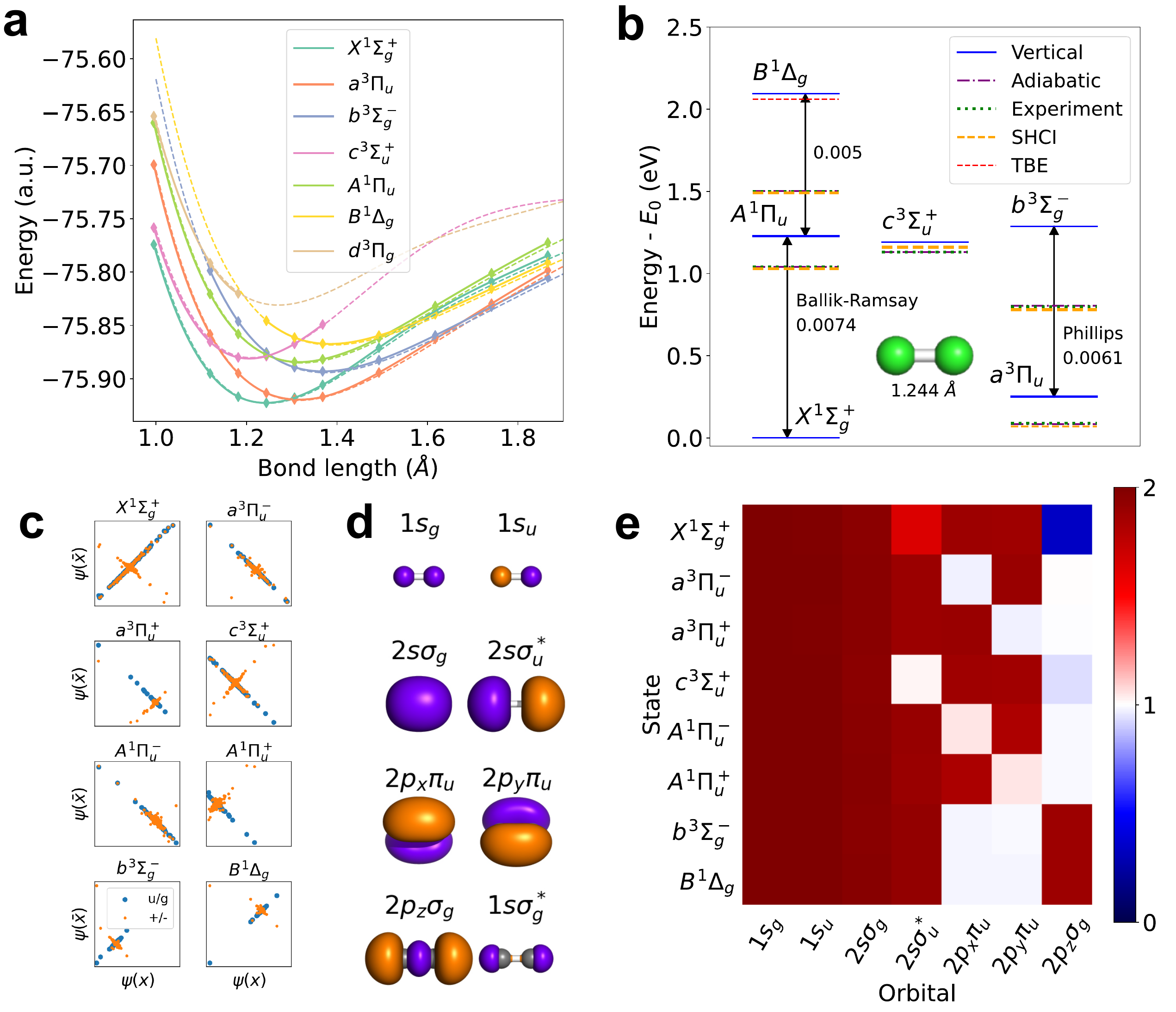}
        \caption{\small \textbf{Excited states of the carbon dimer (C$_2$).} {\bf(a)} Potential energy curves of the low-lying excited states of C$_2$, smoothed by cubic interpolation. The dotted lines are calculated by SHCI \cite{holmes2017excited}, shifted by 0.115 Ha to account for basis set effects. {\bf (b)} The vertical and adiabatic energies of excited states of C$_2$. The green line indicates experimental adiabatic energies\cite{martin1992c2}, the orange line indicates adiabatic energies from SHCI \cite{holmes2017excited} and the red line indicates the vertical energy of the $B^1\Delta_g$ state from QUEST\cite{loos2019reference}. Bright transitions are labelled with their oscillator strength and, when available, their names. {\bf (c)} The symmetries of the different states can be identified by evaluating each single state Ansatz at location $\mathbf{r}$ and $-\mathbf{r}$ for parity symmetry (u/g, blue) or by flipping $\mathbf{r}$ across the x-axis for reflection symmetry (+/--, orange). {\bf (d)} Visualization of the 8 lowest natural orbitals of C$_2$. {\bf (e)} The occupancy of the different natural orbitals for the different excited states of C$_2$, identified from the density matrix of each state. The $a^3\Pi_u$ through $A^1\Pi_u$ states are single excitations while the $b^3\Sigma^-_g$ and $B^1\Delta_g$ states are double excitations. Complete numerical results are given in Tables~\ref{tab:carbon_dimer}, \ref{tab:carbon_dimer_pec} and \ref{tab:carbon_dimer_adiabatic}.}
        \label{fig:carbon-dimer}
\end{figure*}

In addition to computing observable quantities, it is also desirable to be able to say something about the {\em nature} of different states. As a benchmark system for characterizing different states, we study the carbon dimer (C$_2$). Despite its small size, C$_2$ has a complicated electronic structure with many low-lying excited states\cite{phillips1968swan, martin1992c2}. Due to the existence of strong visible bands, C$_2$ is frequently detected in astrophysical measurements, and can be observed in comets rich in organic materials\cite{venkataramani2016optical}. The exact bond order of C$_2$ is still a subject of some controversy -- while molecular orbital theory would classify it as a double bond, valence bond calculations suggest it may be better described as a quadruple bond \cite{shaik2012quadruple}. And C$_2$ is one of the smallest molecules to have low-lying double excitations, a class of excited state which other methods often struggle with \cite{loos2019reference}. Correctly reconstructing the potential energy curves for different low-lying states requires correctly characterizing these different states at different geometries.

We compute the 8 lowest-lying states of C$_2$ at several different bond lengths using the Psiformer Ansatz, and present the results in Figs.~\ref{fig:carbon-dimer}. We classify the different states by computing their spin magnitude and their parity and reflection symmetry. We do not compute the orbital angular momentum, but confirm that we see the expected degeneracy, for instance $\Pi$ states are doubly degenerate. In combination with reference energies from semistochastic heat-bath configuration interaction (SHCI) \cite{holmes2017excited}, we were able to match all computed energies to their respective states. The oscillator strengths at equilibrium show several bright transitions, which we show in Fig.~\ref{fig:carbon-dimer}b. Due to the degeneracy of the $\Pi$ states, we add the oscillator strengths together to give the total strength. We correctly identify the Phillips and Ballik-Ramsay systems\cite{ballik19633, ballik1963extension}, as well as the unnamed $B^1\Delta_g\rightarrow A^1\Pi_u$ transition. We also find that the energy of the $B^1\Delta_g$ energy closely matches the TBE in QUEST\cite{loos2019reference}. The $A^1\Pi_u$, $c^3\Sigma^+_u$ and $b^3\Sigma^-_g$ states all have nearly the same energy at equilibrium, so correctly identifying the oscillator strengths for these transitions is very challenging.

We find that our potential energy curves closely match those computed by SHCI. The vertical energies match SHCI to within chemical accuracy at all bond lengths, and the absolute energy of a state drifts by about 5 mHa/\AA\, relative to SHCI as the bond length changes. We also compute the minimum energy along each interpolated potential energy curve to estimate the {\em adiabatic} energy of the excitations, and find excellent agreement with experiment\cite{martin1992c2} -- while SHCI had a maximum error of 0.03 eV and mean absolute error (MAE) of 0.02 eV on the first five excitations, our maximum error was 7 meV and MAE was 4 meV, a roughly fourfold improvement. The SHCI residual errors are correlated with the equilibrium bond length, so this 5 mHa/\AA\, drift explains most of the difference in the results, which is strong evidence that the NES-VMC curve is closer to the ground truth than SHCI.

To better understand the nature of each state, we compute the occupancy of the different natural orbitals. We first compute the one-electron reduced density matrix (1-RDM) for each single-state Ansatz in a large basis set and then diagonalize these matrices to find the natural orbitals, as described in more detail in Sec.~\ref{sec:dm_and_no}. In this case, the natural orbitals closely match the Hartree-Fock molecular orbitals, so the 1-RDMs are nearly diagonal. We see in Fig.~\ref{fig:carbon-dimer}e that all states above the ground state involve excitation of electrons into the $2p_z\sigma_g$ orbital. The $\Pi$ states are well-described by single excitations from one of the $2p\pi_u$ orbitals while the $c^3\Sigma^+_u$ state promotes an electron from the $2s\sigma_u^*$ orbital. Finally, both the $b^3\Sigma^-_g$ and $B^1\Delta_g$ states are double excitations of the $2p\pi_u$ electrons into the $2s\sigma_u^*$ orbital. Not only is NES-VMC able to predict double excitation energies correctly, but by having an explicit functional form for the wavefunction Ansatz, we can compute quantities which allow us to derive insight about the nature of excitations.

\section{Twisted Ethylene}

\begin{figure*}
     \centering
     \includegraphics[width=\textwidth]{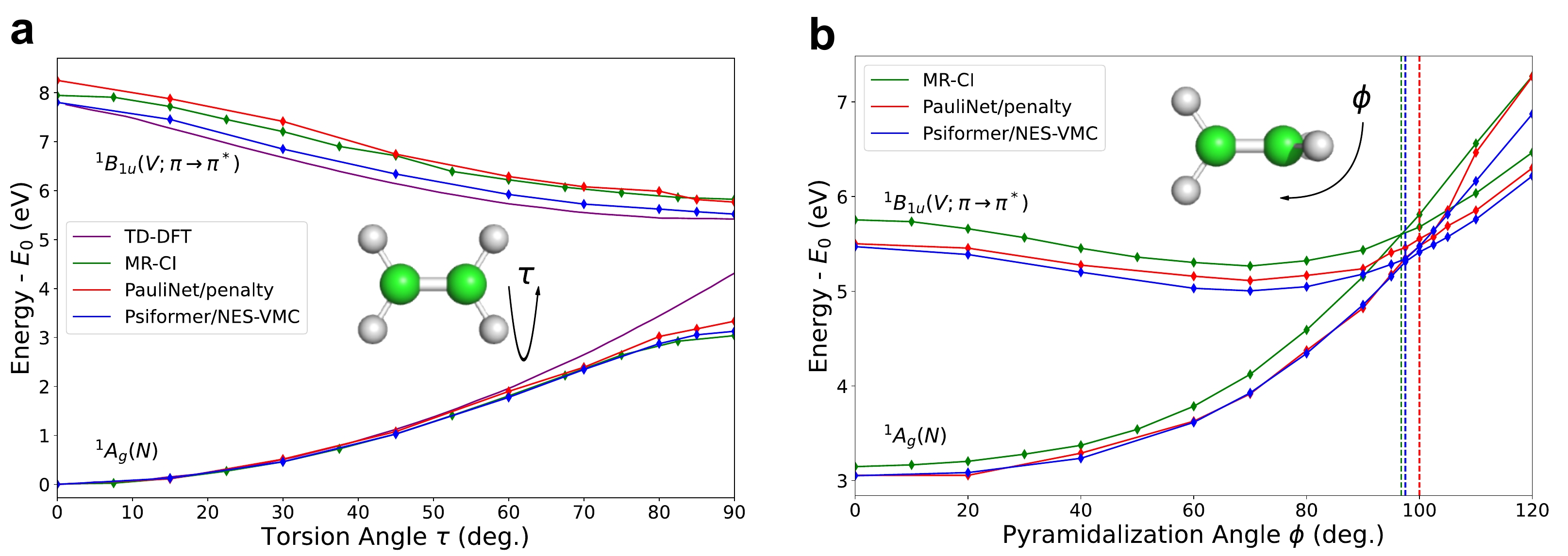}
        \caption{\textbf{Excited states and conical intersection of ethylene (C$_2$H$_4$).} {\bf (a)} Potential energy curve of the first two singlet states of ethylene under torsion around the C-C bond. Our results (blue) are compared against TD-DFT\cite{malis2020trajectory} (purple), MR-CI\cite{barbatti2004photochemistry} (green) and a penalty method used with the PauliNet, without the variance matching correction\cite{entwistle2023electronic} (red). {\bf (b)} Potential energy curve of the first two singlet states of ethylene under pyramidalization of the C-H bonds. The best estimate of the location of the conical intersection of the V and N states for each method is given by the vertical line. Our method is in close agreement with MR-CI up to a constant shift, and agrees with the location of the conical intersection better than the PauliNet penalty method. Note that the $\phi=0$ geometry in Fig.~\ref{fig:ethene}b differs slightly from the $\tau=90$ geometry in Fig.~\ref{fig:ethene}a, as in Barbatti {\em et al.} \cite{barbatti2004photochemistry}. All results are normalized so that the ground state energy at the equilibrium geometry is 0. Complete numerical results are given in Table~\ref{tab:ethene}.}
        \label{fig:ethene}
\end{figure*}

The excited states of ethylene (C$_2$H$_4$) across its potential energy surface present a challenging benchmark problem for many methods. As the carbon double bond is twisted, an avoided crossing occurs when the torsion angle is 90$^\circ$. Even for ground state calculations, DFT and single-reference coupled cluster calculations predict an unphysical cusp at this location\cite{krylov1998size}. Starting from the 90$^\circ$ torsion and bending the hydrogen atoms on one side inward (so-called ``pyramidalization"), ethylene undergoes a conical intersection where the ground state transitions from a $\pi$ to $\pi^*$ highest occupied orbital (the $N$ and $V$ states, with term symbols $^1A_g$ and $^1B_{1u}$). Modeling this intersection requires multireference methods, and while time-dependent density functional theory (TD-DFT) struggles with this system\cite{barbatti2016surface}, multireference configuration interaction (MR-CI) methods describe it well\cite{barbatti2004photochemistry}.

We compute the excited states of ethylene as the torsion angle is varied from 0$^\circ$ to 90$^\circ$, followed by variation of the pyramidalization angle from 0$^\circ$ to 120$^\circ$, enough to include the conical intersection of the $N$ and $V$ states. We use the geometry from previous studies\cite{barbatti2004photochemistry}. Results are shown in Fig.~\ref{fig:ethene}. There are also several low-lying triplet states of ethene, the $^3B_{1u}$ and $^3B_{3u}$ states, and so we calculated $K=3$ excited states for all geometries, which we found was enough to find two singlet states for all geometries except at equilibrium, where we used $K=5$ and took the highest state, as the $^1B_{3u}$ state has lower energy exclusively at equilibrium. We did not find a significant difference between the FermiNet and Psiformer, and show the Psiformer results here. For comparison, in addition to TD-DFT\cite{malis2020trajectory} and MR-CI, we also compare against the PauliNet penalty method\cite{entwistle2023electronic}.

Qualitatively, the results from NES-VMC closely match MR-CI. The spurious cusp when the torsion angle is 90$^\circ$ is avoided, and the error in the ground state relative to MR-CI is smaller than for the PauliNet penalty method across torsion angles. The non-parallelity error in the $V$ state relative to MR-CI is lower for our method than the PauliNet penalty method, and our predicted location for the conical intersection ($\sim$97.5 degrees) is closer to the MR-CI value ($\sim$96 degrees) than the predicted PauliNet penalty method value ($\sim$100 degrees). There is a nearly constant shift in the energy of the $V$ state on the order of several tenths of an eV relative to MR-CI, and a shift in the energy of the $N$ state which grows as the pyramidalization angle grows. Increasing the number of excited states and using a different Ansatz did not seem to make a difference. We note that when using the equilibrium geometry for ethylene from QUEST in Sec~\ref{sec:oscillator_strengths} as opposed to the geometry from MR-CI, our results agreed with the theoretical best estimates to within chemical accuracy. The overall agreement with experimentally relevant quantities like the location of the conical intersection is in excellent agreement with other highly accurate theoretical studies, and NES-VMC is able to capture the important behavior of this system across the potential energy surface.

\section{Double Excitations}
\label{sec:double_excitations}

\begin{figure*}
     \centering
      \hspace*{-1.6cm}
    \includegraphics[width=1.2\textwidth]{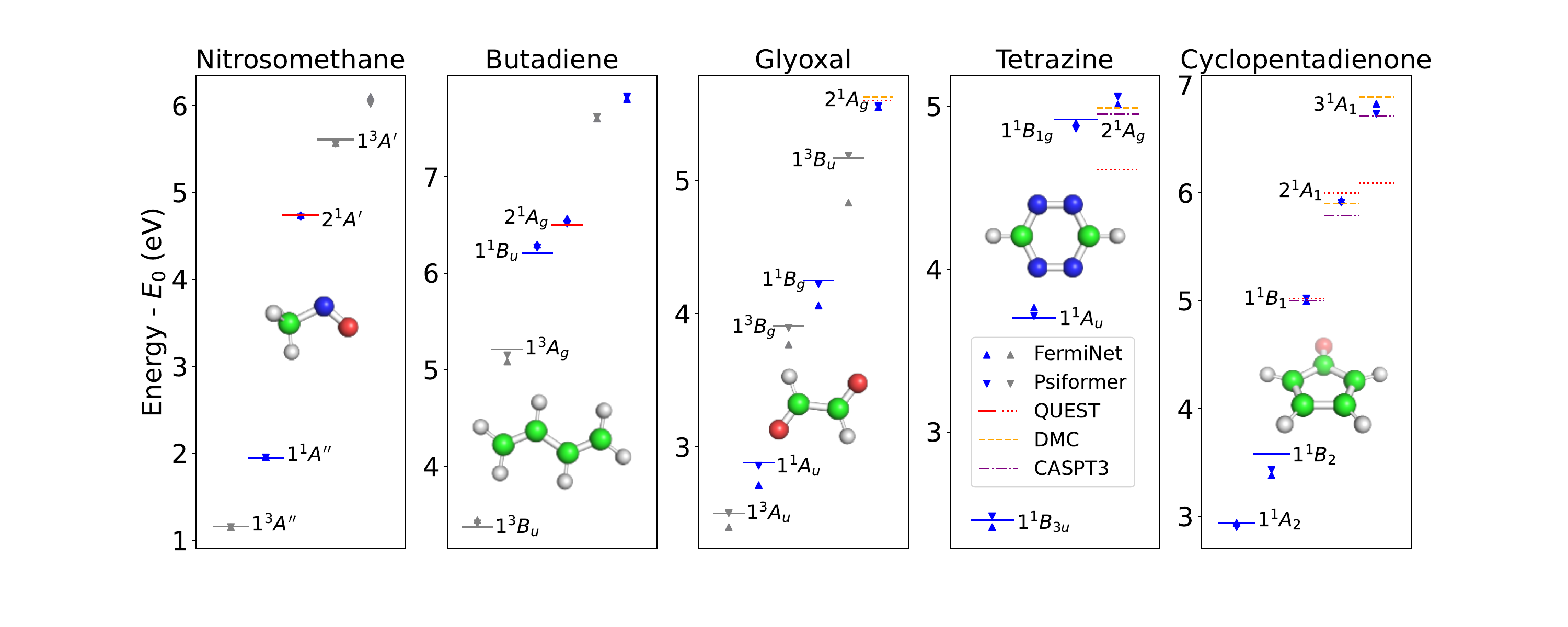}
        \caption{\textbf{Excited states of larger double excitation systems.} NES-VMC with FermiNet and Psiformer for singlet (blue) and triplet (gray) systems are compared against results from the QUEST database\cite{veril2021questdb} (singlet single excitations in blue, triplets in grey, singlet double excitations in red), including ``unsafe" results (dashed lines). For systems where the QUEST results are ``unsafe", more accurate results from DMC (orange) \cite{shepard2022double} or with a CASPT3 correction (purple) \cite{kossoski2024reference} are given. NES-VMC closely matches the more accurate double excitation calculations on the largest and most challenging systems. Complete numerical results are given in Table~\ref{tab:double_excitations}.}
        \label{fig:double_excitations}
\end{figure*}

Accurate calculation of double excitations is known to be far more challenging than single excitations\cite{loos2019reference}. We already demonstrated that NES-VMC is effective at computing the double excitations of nitroxyl and the carbon dimer. To see how well NES-VMC scales to larger systems, we investigated five systems with 24-42 electrons known to have low-lying full or partial double excitations: nitrosomethane, butadiene, glyoxal, tetrazine and cyclopentadienone. For the last two systems, both of which have the same number of electrons as benzene, the original TBEs from QUEST \cite{loos2019reference, veril2021questdb} were known to be unsafe for double excitations, and only very recently did more accurate calculations from QMC \cite{shepard2022double} and CASPT3 \cite{kossoski2024reference} resolve discrepancies as large as almost 1 eV for some states. For these especially challenging two systems, we added an extra term to the Hamiltonian to push up the energy of triplet states so that the excitations of interest could be resolved with a reasonable computational budget (see Sec.~\ref{sec:singlet_targeting}). Results on all systems are shown in Fig.~\ref{fig:double_excitations} and Table~\ref{tab:double_excitations}.

Butadiene is of particular interest, as it is the smallest conjugated organic molecule, a class of molecules whose photochemistry is relevant to vision, photosynthesis, dyes and photovoltaics. The exact ordering of the two lowest-lying singlet transitions, the bright $1^1A_g\rightarrow1^1B_u$ and dark $1^1A_g\rightarrow2^1A_g$ transitions, have been the subject of controversy for many years\cite{buenker1968ab}, only being resolved in the last decade or so\cite{watson2012excited} after extensive study. While the $1^1B_u$ state is a single excitation, the $2^1A_g$ is known to have roughly 30\% double excitation character, making it especially challenging to compute. For the FermiNet and Psiformer, NES-VMC is able to not only correctly predict the ordering of the states, but matches the TBE from QUEST to within 92 and 60 meV respectively for the $1^1B_u$ state and 62 and 9 meV respectively for the $2^1A_g$, a remarkably high degree of agreement for such a notorious system.

On all double excitation systems with the Psiformer, and 4 out of 5 systems with the FermiNet, NES-VMC is in excellent agreement with the best computational results. The one exception for the FermiNet, glyoxal, is at the scale where the FermiNet is known to perform worse than the Psiformer at ground state calculations \cite{von2023self}, so it is not surprising that it struggles on some systems of this size. The FermiNet and Psiformer achieve a mean absolute error relative to the TBE of 15 and 21 meV on nitrosomethane, 84 and 38 meV on butadiene, 167 and 28 meV on glyoxal, 45 and 54 meV on tetrazine, and 92 and 66 meV on cyclopentadienone respectively. For the Psiformer, this is within chemical accuracy (43 meV) for all systems except tetrazine and cyclopentadienone. On tetrazine, the Psiformer is within 0.1 eV of the best estimates of the $2^1A_g$ vertical excitation energy, while the previous TBE in QUEST was off by nearly 1 eV. On cyclopentadienone, even the best current estimates of the $2^1A_1$ and $3^1A_1$ excitation energies disagree by 0.1-0.15 eV, a range which the Psiformer is within. This demonstrates that NES-VMC is among the state of the art in challenging excited state calculations, where even other top methods disagree by more than chemical accuracy.

\section{Benzene}
\label{sec:benzene}

\begin{figure*}
     \centering
    \includegraphics[width=\textwidth]{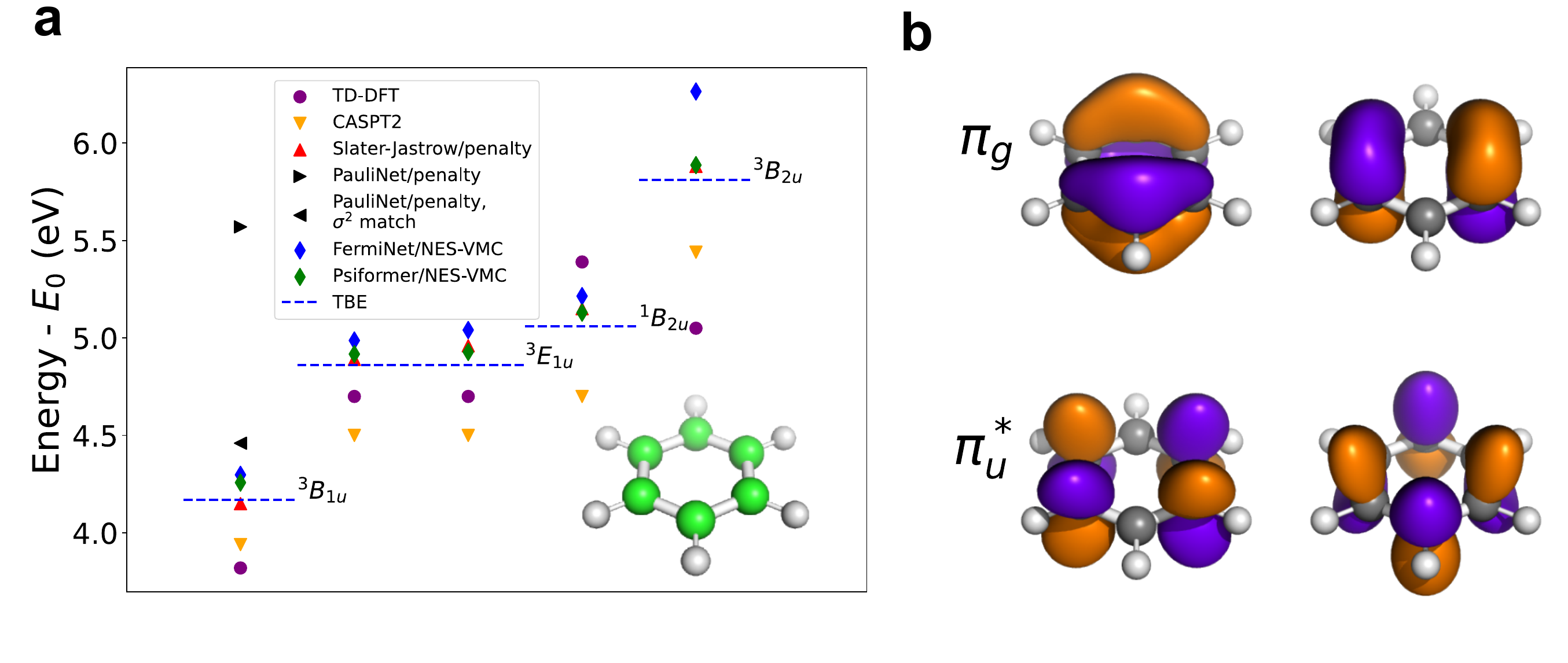}
        \caption{\textbf{Excited states of benzene.} {\bf (a)} Energy levels of benzene. The NES-VMC results (green and blue) are compared against theoretical best estimates from QUEST\cite{loos2020mountaineering, loos2022mountaineering} alongside TD-DFT-PBE0\cite{adamo1999accurate}, CASPT2\cite{roos1992towards}, DMC with a Slater-Jastrow Ansatz and penalty method\cite{pathak2021excited}, and the PauliNet with a penalty method, with and without variance matching\cite{entwistle2023electronic}. NES-VMC with the Psiformer Ansatz is competitive with state-of-the-art methods, and is far more accurate than the only other neural network result, the PauliNet penalty method. Complete numerical results are given in Table~\ref{tab:benzene}. {\bf (b)} The orbitals involved in the excitation of benzene. All excitations computed here are $\pi\rightarrow\pi^*$ excitations. Top row: $\pi$ orbitals occupied in the ground state. Bottom row: $\pi^*$ orbitals occupied in the excited states.}
        \label{fig:benzene}
\end{figure*}

Finally, we applied NES-VMC with both the FermiNet and Psiformer to benzene. While benzene is the same size as tetrazine and cyclopentadienone, it is a common benchmark for medium-sized molecules, so there is more abundant data for us to compare against. For VMC, in addition to the penalty method of Entwistle {\em et al.}\cite{entwistle2023electronic}, there is also the penalty method of Pathak {\em et al.}\cite{pathak2021excited}, which is used with a traditional Slater-Jastrow Ansatz, and uses a different penalty function which allows for unbiased gradients. On top of VMC results and coupled-cluster-based TBEs from QUEST, we also compare against CASPT2\cite{roos1992towards} and TD-DFT with the PBE0 functional\cite{adamo1999accurate}. Results are shown in Fig.~\ref{fig:benzene}, with complex numerical results in Table~\ref{tab:benzene}. For our calculations, we used the same geometry as in QUEST\cite{loos2020mountaineering}.

To better understand the nature of the excitations computed, we inspected the density matrices of the respective states, similarly to the analysis of C$_2$ in Figs.~\ref{fig:carbon-dimer}d and \ref{fig:carbon-dimer}e. The density matrices in the Hartree-Fock basis are nearly diagonal. All five excited states for benzene we computed are single excitations from a $\pi$ to $\pi^*$ orbital, but they are best described by exciting half an electron from two distinct $\pi_g$ orbitals into two distinct $\pi^*_u$ orbitals. These orbitals are visualized in Fig~\ref{fig:benzene}b.

NES-VMC with the Psiformer comes very close to reaching the TBE for all computed states. The FermiNet is not quite as accurate, and struggles with the highest energy $^3B_{2u}$ state. The highest excited state of the FermiNet converges to a mixture of a triplet and singlet state, which suggests that contamination from the $^1B_{1u}$ state is affecting the performance. Like with glyoxal, is not surprising that the Psiformer is better suited for computing vertical excitation energies here. CASPT2 and TD-DFT methods are less accurate across the board, and CASPT2 is generally intermediate in accuracy between TD-DFT and coupled cluster. The penalty method of Pathak {\em et al.} generally reaches comparable levels of accuracy to NES-VMC with the Psiformer. Additionally, the results reported in Pathak {\em et al.} include a diffusion Monte Carlo correction which reduces the error by $\sim$0.1 eV, while NES-VMC does not include any post-processing of VMC results.

The only other neural network approach that has been tried on molecules this large is the penalty method of Entwistle {\em et al.} in combination with the PauliNet Ansatz \cite{entwistle2023electronic}, albeit only for the $^1A_{1g}$ and $^3B_{1u}$ states. Relative to the TBE, the FermiNet and Psiformer errors on the $^1A_{1g}\rightarrow^3B_{1u}$ transition are 0.126 and 0.088 eV, respectively. Without variance matching, the PauliNet penalty method error is 1.4 eV, and even with variance matching the error is only reduced to 0.29 eV, significantly worse than the Psiformer. This means that NES-VMC is the first method used in conjunction with a neural network Ansatz to compute excited state properties which can reach high accuracy on larger molecules.

\section{Conclusions and Outlook}

We have presented a novel method for calculating excited state properties of quantum systems by variational quantum Monte Carlo (VMC), the natural excited states method (NES-VMC). NES-VMC has no free parameters to tune, and allows for unbiased estimation of energies and gradients, by reformulating a multi-state approach as the problem of finding the ground state of an extended system. In much the same way that sampling from $\psi^2$ considerably simplifies the computation of ground state properties by VMC, NES-VMC provides a simple variational principle for computing excited state properties. Additionally, it dovetails well with recent work on neural network Ans{\"a}tze for many-body systems.

We have demonstrated the effectiveness of NES-VMC on a number of benchmark problems ranging from small atoms and molecules up to benzene-scale molecules. In all cases, NES-VMC is competitive with theoretical best estimates for energies and oscillator strengths, and can capture the behavior of very challenging double excitations and conical intersections. It enables unprecedented accuracy for neural network Ans{\"a}tze as system size scales. The optimized Ansatz can be used in downstream analyses to characterize the nature of the electronic structure of different excited states. NES-VMC is as effective as any other method for computing excited states with QMC that we have been able to compare against, and is so far the {\em only} effective method for neural network Ans{\"a}tze on larger systems, with the added benefit of simplicity and generality.

While we focused on applications using neural network Ans{\"a}tze, which can be computationally expensive, classic Ans{\"a}tze like the Slater-Jastrow Ansatz can be scaled to much larger systems\cite{filippi2009absorption}. As the system size grows, applying different weighting to the different states is often needed to get good performance with existing QMC methods\cite{cuzzocrea2022reference}. It remains to be seen if the same holds true for NES-VMC. Although our results suggest that more accurate Ans{\"a}tze are quite important for achieving good performance, we look forward to finding out how well NES-VMC works in conjunction with these classic Ans{\"a}tze on large problems.

Finally, while our experiments in this paper focused on electronic excitations of molecular systems, NES-VMC is fully general and can be applied to {\em any} quantum Hamiltonian. Vibronic couplings, where electronic and vibrational excitations interact, are critical for understanding the behavior of molecules near avoided crossings and conical intersections \cite{azumi1977does}. These could be investigated with neural networks by combining NES-VMC with recent work on fitting multiple geometries of molecular systems with a single neural network Ansatz \cite{gao2022ab, scherbela2024towards}. Going even further, vibronic couplings could be studied by applying NES-VMC to Hamiltonians beyond the Born-Oppenheimer approximation which deal with nuclear positions in a fully quantum manner. Beyond molecular systems, excited state calculations with QMC are an important tool for studying nuclear physics\cite{carlson2015quantum}, optical band gaps in condensed matter physics\cite{hunt2018quantum, zhao2019variational}, many properties of spin systems, as well as time dynamics and finite temperature phenomena. Neural network Ans{\"a}tze have already been applied to ground state calculations in some of these domains\cite{yang2023deep, li2022ab}. We are excited to see how NES-VMC and deep neural networks can be applied to many of the most challenging open problems in many-body quantum mechanics in the future.

\bibliography{main}

\begin{thebibliography}{92}%
\makeatletter
\providecommand \@ifxundefined [1]{%
 \@ifx{#1\undefined}
}%
\providecommand \@ifnum [1]{%
 \ifnum #1\expandafter \@firstoftwo
 \else \expandafter \@secondoftwo
 \fi
}%
\providecommand \@ifx [1]{%
 \ifx #1\expandafter \@firstoftwo
 \else \expandafter \@secondoftwo
 \fi
}%
\providecommand \natexlab [1]{#1}%
\providecommand \enquote  [1]{``#1''}%
\providecommand \bibnamefont  [1]{#1}%
\providecommand \bibfnamefont [1]{#1}%
\providecommand \citenamefont [1]{#1}%
\providecommand \href@noop [0]{\@secondoftwo}%
\providecommand \href [0]{\begingroup \@sanitize@url \@href}%
\providecommand \@href[1]{\@@startlink{#1}\@@href}%
\providecommand \@@href[1]{\endgroup#1\@@endlink}%
\providecommand \@sanitize@url [0]{\catcode `\\12\catcode `\$12\catcode
  `\&12\catcode `\#12\catcode `\^12\catcode `\_12\catcode `\%12\relax}%
\providecommand \@@startlink[1]{}%
\providecommand \@@endlink[0]{}%
\providecommand \url  [0]{\begingroup\@sanitize@url \@url }%
\providecommand \@url [1]{\endgroup\@href {#1}{\urlprefix }}%
\providecommand \urlprefix  [0]{URL }%
\providecommand \Eprint [0]{\href }%
\providecommand \doibase [0]{http://dx.doi.org/}%
\providecommand \selectlanguage [0]{\@gobble}%
\providecommand \bibinfo  [0]{\@secondoftwo}%
\providecommand \bibfield  [0]{\@secondoftwo}%
\providecommand \translation [1]{[#1]}%
\providecommand \BibitemOpen [0]{}%
\providecommand \bibitemStop [0]{}%
\providecommand \bibitemNoStop [0]{.\EOS\space}%
\providecommand \EOS [0]{\spacefactor3000\relax}%
\providecommand \BibitemShut  [1]{\csname bibitem#1\endcsname}%
\let\auto@bib@innerbib\@empty
\bibitem [{\citenamefont {Brus}(1984)}]{brus1984electron}%
  \BibitemOpen
  \bibfield  {author} {\bibinfo {author} {\bibfnamefont {L.~E.}\ \bibnamefont
  {Brus}},\ }\href@noop {} {\bibfield  {journal} {\bibinfo  {journal} {The
  Journal of Chemical Physics}\ }\textbf {\bibinfo {volume} {80}},\ \bibinfo
  {pages} {4403} (\bibinfo {year} {1984})}\BibitemShut {NoStop}%
\bibitem [{\citenamefont {Polli}\ \emph {et~al.}(2010)\citenamefont {Polli},
  \citenamefont {Alto{\`e}}, \citenamefont {Weingart}, \citenamefont
  {Spillane}, \citenamefont {Manzoni}, \citenamefont {Brida}, \citenamefont
  {Tomasello}, \citenamefont {Orlandi}, \citenamefont {Kukura}, \citenamefont
  {Mathies}, \citenamefont {Garavelli},\ and\ \citenamefont
  {Cerullo}}]{polli2010conical}%
  \BibitemOpen
  \bibfield  {author} {\bibinfo {author} {\bibfnamefont {D.}~\bibnamefont
  {Polli}}, \bibinfo {author} {\bibfnamefont {P.}~\bibnamefont {Alto{\`e}}},
  \bibinfo {author} {\bibfnamefont {O.}~\bibnamefont {Weingart}}, \bibinfo
  {author} {\bibfnamefont {K.~M.}\ \bibnamefont {Spillane}}, \bibinfo {author}
  {\bibfnamefont {C.}~\bibnamefont {Manzoni}}, \bibinfo {author} {\bibfnamefont
  {D.}~\bibnamefont {Brida}}, \bibinfo {author} {\bibfnamefont
  {G.}~\bibnamefont {Tomasello}}, \bibinfo {author} {\bibfnamefont
  {G.}~\bibnamefont {Orlandi}}, \bibinfo {author} {\bibfnamefont
  {P.}~\bibnamefont {Kukura}}, \bibinfo {author} {\bibfnamefont {R.~A.}\
  \bibnamefont {Mathies}}, \bibinfo {author} {\bibfnamefont {M.}~\bibnamefont
  {Garavelli}}, \ and\ \bibinfo {author} {\bibfnamefont {G.}~\bibnamefont
  {Cerullo}},\ }\href@noop {} {\bibfield  {journal} {\bibinfo  {journal}
  {Nature}\ }\textbf {\bibinfo {volume} {467}},\ \bibinfo {pages} {440}
  (\bibinfo {year} {2010})}\BibitemShut {NoStop}%
\bibitem [{\citenamefont {Prier}\ \emph {et~al.}(2013)\citenamefont {Prier},
  \citenamefont {Rankic},\ and\ \citenamefont {MacMillan}}]{prier2013visible}%
  \BibitemOpen
  \bibfield  {author} {\bibinfo {author} {\bibfnamefont {C.~K.}\ \bibnamefont
  {Prier}}, \bibinfo {author} {\bibfnamefont {D.~A.}\ \bibnamefont {Rankic}}, \
  and\ \bibinfo {author} {\bibfnamefont {D.~W.}\ \bibnamefont {MacMillan}},\
  }\href@noop {} {\bibfield  {journal} {\bibinfo  {journal} {Chemical Reviews}\
  }\textbf {\bibinfo {volume} {113}},\ \bibinfo {pages} {5322} (\bibinfo {year}
  {2013})}\BibitemShut {NoStop}%
\bibitem [{\citenamefont {Chiara}\ \emph {et~al.}(2018)\citenamefont {Chiara},
  \citenamefont {Carroll}, \citenamefont {Carpenter}, \citenamefont {Greene},
  \citenamefont {Hartley}, \citenamefont {Janssens}, \citenamefont {Lane},
  \citenamefont {Marsh}, \citenamefont {Matters}, \citenamefont {Polasik},
  \citenamefont {Rzadkiewicz}, \citenamefont {Seweryniak}, \citenamefont {Zhu},
  \citenamefont {Bottoni}, \citenamefont {Hayes},\ and\ \citenamefont
  {Karamian}}]{chiara2018isomer}%
  \BibitemOpen
  \bibfield  {author} {\bibinfo {author} {\bibfnamefont {C.~J.}\ \bibnamefont
  {Chiara}}, \bibinfo {author} {\bibfnamefont {J.~J.}\ \bibnamefont {Carroll}},
  \bibinfo {author} {\bibfnamefont {M.~P.}\ \bibnamefont {Carpenter}}, \bibinfo
  {author} {\bibfnamefont {J.~P.}\ \bibnamefont {Greene}}, \bibinfo {author}
  {\bibfnamefont {D.~J.}\ \bibnamefont {Hartley}}, \bibinfo {author}
  {\bibfnamefont {R.~V.~F.}\ \bibnamefont {Janssens}}, \bibinfo {author}
  {\bibfnamefont {G.~J.}\ \bibnamefont {Lane}}, \bibinfo {author}
  {\bibfnamefont {J.~C.}\ \bibnamefont {Marsh}}, \bibinfo {author}
  {\bibfnamefont {D.~A.}\ \bibnamefont {Matters}}, \bibinfo {author}
  {\bibfnamefont {M.}~\bibnamefont {Polasik}}, \bibinfo {author} {\bibfnamefont
  {J.}~\bibnamefont {Rzadkiewicz}}, \bibinfo {author} {\bibfnamefont
  {D.}~\bibnamefont {Seweryniak}}, \bibinfo {author} {\bibfnamefont
  {S.}~\bibnamefont {Zhu}}, \bibinfo {author} {\bibfnamefont {S.}~\bibnamefont
  {Bottoni}}, \bibinfo {author} {\bibfnamefont {A.~B.}\ \bibnamefont {Hayes}},
  \ and\ \bibinfo {author} {\bibfnamefont {S.~A.}\ \bibnamefont {Karamian}},\
  }\href@noop {} {\bibfield  {journal} {\bibinfo  {journal} {Nature}\ }\textbf
  {\bibinfo {volume} {554}},\ \bibinfo {pages} {216} (\bibinfo {year}
  {2018})}\BibitemShut {NoStop}%
\bibitem [{\citenamefont {Westermayr}\ and\ \citenamefont
  {Marquetand}(2020)}]{westermayr2020machine}%
  \BibitemOpen
  \bibfield  {author} {\bibinfo {author} {\bibfnamefont {J.}~\bibnamefont
  {Westermayr}}\ and\ \bibinfo {author} {\bibfnamefont {P.}~\bibnamefont
  {Marquetand}},\ }\href@noop {} {\bibfield  {journal} {\bibinfo  {journal}
  {Chemical Reviews}\ }\textbf {\bibinfo {volume} {121}},\ \bibinfo {pages}
  {9873} (\bibinfo {year} {2020})}\BibitemShut {NoStop}%
\bibitem [{\citenamefont {Carleo}\ and\ \citenamefont
  {Troyer}(2017)}]{carleo2017solving}%
  \BibitemOpen
  \bibfield  {author} {\bibinfo {author} {\bibfnamefont {G.}~\bibnamefont
  {Carleo}}\ and\ \bibinfo {author} {\bibfnamefont {M.}~\bibnamefont
  {Troyer}},\ }\href@noop {} {\bibfield  {journal} {\bibinfo  {journal}
  {Science}\ }\textbf {\bibinfo {volume} {355}},\ \bibinfo {pages} {602}
  (\bibinfo {year} {2017})}\BibitemShut {NoStop}%
\bibitem [{\citenamefont {Hermann}\ \emph {et~al.}(2023)\citenamefont
  {Hermann}, \citenamefont {Spencer}, \citenamefont {Choo}, \citenamefont
  {Mezzacapo}, \citenamefont {Foulkes}, \citenamefont {Pfau}, \citenamefont
  {Carleo},\ and\ \citenamefont {No{\'e}}}]{hermann2023ab}%
  \BibitemOpen
  \bibfield  {author} {\bibinfo {author} {\bibfnamefont {J.}~\bibnamefont
  {Hermann}}, \bibinfo {author} {\bibfnamefont {J.}~\bibnamefont {Spencer}},
  \bibinfo {author} {\bibfnamefont {K.}~\bibnamefont {Choo}}, \bibinfo {author}
  {\bibfnamefont {A.}~\bibnamefont {Mezzacapo}}, \bibinfo {author}
  {\bibfnamefont {W.}~\bibnamefont {Foulkes}}, \bibinfo {author} {\bibfnamefont
  {D.}~\bibnamefont {Pfau}}, \bibinfo {author} {\bibfnamefont {G.}~\bibnamefont
  {Carleo}}, \ and\ \bibinfo {author} {\bibfnamefont {F.}~\bibnamefont
  {No{\'e}}},\ }\href@noop {} {\bibfield  {journal} {\bibinfo  {journal}
  {Nature Reviews Chemistry}\ }\textbf {\bibinfo {volume} {7}} (\bibinfo {year}
  {2023})}\BibitemShut {NoStop}%
\bibitem [{\citenamefont {Foulkes}\ \emph {et~al.}(2001)\citenamefont
  {Foulkes}, \citenamefont {Mitas}, \citenamefont {Needs},\ and\ \citenamefont
  {Rajagopal}}]{foulkes2001quantum}%
  \BibitemOpen
  \bibfield  {author} {\bibinfo {author} {\bibfnamefont {W.~M.~C.}\
  \bibnamefont {Foulkes}}, \bibinfo {author} {\bibfnamefont {L.}~\bibnamefont
  {Mitas}}, \bibinfo {author} {\bibfnamefont {R.~J.}\ \bibnamefont {Needs}}, \
  and\ \bibinfo {author} {\bibfnamefont {G.}~\bibnamefont {Rajagopal}},\
  }\href@noop {} {\bibfield  {journal} {\bibinfo  {journal} {Reviews of Modern
  Physics}\ }\textbf {\bibinfo {volume} {73}},\ \bibinfo {pages} {33} (\bibinfo
  {year} {2001})}\BibitemShut {NoStop}%
\bibitem [{\citenamefont {Carlson}\ \emph {et~al.}(2015)\citenamefont
  {Carlson}, \citenamefont {Gandolfi}, \citenamefont {Pederiva}, \citenamefont
  {Pieper}, \citenamefont {Schiavilla}, \citenamefont {Schmidt},\ and\
  \citenamefont {Wiringa}}]{carlson2015quantum}%
  \BibitemOpen
  \bibfield  {author} {\bibinfo {author} {\bibfnamefont {J.}~\bibnamefont
  {Carlson}}, \bibinfo {author} {\bibfnamefont {S.}~\bibnamefont {Gandolfi}},
  \bibinfo {author} {\bibfnamefont {F.}~\bibnamefont {Pederiva}}, \bibinfo
  {author} {\bibfnamefont {S.~C.}\ \bibnamefont {Pieper}}, \bibinfo {author}
  {\bibfnamefont {R.}~\bibnamefont {Schiavilla}}, \bibinfo {author}
  {\bibfnamefont {K.}~\bibnamefont {Schmidt}}, \ and\ \bibinfo {author}
  {\bibfnamefont {R.~B.}\ \bibnamefont {Wiringa}},\ }\href@noop {} {\bibfield
  {journal} {\bibinfo  {journal} {Reviews of Modern Physics}\ }\textbf
  {\bibinfo {volume} {87}},\ \bibinfo {pages} {1067} (\bibinfo {year}
  {2015})}\BibitemShut {NoStop}%
\bibitem [{\citenamefont {Choo}\ \emph {et~al.}(2018)\citenamefont {Choo},
  \citenamefont {Carleo}, \citenamefont {Regnault},\ and\ \citenamefont
  {Neupert}}]{choo2018symmetries}%
  \BibitemOpen
  \bibfield  {author} {\bibinfo {author} {\bibfnamefont {K.}~\bibnamefont
  {Choo}}, \bibinfo {author} {\bibfnamefont {G.}~\bibnamefont {Carleo}},
  \bibinfo {author} {\bibfnamefont {N.}~\bibnamefont {Regnault}}, \ and\
  \bibinfo {author} {\bibfnamefont {T.}~\bibnamefont {Neupert}},\ }\href@noop
  {} {\bibfield  {journal} {\bibinfo  {journal} {Physical Review Letters}\
  }\textbf {\bibinfo {volume} {121}},\ \bibinfo {pages} {167204} (\bibinfo
  {year} {2018})}\BibitemShut {NoStop}%
\bibitem [{\citenamefont {Entwistle}\ \emph {et~al.}(2023)\citenamefont
  {Entwistle}, \citenamefont {Sch{\"a}tzle}, \citenamefont {Erdman},
  \citenamefont {Hermann},\ and\ \citenamefont
  {No{\'e}}}]{entwistle2023electronic}%
  \BibitemOpen
  \bibfield  {author} {\bibinfo {author} {\bibfnamefont {M.}~\bibnamefont
  {Entwistle}}, \bibinfo {author} {\bibfnamefont {Z.}~\bibnamefont
  {Sch{\"a}tzle}}, \bibinfo {author} {\bibfnamefont {P.~A.}\ \bibnamefont
  {Erdman}}, \bibinfo {author} {\bibfnamefont {J.}~\bibnamefont {Hermann}}, \
  and\ \bibinfo {author} {\bibfnamefont {F.}~\bibnamefont {No{\'e}}},\
  }\href@noop {} {\bibfield  {journal} {\bibinfo  {journal} {Nature
  Communications}\ }\textbf {\bibinfo {volume} {14}},\ \bibinfo {pages} {274}
  (\bibinfo {year} {2023})}\BibitemShut {NoStop}%
\bibitem [{\citenamefont {Kwon}\ \emph {et~al.}(1993)\citenamefont {Kwon},
  \citenamefont {Ceperley},\ and\ \citenamefont {Martin}}]{kwon1993effects}%
  \BibitemOpen
  \bibfield  {author} {\bibinfo {author} {\bibfnamefont {Y.}~\bibnamefont
  {Kwon}}, \bibinfo {author} {\bibfnamefont {D.}~\bibnamefont {Ceperley}}, \
  and\ \bibinfo {author} {\bibfnamefont {R.~M.}\ \bibnamefont {Martin}},\
  }\href@noop {} {\bibfield  {journal} {\bibinfo  {journal} {Physical Review
  B}\ }\textbf {\bibinfo {volume} {48}},\ \bibinfo {pages} {12037} (\bibinfo
  {year} {1993})}\BibitemShut {NoStop}%
\bibitem [{\citenamefont {Bajdich}\ \emph {et~al.}(2008)\citenamefont
  {Bajdich}, \citenamefont {Mitas}, \citenamefont {Wagner},\ and\ \citenamefont
  {Schmidt}}]{bajdich2008pfaffian}%
  \BibitemOpen
  \bibfield  {author} {\bibinfo {author} {\bibfnamefont {M.}~\bibnamefont
  {Bajdich}}, \bibinfo {author} {\bibfnamefont {L.}~\bibnamefont {Mitas}},
  \bibinfo {author} {\bibfnamefont {L.}~\bibnamefont {Wagner}}, \ and\ \bibinfo
  {author} {\bibfnamefont {K.}~\bibnamefont {Schmidt}},\ }\href@noop {}
  {\bibfield  {journal} {\bibinfo  {journal} {Physical Review B}\ }\textbf
  {\bibinfo {volume} {77}},\ \bibinfo {pages} {115112} (\bibinfo {year}
  {2008})}\BibitemShut {NoStop}%
\bibitem [{\citenamefont {Sorella}(1998)}]{sorella1998green}%
  \BibitemOpen
  \bibfield  {author} {\bibinfo {author} {\bibfnamefont {S.}~\bibnamefont
  {Sorella}},\ }\href@noop {} {\bibfield  {journal} {\bibinfo  {journal}
  {Physical Review Letters}\ }\textbf {\bibinfo {volume} {80}},\ \bibinfo
  {pages} {4558} (\bibinfo {year} {1998})}\BibitemShut {NoStop}%
\bibitem [{\citenamefont {Toulouse}\ and\ \citenamefont
  {Umrigar}(2007)}]{toulouse2007optimization}%
  \BibitemOpen
  \bibfield  {author} {\bibinfo {author} {\bibfnamefont {J.}~\bibnamefont
  {Toulouse}}\ and\ \bibinfo {author} {\bibfnamefont {C.~J.}\ \bibnamefont
  {Umrigar}},\ }\href@noop {} {\bibfield  {journal} {\bibinfo  {journal} {The
  Journal of Chemical Physics}\ }\textbf {\bibinfo {volume} {126}},\ \bibinfo
  {pages} {084102} (\bibinfo {year} {2007})}\BibitemShut {NoStop}%
\bibitem [{\citenamefont {Zhao}\ and\ \citenamefont
  {Neuscamman}(2019)}]{zhao2019variational}%
  \BibitemOpen
  \bibfield  {author} {\bibinfo {author} {\bibfnamefont {L.}~\bibnamefont
  {Zhao}}\ and\ \bibinfo {author} {\bibfnamefont {E.}~\bibnamefont
  {Neuscamman}},\ }\href@noop {} {\bibfield  {journal} {\bibinfo  {journal}
  {Physical Review Letters}\ }\textbf {\bibinfo {volume} {123}},\ \bibinfo
  {pages} {036402} (\bibinfo {year} {2019})}\BibitemShut {NoStop}%
\bibitem [{\citenamefont {Otis}\ and\ \citenamefont
  {Neuscamman}(2023)}]{otis2023promising}%
  \BibitemOpen
  \bibfield  {author} {\bibinfo {author} {\bibfnamefont {L.}~\bibnamefont
  {Otis}}\ and\ \bibinfo {author} {\bibfnamefont {E.}~\bibnamefont
  {Neuscamman}},\ }\href@noop {} {\bibfield  {journal} {\bibinfo  {journal}
  {Wiley Interdisciplinary Reviews: Computational Molecular Science}\ ,\
  \bibinfo {pages} {e1659}} (\bibinfo {year} {2023})}\BibitemShut {NoStop}%
\bibitem [{\citenamefont {Zimmerman}\ \emph {et~al.}(2009)\citenamefont
  {Zimmerman}, \citenamefont {Toulouse}, \citenamefont {Zhang}, \citenamefont
  {Musgrave},\ and\ \citenamefont {Umrigar}}]{zimmerman2009excited}%
  \BibitemOpen
  \bibfield  {author} {\bibinfo {author} {\bibfnamefont {P.~M.}\ \bibnamefont
  {Zimmerman}}, \bibinfo {author} {\bibfnamefont {J.}~\bibnamefont {Toulouse}},
  \bibinfo {author} {\bibfnamefont {Z.}~\bibnamefont {Zhang}}, \bibinfo
  {author} {\bibfnamefont {C.~B.}\ \bibnamefont {Musgrave}}, \ and\ \bibinfo
  {author} {\bibfnamefont {C.~J.}\ \bibnamefont {Umrigar}},\ }\href@noop {}
  {\bibfield  {journal} {\bibinfo  {journal} {The Journal of Chemical Physics}\
  }\textbf {\bibinfo {volume} {131}},\ \bibinfo {pages} {124103} (\bibinfo
  {year} {2009})}\BibitemShut {NoStop}%
\bibitem [{\citenamefont {Pathak}\ \emph {et~al.}(2021)\citenamefont {Pathak},
  \citenamefont {Busemeyer}, \citenamefont {Rodrigues},\ and\ \citenamefont
  {Wagner}}]{pathak2021excited}%
  \BibitemOpen
  \bibfield  {author} {\bibinfo {author} {\bibfnamefont {S.}~\bibnamefont
  {Pathak}}, \bibinfo {author} {\bibfnamefont {B.}~\bibnamefont {Busemeyer}},
  \bibinfo {author} {\bibfnamefont {J.~N.}\ \bibnamefont {Rodrigues}}, \ and\
  \bibinfo {author} {\bibfnamefont {L.~K.}\ \bibnamefont {Wagner}},\
  }\href@noop {} {\bibfield  {journal} {\bibinfo  {journal} {The Journal of
  Chemical Physics}\ }\textbf {\bibinfo {volume} {154}},\ \bibinfo {pages}
  {034101} (\bibinfo {year} {2021})}\BibitemShut {NoStop}%
\bibitem [{\citenamefont {Wheeler}\ \emph {et~al.}(2024)\citenamefont
  {Wheeler}, \citenamefont {Kleiner},\ and\ \citenamefont
  {Wagner}}]{wheeler2023ensemble}%
  \BibitemOpen
  \bibfield  {author} {\bibinfo {author} {\bibfnamefont {W.~A.}\ \bibnamefont
  {Wheeler}}, \bibinfo {author} {\bibfnamefont {K.~G.}\ \bibnamefont
  {Kleiner}}, \ and\ \bibinfo {author} {\bibfnamefont {L.~K.}\ \bibnamefont
  {Wagner}},\ }\href@noop {} {\bibfield  {journal} {\bibinfo  {journal}
  {Electronic Structure}\ }\textbf {\bibinfo {volume} {6}},\ \bibinfo {pages}
  {025001} (\bibinfo {year} {2024})}\BibitemShut {NoStop}%
\bibitem [{\citenamefont {Schautz}\ and\ \citenamefont
  {Filippi}(2004)}]{schautz2004optimized}%
  \BibitemOpen
  \bibfield  {author} {\bibinfo {author} {\bibfnamefont {F.}~\bibnamefont
  {Schautz}}\ and\ \bibinfo {author} {\bibfnamefont {C.}~\bibnamefont
  {Filippi}},\ }\href@noop {} {\bibfield  {journal} {\bibinfo  {journal} {The
  Journal of Chemical Physics}\ }\textbf {\bibinfo {volume} {120}},\ \bibinfo
  {pages} {10931} (\bibinfo {year} {2004})}\BibitemShut {NoStop}%
\bibitem [{\citenamefont {Cordova}\ \emph {et~al.}(2007)\citenamefont
  {Cordova}, \citenamefont {Doriol}, \citenamefont {Ipatov}, \citenamefont
  {Casida}, \citenamefont {Filippi},\ and\ \citenamefont
  {Vela}}]{cordova2007troubleshooting}%
  \BibitemOpen
  \bibfield  {author} {\bibinfo {author} {\bibfnamefont {F.}~\bibnamefont
  {Cordova}}, \bibinfo {author} {\bibfnamefont {L.~J.}\ \bibnamefont {Doriol}},
  \bibinfo {author} {\bibfnamefont {A.}~\bibnamefont {Ipatov}}, \bibinfo
  {author} {\bibfnamefont {M.~E.}\ \bibnamefont {Casida}}, \bibinfo {author}
  {\bibfnamefont {C.}~\bibnamefont {Filippi}}, \ and\ \bibinfo {author}
  {\bibfnamefont {A.}~\bibnamefont {Vela}},\ }\href@noop {} {\bibfield
  {journal} {\bibinfo  {journal} {The Journal of Chemical Physics}\ }\textbf
  {\bibinfo {volume} {127}} (\bibinfo {year} {2007})}\BibitemShut {NoStop}%
\bibitem [{\citenamefont {Filippi}\ \emph {et~al.}(2009)\citenamefont
  {Filippi}, \citenamefont {Zaccheddu},\ and\ \citenamefont
  {Buda}}]{filippi2009absorption}%
  \BibitemOpen
  \bibfield  {author} {\bibinfo {author} {\bibfnamefont {C.}~\bibnamefont
  {Filippi}}, \bibinfo {author} {\bibfnamefont {M.}~\bibnamefont {Zaccheddu}},
  \ and\ \bibinfo {author} {\bibfnamefont {F.}~\bibnamefont {Buda}},\
  }\href@noop {} {\bibfield  {journal} {\bibinfo  {journal} {Journal of
  Chemical Theory and Computation}\ }\textbf {\bibinfo {volume} {5}},\ \bibinfo
  {pages} {2074} (\bibinfo {year} {2009})}\BibitemShut {NoStop}%
\bibitem [{\citenamefont {Cuzzocrea}\ \emph {et~al.}(2020)\citenamefont
  {Cuzzocrea}, \citenamefont {Scemama}, \citenamefont {Briels}, \citenamefont
  {Moroni},\ and\ \citenamefont {Filippi}}]{cuzzocrea2020variational}%
  \BibitemOpen
  \bibfield  {author} {\bibinfo {author} {\bibfnamefont {A.}~\bibnamefont
  {Cuzzocrea}}, \bibinfo {author} {\bibfnamefont {A.}~\bibnamefont {Scemama}},
  \bibinfo {author} {\bibfnamefont {W.~J.}\ \bibnamefont {Briels}}, \bibinfo
  {author} {\bibfnamefont {S.}~\bibnamefont {Moroni}}, \ and\ \bibinfo {author}
  {\bibfnamefont {C.}~\bibnamefont {Filippi}},\ }\href@noop {} {\bibfield
  {journal} {\bibinfo  {journal} {Journal of Chemical Theory and Computation}\
  }\textbf {\bibinfo {volume} {16}},\ \bibinfo {pages} {4203} (\bibinfo {year}
  {2020})}\BibitemShut {NoStop}%
\bibitem [{\citenamefont {Dash}\ \emph {et~al.}(2021)\citenamefont {Dash},
  \citenamefont {Moroni}, \citenamefont {Filippi},\ and\ \citenamefont
  {Scemama}}]{dash2021tailoring}%
  \BibitemOpen
  \bibfield  {author} {\bibinfo {author} {\bibfnamefont {M.}~\bibnamefont
  {Dash}}, \bibinfo {author} {\bibfnamefont {S.}~\bibnamefont {Moroni}},
  \bibinfo {author} {\bibfnamefont {C.}~\bibnamefont {Filippi}}, \ and\
  \bibinfo {author} {\bibfnamefont {A.}~\bibnamefont {Scemama}},\ }\href@noop
  {} {\bibfield  {journal} {\bibinfo  {journal} {Journal of Chemical Theory and
  Computation}\ }\textbf {\bibinfo {volume} {17}},\ \bibinfo {pages} {3426}
  (\bibinfo {year} {2021})}\BibitemShut {NoStop}%
\bibitem [{\citenamefont {Luo}\ and\ \citenamefont
  {Clark}(2019)}]{luo2019backflow}%
  \BibitemOpen
  \bibfield  {author} {\bibinfo {author} {\bibfnamefont {D.}~\bibnamefont
  {Luo}}\ and\ \bibinfo {author} {\bibfnamefont {B.~K.}\ \bibnamefont
  {Clark}},\ }\href@noop {} {\bibfield  {journal} {\bibinfo  {journal}
  {Physical Review Letters}\ }\textbf {\bibinfo {volume} {122}},\ \bibinfo
  {pages} {226401} (\bibinfo {year} {2019})}\BibitemShut {NoStop}%
\bibitem [{\citenamefont {Pfau}\ \emph {et~al.}(2020)\citenamefont {Pfau},
  \citenamefont {Spencer}, \citenamefont {Matthews},\ and\ \citenamefont
  {Foulkes}}]{pfau2020ab}%
  \BibitemOpen
  \bibfield  {author} {\bibinfo {author} {\bibfnamefont {D.}~\bibnamefont
  {Pfau}}, \bibinfo {author} {\bibfnamefont {J.~S.}\ \bibnamefont {Spencer}},
  \bibinfo {author} {\bibfnamefont {A.~G.}\ \bibnamefont {Matthews}}, \ and\
  \bibinfo {author} {\bibfnamefont {W.~M.~C.}\ \bibnamefont {Foulkes}},\
  }\href@noop {} {\bibfield  {journal} {\bibinfo  {journal} {Physical Review
  Research}\ }\textbf {\bibinfo {volume} {2}},\ \bibinfo {pages} {033429}
  (\bibinfo {year} {2020})}\BibitemShut {NoStop}%
\bibitem [{\citenamefont {Hermann}\ \emph {et~al.}(2020)\citenamefont
  {Hermann}, \citenamefont {Sch{\"a}tzle},\ and\ \citenamefont
  {No{\'e}}}]{hermann2020deep}%
  \BibitemOpen
  \bibfield  {author} {\bibinfo {author} {\bibfnamefont {J.}~\bibnamefont
  {Hermann}}, \bibinfo {author} {\bibfnamefont {Z.}~\bibnamefont
  {Sch{\"a}tzle}}, \ and\ \bibinfo {author} {\bibfnamefont {F.}~\bibnamefont
  {No{\'e}}},\ }\href@noop {} {\bibfield  {journal} {\bibinfo  {journal}
  {Nature Chemistry}\ }\textbf {\bibinfo {volume} {12}},\ \bibinfo {pages}
  {891} (\bibinfo {year} {2020})}\BibitemShut {NoStop}%
\bibitem [{\citenamefont {Dorando}\ \emph {et~al.}(2007)\citenamefont
  {Dorando}, \citenamefont {Hachmann},\ and\ \citenamefont
  {Chan}}]{dorando2007targeted}%
  \BibitemOpen
  \bibfield  {author} {\bibinfo {author} {\bibfnamefont {J.~J.}\ \bibnamefont
  {Dorando}}, \bibinfo {author} {\bibfnamefont {J.}~\bibnamefont {Hachmann}}, \
  and\ \bibinfo {author} {\bibfnamefont {G.~K.}\ \bibnamefont {Chan}},\
  }\href@noop {} {\bibfield  {journal} {\bibinfo  {journal} {The Journal of
  Chemical Physics}\ }\textbf {\bibinfo {volume} {127}} (\bibinfo {year}
  {2007})}\BibitemShut {NoStop}%
\bibitem [{\citenamefont {Lewin}(2008)}]{lewin2008computation}%
  \BibitemOpen
  \bibfield  {author} {\bibinfo {author} {\bibfnamefont {M.}~\bibnamefont
  {Lewin}},\ }\href@noop {} {\bibfield  {journal} {\bibinfo  {journal} {Journal
  of Mathematical Chemistry}\ }\textbf {\bibinfo {volume} {44}},\ \bibinfo
  {pages} {967} (\bibinfo {year} {2008})}\BibitemShut {NoStop}%
\bibitem [{\citenamefont {Bottou}\ and\ \citenamefont
  {Bousquet}(2007)}]{bottou2007tradeoffs}%
  \BibitemOpen
  \bibfield  {author} {\bibinfo {author} {\bibfnamefont {L.}~\bibnamefont
  {Bottou}}\ and\ \bibinfo {author} {\bibfnamefont {O.}~\bibnamefont
  {Bousquet}},\ }\href@noop {} {\bibfield  {journal} {\bibinfo  {journal}
  {Advances in Neural Information Processing Systems (NeurIPS)}\ }\textbf
  {\bibinfo {volume} {20}} (\bibinfo {year} {2007})}\BibitemShut {NoStop}%
\bibitem [{\citenamefont {Ceperley}\ and\ \citenamefont
  {Bernu}(1988)}]{ceperley1988calculation}%
  \BibitemOpen
  \bibfield  {author} {\bibinfo {author} {\bibfnamefont {D.~M.}\ \bibnamefont
  {Ceperley}}\ and\ \bibinfo {author} {\bibfnamefont {B.}~\bibnamefont
  {Bernu}},\ }\href@noop {} {\bibfield  {journal} {\bibinfo  {journal} {The
  Journal of Chemical Physics}\ }\textbf {\bibinfo {volume} {89}},\ \bibinfo
  {pages} {6316} (\bibinfo {year} {1988})}\BibitemShut {NoStop}%
\bibitem [{\citenamefont {Nightingale}\ and\ \citenamefont
  {Melik-Alaverdian}(2001)}]{nightingale2001optimization}%
  \BibitemOpen
  \bibfield  {author} {\bibinfo {author} {\bibfnamefont {M.~P.}\ \bibnamefont
  {Nightingale}}\ and\ \bibinfo {author} {\bibfnamefont {V.}~\bibnamefont
  {Melik-Alaverdian}},\ }\href@noop {} {\bibfield  {journal} {\bibinfo
  {journal} {Physical Review Letters}\ }\textbf {\bibinfo {volume} {87}},\
  \bibinfo {pages} {043401} (\bibinfo {year} {2001})}\BibitemShut {NoStop}%
\bibitem [{\citenamefont {Carlson}\ \emph {et~al.}(1984)\citenamefont
  {Carlson}, \citenamefont {Pandharipande},\ and\ \citenamefont
  {Wiringa}}]{carlson1984variational}%
  \BibitemOpen
  \bibfield  {author} {\bibinfo {author} {\bibfnamefont {J.}~\bibnamefont
  {Carlson}}, \bibinfo {author} {\bibfnamefont {V.~R.}\ \bibnamefont
  {Pandharipande}}, \ and\ \bibinfo {author} {\bibfnamefont {R.~B.}\
  \bibnamefont {Wiringa}},\ }\href@noop {} {\bibfield  {journal} {\bibinfo
  {journal} {Nuclear Physics A}\ }\textbf {\bibinfo {volume} {424}},\ \bibinfo
  {pages} {47} (\bibinfo {year} {1984})}\BibitemShut {NoStop}%
\bibitem [{\citenamefont {Robbins}\ and\ \citenamefont
  {Monro}(1951)}]{robbins1951stochastic}%
  \BibitemOpen
  \bibfield  {author} {\bibinfo {author} {\bibfnamefont {H.}~\bibnamefont
  {Robbins}}\ and\ \bibinfo {author} {\bibfnamefont {S.}~\bibnamefont
  {Monro}},\ }\href@noop {} {\bibfield  {journal} {\bibinfo  {journal} {The
  Annals of Mathematical Statistics}\ ,\ \bibinfo {pages} {400}} (\bibinfo
  {year} {1951})}\BibitemShut {NoStop}%
\bibitem [{\citenamefont {von Glehn}\ \emph {et~al.}(2023)\citenamefont {von
  Glehn}, \citenamefont {Spencer},\ and\ \citenamefont {Pfau}}]{von2023self}%
  \BibitemOpen
  \bibfield  {author} {\bibinfo {author} {\bibfnamefont {I.}~\bibnamefont {von
  Glehn}}, \bibinfo {author} {\bibfnamefont {J.~S.}\ \bibnamefont {Spencer}}, \
  and\ \bibinfo {author} {\bibfnamefont {D.}~\bibnamefont {Pfau}},\ }\href@noop
  {} {\bibfield  {journal} {\bibinfo  {journal} {The 11th International
  Conference on Learning Representations (ICLR)}\ } (\bibinfo {year}
  {2023})}\BibitemShut {NoStop}%
\bibitem [{\citenamefont {Sansonetti}\ and\ \citenamefont
  {Martin}(2005)}]{sansonetti2005handbook}%
  \BibitemOpen
  \bibfield  {author} {\bibinfo {author} {\bibfnamefont {J.~E.}\ \bibnamefont
  {Sansonetti}}\ and\ \bibinfo {author} {\bibfnamefont {W.~C.}\ \bibnamefont
  {Martin}},\ }\href@noop {} {\bibfield  {journal} {\bibinfo  {journal}
  {Journal of Physical and Chemical Reference Data}\ }\textbf {\bibinfo
  {volume} {34}},\ \bibinfo {pages} {1559} (\bibinfo {year}
  {2005})}\BibitemShut {NoStop}%
\bibitem [{\citenamefont {Chrayteh}\ \emph {et~al.}(2020)\citenamefont
  {Chrayteh}, \citenamefont {Blondel}, \citenamefont {Loos},\ and\
  \citenamefont {Jacquemin}}]{chrayteh2020mountaineering}%
  \BibitemOpen
  \bibfield  {author} {\bibinfo {author} {\bibfnamefont {A.}~\bibnamefont
  {Chrayteh}}, \bibinfo {author} {\bibfnamefont {A.}~\bibnamefont {Blondel}},
  \bibinfo {author} {\bibfnamefont {P.-F.}\ \bibnamefont {Loos}}, \ and\
  \bibinfo {author} {\bibfnamefont {D.}~\bibnamefont {Jacquemin}},\ }\href@noop
  {} {\bibfield  {journal} {\bibinfo  {journal} {Journal of Chemical Theory and
  Computation}\ }\textbf {\bibinfo {volume} {17}},\ \bibinfo {pages} {416}
  (\bibinfo {year} {2020})}\BibitemShut {NoStop}%
\bibitem [{\citenamefont {V{\'e}ril}\ \emph {et~al.}(2021)\citenamefont
  {V{\'e}ril}, \citenamefont {Scemama}, \citenamefont {Caffarel}, \citenamefont
  {Lipparini}, \citenamefont {Boggio-Pasqua}, \citenamefont {Jacquemin},\ and\
  \citenamefont {Loos}}]{veril2021questdb}%
  \BibitemOpen
  \bibfield  {author} {\bibinfo {author} {\bibfnamefont {M.}~\bibnamefont
  {V{\'e}ril}}, \bibinfo {author} {\bibfnamefont {A.}~\bibnamefont {Scemama}},
  \bibinfo {author} {\bibfnamefont {M.}~\bibnamefont {Caffarel}}, \bibinfo
  {author} {\bibfnamefont {F.}~\bibnamefont {Lipparini}}, \bibinfo {author}
  {\bibfnamefont {M.}~\bibnamefont {Boggio-Pasqua}}, \bibinfo {author}
  {\bibfnamefont {D.}~\bibnamefont {Jacquemin}}, \ and\ \bibinfo {author}
  {\bibfnamefont {P.-F.}\ \bibnamefont {Loos}},\ }\href@noop {} {\bibfield
  {journal} {\bibinfo  {journal} {Wiley Interdisciplinary Reviews:
  Computational Molecular Science}\ }\textbf {\bibinfo {volume} {11}},\
  \bibinfo {pages} {e1517} (\bibinfo {year} {2021})}\BibitemShut {NoStop}%
\bibitem [{\citenamefont {Loos}\ \emph {et~al.}(2018)\citenamefont {Loos},
  \citenamefont {Scemama}, \citenamefont {Blondel}, \citenamefont {Garniron},
  \citenamefont {Caffarel},\ and\ \citenamefont
  {Jacquemin}}]{loos2018mountaineering}%
  \BibitemOpen
  \bibfield  {author} {\bibinfo {author} {\bibfnamefont {P.-F.}\ \bibnamefont
  {Loos}}, \bibinfo {author} {\bibfnamefont {A.}~\bibnamefont {Scemama}},
  \bibinfo {author} {\bibfnamefont {A.}~\bibnamefont {Blondel}}, \bibinfo
  {author} {\bibfnamefont {Y.}~\bibnamefont {Garniron}}, \bibinfo {author}
  {\bibfnamefont {M.}~\bibnamefont {Caffarel}}, \ and\ \bibinfo {author}
  {\bibfnamefont {D.}~\bibnamefont {Jacquemin}},\ }\href@noop {} {\bibfield
  {journal} {\bibinfo  {journal} {Journal of Chemical Theory and Computation}\
  }\textbf {\bibinfo {volume} {14}},\ \bibinfo {pages} {4360} (\bibinfo {year}
  {2018})}\BibitemShut {NoStop}%
\bibitem [{\citenamefont {Loos}\ \emph {et~al.}(2019)\citenamefont {Loos},
  \citenamefont {Boggio-Pasqua}, \citenamefont {Scemama}, \citenamefont
  {Caffarel},\ and\ \citenamefont {Jacquemin}}]{loos2019reference}%
  \BibitemOpen
  \bibfield  {author} {\bibinfo {author} {\bibfnamefont {P.-F.}\ \bibnamefont
  {Loos}}, \bibinfo {author} {\bibfnamefont {M.}~\bibnamefont {Boggio-Pasqua}},
  \bibinfo {author} {\bibfnamefont {A.}~\bibnamefont {Scemama}}, \bibinfo
  {author} {\bibfnamefont {M.}~\bibnamefont {Caffarel}}, \ and\ \bibinfo
  {author} {\bibfnamefont {D.}~\bibnamefont {Jacquemin}},\ }\href@noop {}
  {\bibfield  {journal} {\bibinfo  {journal} {Journal of Chemical Theory and
  Computation}\ }\textbf {\bibinfo {volume} {15}},\ \bibinfo {pages} {1939}
  (\bibinfo {year} {2019})}\BibitemShut {NoStop}%
\bibitem [{\citenamefont {Loos}\ \emph
  {et~al.}(2020{\natexlab{a}})\citenamefont {Loos}, \citenamefont {Lipparini},
  \citenamefont {Boggio-Pasqua}, \citenamefont {Scemama},\ and\ \citenamefont
  {Jacquemin}}]{loos2020mountaineering}%
  \BibitemOpen
  \bibfield  {author} {\bibinfo {author} {\bibfnamefont {P.-F.}\ \bibnamefont
  {Loos}}, \bibinfo {author} {\bibfnamefont {F.}~\bibnamefont {Lipparini}},
  \bibinfo {author} {\bibfnamefont {M.}~\bibnamefont {Boggio-Pasqua}}, \bibinfo
  {author} {\bibfnamefont {A.}~\bibnamefont {Scemama}}, \ and\ \bibinfo
  {author} {\bibfnamefont {D.}~\bibnamefont {Jacquemin}},\ }\href@noop {}
  {\bibfield  {journal} {\bibinfo  {journal} {Journal of Chemical Theory and
  Computation}\ }\textbf {\bibinfo {volume} {16}},\ \bibinfo {pages} {1711}
  (\bibinfo {year} {2020}{\natexlab{a}})}\BibitemShut {NoStop}%
\bibitem [{\citenamefont {Loos}\ \emph
  {et~al.}(2020{\natexlab{b}})\citenamefont {Loos}, \citenamefont {Scemama},
  \citenamefont {Boggio-Pasqua},\ and\ \citenamefont
  {Jacquemin}}]{loos2020mountaineeringb}%
  \BibitemOpen
  \bibfield  {author} {\bibinfo {author} {\bibfnamefont {P.-F.}\ \bibnamefont
  {Loos}}, \bibinfo {author} {\bibfnamefont {A.}~\bibnamefont {Scemama}},
  \bibinfo {author} {\bibfnamefont {M.}~\bibnamefont {Boggio-Pasqua}}, \ and\
  \bibinfo {author} {\bibfnamefont {D.}~\bibnamefont {Jacquemin}},\ }\href@noop
  {} {\bibfield  {journal} {\bibinfo  {journal} {Journal of Chemical Theory and
  Computation}\ }\textbf {\bibinfo {volume} {16}},\ \bibinfo {pages} {3720}
  (\bibinfo {year} {2020}{\natexlab{b}})}\BibitemShut {NoStop}%
\bibitem [{\citenamefont {Loos}\ \emph {et~al.}(2021)\citenamefont {Loos},
  \citenamefont {Comin}, \citenamefont {Blase},\ and\ \citenamefont
  {Jacquemin}}]{loos2021reference}%
  \BibitemOpen
  \bibfield  {author} {\bibinfo {author} {\bibfnamefont {P.-F.}\ \bibnamefont
  {Loos}}, \bibinfo {author} {\bibfnamefont {M.}~\bibnamefont {Comin}},
  \bibinfo {author} {\bibfnamefont {X.}~\bibnamefont {Blase}}, \ and\ \bibinfo
  {author} {\bibfnamefont {D.}~\bibnamefont {Jacquemin}},\ }\href@noop {}
  {\bibfield  {journal} {\bibinfo  {journal} {Journal of Chemical Theory and
  Computation}\ }\textbf {\bibinfo {volume} {17}},\ \bibinfo {pages} {3666}
  (\bibinfo {year} {2021})}\BibitemShut {NoStop}%
\bibitem [{\citenamefont {Loos}\ and\ \citenamefont
  {Jacquemin}(2021)}]{loos2021mountaineering}%
  \BibitemOpen
  \bibfield  {author} {\bibinfo {author} {\bibfnamefont {P.-F.}\ \bibnamefont
  {Loos}}\ and\ \bibinfo {author} {\bibfnamefont {D.}~\bibnamefont
  {Jacquemin}},\ }\href@noop {} {\bibfield  {journal} {\bibinfo  {journal} {The
  Journal of Physical Chemistry A}\ }\textbf {\bibinfo {volume} {125}},\
  \bibinfo {pages} {10174} (\bibinfo {year} {2021})}\BibitemShut {NoStop}%
\bibitem [{\citenamefont {Loos}\ \emph {et~al.}(2022)\citenamefont {Loos},
  \citenamefont {Lipparini}, \citenamefont {Matthews}, \citenamefont
  {Blondel},\ and\ \citenamefont {Jacquemin}}]{loos2022mountaineering}%
  \BibitemOpen
  \bibfield  {author} {\bibinfo {author} {\bibfnamefont {P.-F.}\ \bibnamefont
  {Loos}}, \bibinfo {author} {\bibfnamefont {F.}~\bibnamefont {Lipparini}},
  \bibinfo {author} {\bibfnamefont {D.~A.}\ \bibnamefont {Matthews}}, \bibinfo
  {author} {\bibfnamefont {A.}~\bibnamefont {Blondel}}, \ and\ \bibinfo
  {author} {\bibfnamefont {D.}~\bibnamefont {Jacquemin}},\ }\href@noop {}
  {\bibfield  {journal} {\bibinfo  {journal} {Journal of Chemical Theory and
  Computation}\ }\textbf {\bibinfo {volume} {18}},\ \bibinfo {pages} {4418}
  (\bibinfo {year} {2022})}\BibitemShut {NoStop}%
\bibitem [{\citenamefont {Crossley}(1984)}]{crossley1984fifteen}%
  \BibitemOpen
  \bibfield  {author} {\bibinfo {author} {\bibfnamefont {R.}~\bibnamefont
  {Crossley}},\ }\href@noop {} {\bibfield  {journal} {\bibinfo  {journal}
  {Physica Scripta}\ }\textbf {\bibinfo {volume} {1984}},\ \bibinfo {pages}
  {117} (\bibinfo {year} {1984})}\BibitemShut {NoStop}%
\bibitem [{\citenamefont {Holmes}\ \emph {et~al.}(2017)\citenamefont {Holmes},
  \citenamefont {Umrigar},\ and\ \citenamefont {Sharma}}]{holmes2017excited}%
  \BibitemOpen
  \bibfield  {author} {\bibinfo {author} {\bibfnamefont {A.~A.}\ \bibnamefont
  {Holmes}}, \bibinfo {author} {\bibfnamefont {C.~J.}\ \bibnamefont {Umrigar}},
  \ and\ \bibinfo {author} {\bibfnamefont {S.}~\bibnamefont {Sharma}},\
  }\href@noop {} {\bibfield  {journal} {\bibinfo  {journal} {The Journal of
  Chemical Physics}\ }\textbf {\bibinfo {volume} {147}} (\bibinfo {year}
  {2017})}\BibitemShut {NoStop}%
\bibitem [{\citenamefont {Martin}(1992)}]{martin1992c2}%
  \BibitemOpen
  \bibfield  {author} {\bibinfo {author} {\bibfnamefont {M.}~\bibnamefont
  {Martin}},\ }\href@noop {} {\bibfield  {journal} {\bibinfo  {journal}
  {Journal of Photochemistry and Photobiology A: Chemistry}\ }\textbf {\bibinfo
  {volume} {66}},\ \bibinfo {pages} {263} (\bibinfo {year} {1992})}\BibitemShut
  {NoStop}%
\bibitem [{\citenamefont {Phillips}\ and\ \citenamefont
  {Davis}(1968)}]{phillips1968swan}%
  \BibitemOpen
  \bibfield  {author} {\bibinfo {author} {\bibfnamefont {J.~G.}\ \bibnamefont
  {Phillips}}\ and\ \bibinfo {author} {\bibfnamefont {S.~P.}\ \bibnamefont
  {Davis}},\ }\href@noop {} {\emph {\bibinfo {title} {{The Swan System of the
  C$_2$ Molecule: The Spectrum of the HgH Molecule}}}},\ Vol.~\bibinfo {volume}
  {2}\ (\bibinfo  {publisher} {Univ of California Press},\ \bibinfo {year}
  {1968})\BibitemShut {NoStop}%
\bibitem [{\citenamefont {Venkataramani}\ \emph {et~al.}(2016)\citenamefont
  {Venkataramani}, \citenamefont {Ghetiya}, \citenamefont {Ganesh},
  \citenamefont {Joshi}, \citenamefont {Agnihotri},\ and\ \citenamefont
  {Baliyan}}]{venkataramani2016optical}%
  \BibitemOpen
  \bibfield  {author} {\bibinfo {author} {\bibfnamefont {K.}~\bibnamefont
  {Venkataramani}}, \bibinfo {author} {\bibfnamefont {S.}~\bibnamefont
  {Ghetiya}}, \bibinfo {author} {\bibfnamefont {S.}~\bibnamefont {Ganesh}},
  \bibinfo {author} {\bibfnamefont {U.}~\bibnamefont {Joshi}}, \bibinfo
  {author} {\bibfnamefont {V.~K.}\ \bibnamefont {Agnihotri}}, \ and\ \bibinfo
  {author} {\bibfnamefont {K.~S.}\ \bibnamefont {Baliyan}},\ }\href@noop {}
  {\bibfield  {journal} {\bibinfo  {journal} {Monthly Notices of the Royal
  Astronomical Society}\ }\textbf {\bibinfo {volume} {463}},\ \bibinfo {pages}
  {2137} (\bibinfo {year} {2016})}\BibitemShut {NoStop}%
\bibitem [{\citenamefont {Shaik}\ \emph {et~al.}(2012)\citenamefont {Shaik},
  \citenamefont {Danovich}, \citenamefont {Wu}, \citenamefont {Su},
  \citenamefont {Rzepa},\ and\ \citenamefont {Hiberty}}]{shaik2012quadruple}%
  \BibitemOpen
  \bibfield  {author} {\bibinfo {author} {\bibfnamefont {S.}~\bibnamefont
  {Shaik}}, \bibinfo {author} {\bibfnamefont {D.}~\bibnamefont {Danovich}},
  \bibinfo {author} {\bibfnamefont {W.}~\bibnamefont {Wu}}, \bibinfo {author}
  {\bibfnamefont {P.}~\bibnamefont {Su}}, \bibinfo {author} {\bibfnamefont
  {H.~S.}\ \bibnamefont {Rzepa}}, \ and\ \bibinfo {author} {\bibfnamefont
  {P.~C.}\ \bibnamefont {Hiberty}},\ }\href@noop {} {\bibfield  {journal}
  {\bibinfo  {journal} {Nature Chemistry}\ }\textbf {\bibinfo {volume} {4}},\
  \bibinfo {pages} {195} (\bibinfo {year} {2012})}\BibitemShut {NoStop}%
\bibitem [{\citenamefont {Ballik}\ and\ \citenamefont
  {Ramsay}(1963{\natexlab{a}})}]{ballik19633}%
  \BibitemOpen
  \bibfield  {author} {\bibinfo {author} {\bibfnamefont {E.}~\bibnamefont
  {Ballik}}\ and\ \bibinfo {author} {\bibfnamefont {D.}~\bibnamefont
  {Ramsay}},\ }\href@noop {} {\bibfield  {journal} {\bibinfo  {journal} {The
  Astrophysical Journal}\ }\textbf {\bibinfo {volume} {137}},\ \bibinfo {pages}
  {61} (\bibinfo {year} {1963}{\natexlab{a}})}\BibitemShut {NoStop}%
\bibitem [{\citenamefont {Ballik}\ and\ \citenamefont
  {Ramsay}(1963{\natexlab{b}})}]{ballik1963extension}%
  \BibitemOpen
  \bibfield  {author} {\bibinfo {author} {\bibfnamefont {E.}~\bibnamefont
  {Ballik}}\ and\ \bibinfo {author} {\bibfnamefont {D.}~\bibnamefont
  {Ramsay}},\ }\href@noop {} {\bibfield  {journal} {\bibinfo  {journal} {The
  Astrophysical Journal}\ }\textbf {\bibinfo {volume} {137}},\ \bibinfo {pages}
  {84} (\bibinfo {year} {1963}{\natexlab{b}})}\BibitemShut {NoStop}%
\bibitem [{\citenamefont {Mališ}\ and\ \citenamefont
  {Luber}(2020)}]{malis2020trajectory}%
  \BibitemOpen
  \bibfield  {author} {\bibinfo {author} {\bibfnamefont {M.}~\bibnamefont
  {Mališ}}\ and\ \bibinfo {author} {\bibfnamefont {S.}~\bibnamefont {Luber}},\
  }\href@noop {} {\bibfield  {journal} {\bibinfo  {journal} {Journal of
  Chemical Theory and Computation}\ }\textbf {\bibinfo {volume} {16}},\
  \bibinfo {pages} {4071} (\bibinfo {year} {2020})}\BibitemShut {NoStop}%
\bibitem [{\citenamefont {Barbatti}\ \emph {et~al.}(2004)\citenamefont
  {Barbatti}, \citenamefont {Paier},\ and\ \citenamefont
  {Lischka}}]{barbatti2004photochemistry}%
  \BibitemOpen
  \bibfield  {author} {\bibinfo {author} {\bibfnamefont {M.}~\bibnamefont
  {Barbatti}}, \bibinfo {author} {\bibfnamefont {J.}~\bibnamefont {Paier}}, \
  and\ \bibinfo {author} {\bibfnamefont {H.}~\bibnamefont {Lischka}},\
  }\href@noop {} {\bibfield  {journal} {\bibinfo  {journal} {The Journal of
  Chemical Physics}\ }\textbf {\bibinfo {volume} {121}},\ \bibinfo {pages}
  {11614} (\bibinfo {year} {2004})}\BibitemShut {NoStop}%
\bibitem [{\citenamefont {Krylov}\ \emph {et~al.}(1998)\citenamefont {Krylov},
  \citenamefont {Sherrill}, \citenamefont {Byrd},\ and\ \citenamefont
  {Head-Gordon}}]{krylov1998size}%
  \BibitemOpen
  \bibfield  {author} {\bibinfo {author} {\bibfnamefont {A.~I.}\ \bibnamefont
  {Krylov}}, \bibinfo {author} {\bibfnamefont {C.~D.}\ \bibnamefont
  {Sherrill}}, \bibinfo {author} {\bibfnamefont {E.~F.}\ \bibnamefont {Byrd}},
  \ and\ \bibinfo {author} {\bibfnamefont {M.}~\bibnamefont {Head-Gordon}},\
  }\href@noop {} {\bibfield  {journal} {\bibinfo  {journal} {The Journal of
  Chemical Physics}\ }\textbf {\bibinfo {volume} {109}},\ \bibinfo {pages}
  {10669} (\bibinfo {year} {1998})}\BibitemShut {NoStop}%
\bibitem [{\citenamefont {Barbatti}\ and\ \citenamefont
  {Crespo-Otero}(2016)}]{barbatti2016surface}%
  \BibitemOpen
  \bibfield  {author} {\bibinfo {author} {\bibfnamefont {M.}~\bibnamefont
  {Barbatti}}\ and\ \bibinfo {author} {\bibfnamefont {R.}~\bibnamefont
  {Crespo-Otero}},\ }\href@noop {} {\bibfield  {journal} {\bibinfo  {journal}
  {Density-Functional Methods for Excited States}\ ,\ \bibinfo {pages} {415}}
  (\bibinfo {year} {2016})}\BibitemShut {NoStop}%
\bibitem [{\citenamefont {Shepard}\ \emph {et~al.}(2022)\citenamefont
  {Shepard}, \citenamefont {Panad{\'e}s-Barrueta}, \citenamefont {Moroni},
  \citenamefont {Scemama},\ and\ \citenamefont {Filippi}}]{shepard2022double}%
  \BibitemOpen
  \bibfield  {author} {\bibinfo {author} {\bibfnamefont {S.}~\bibnamefont
  {Shepard}}, \bibinfo {author} {\bibfnamefont {R.~L.}\ \bibnamefont
  {Panad{\'e}s-Barrueta}}, \bibinfo {author} {\bibfnamefont {S.}~\bibnamefont
  {Moroni}}, \bibinfo {author} {\bibfnamefont {A.}~\bibnamefont {Scemama}}, \
  and\ \bibinfo {author} {\bibfnamefont {C.}~\bibnamefont {Filippi}},\
  }\href@noop {} {\bibfield  {journal} {\bibinfo  {journal} {Journal of
  Chemical Theory and Computation}\ }\textbf {\bibinfo {volume} {18}},\
  \bibinfo {pages} {6722} (\bibinfo {year} {2022})}\BibitemShut {NoStop}%
\bibitem [{\citenamefont {Kossoski}\ \emph {et~al.}(2024)\citenamefont
  {Kossoski}, \citenamefont {Boggio-Pasqua}, \citenamefont {Loos},\ and\
  \citenamefont {Jacquemin}}]{kossoski2024reference}%
  \BibitemOpen
  \bibfield  {author} {\bibinfo {author} {\bibfnamefont {F.}~\bibnamefont
  {Kossoski}}, \bibinfo {author} {\bibfnamefont {M.}~\bibnamefont
  {Boggio-Pasqua}}, \bibinfo {author} {\bibfnamefont {P.-F.}\ \bibnamefont
  {Loos}}, \ and\ \bibinfo {author} {\bibfnamefont {D.}~\bibnamefont
  {Jacquemin}},\ }\href@noop {} {\bibfield  {journal} {\bibinfo  {journal}
  {arXiv preprint arXiv:2403.19597}\ } (\bibinfo {year} {2024})}\BibitemShut
  {NoStop}%
\bibitem [{\citenamefont {Buenker}\ and\ \citenamefont
  {Whitten}(1968)}]{buenker1968ab}%
  \BibitemOpen
  \bibfield  {author} {\bibinfo {author} {\bibfnamefont {R.~J.}\ \bibnamefont
  {Buenker}}\ and\ \bibinfo {author} {\bibfnamefont {J.~L.}\ \bibnamefont
  {Whitten}},\ }\href@noop {} {\bibfield  {journal} {\bibinfo  {journal} {The
  Journal of Chemical Physics}\ }\textbf {\bibinfo {volume} {49}},\ \bibinfo
  {pages} {5381} (\bibinfo {year} {1968})}\BibitemShut {NoStop}%
\bibitem [{\citenamefont {Watson}\ and\ \citenamefont
  {Chan}(2012)}]{watson2012excited}%
  \BibitemOpen
  \bibfield  {author} {\bibinfo {author} {\bibfnamefont {M.~A.}\ \bibnamefont
  {Watson}}\ and\ \bibinfo {author} {\bibfnamefont {G.~K.-L.}\ \bibnamefont
  {Chan}},\ }\href@noop {} {\bibfield  {journal} {\bibinfo  {journal} {Journal
  of Chemical Theory and Computation}\ }\textbf {\bibinfo {volume} {8}},\
  \bibinfo {pages} {4013} (\bibinfo {year} {2012})}\BibitemShut {NoStop}%
\bibitem [{\citenamefont {Adamo}\ \emph {et~al.}(1999)\citenamefont {Adamo},
  \citenamefont {Scuseria},\ and\ \citenamefont {Barone}}]{adamo1999accurate}%
  \BibitemOpen
  \bibfield  {author} {\bibinfo {author} {\bibfnamefont {C.}~\bibnamefont
  {Adamo}}, \bibinfo {author} {\bibfnamefont {G.~E.}\ \bibnamefont {Scuseria}},
  \ and\ \bibinfo {author} {\bibfnamefont {V.}~\bibnamefont {Barone}},\
  }\href@noop {} {\bibfield  {journal} {\bibinfo  {journal} {The Journal of
  chemical physics}\ }\textbf {\bibinfo {volume} {111}},\ \bibinfo {pages}
  {2889} (\bibinfo {year} {1999})}\BibitemShut {NoStop}%
\bibitem [{\citenamefont {Roos}\ \emph {et~al.}(1992)\citenamefont {Roos},
  \citenamefont {Andersson},\ and\ \citenamefont
  {F{\"u}lscher}}]{roos1992towards}%
  \BibitemOpen
  \bibfield  {author} {\bibinfo {author} {\bibfnamefont {B.~O.}\ \bibnamefont
  {Roos}}, \bibinfo {author} {\bibfnamefont {K.}~\bibnamefont {Andersson}}, \
  and\ \bibinfo {author} {\bibfnamefont {M.~P.}\ \bibnamefont {F{\"u}lscher}},\
  }\href@noop {} {\bibfield  {journal} {\bibinfo  {journal} {Chemical Physics
  Letters}\ }\textbf {\bibinfo {volume} {192}},\ \bibinfo {pages} {5} (\bibinfo
  {year} {1992})}\BibitemShut {NoStop}%
\bibitem [{\citenamefont {Cuzzocrea}\ \emph {et~al.}(2022)\citenamefont
  {Cuzzocrea}, \citenamefont {Moroni}, \citenamefont {Scemama},\ and\
  \citenamefont {Filippi}}]{cuzzocrea2022reference}%
  \BibitemOpen
  \bibfield  {author} {\bibinfo {author} {\bibfnamefont {A.}~\bibnamefont
  {Cuzzocrea}}, \bibinfo {author} {\bibfnamefont {S.}~\bibnamefont {Moroni}},
  \bibinfo {author} {\bibfnamefont {A.}~\bibnamefont {Scemama}}, \ and\
  \bibinfo {author} {\bibfnamefont {C.}~\bibnamefont {Filippi}},\ }\href@noop
  {} {\bibfield  {journal} {\bibinfo  {journal} {Journal of Chemical Theory and
  Computation}\ }\textbf {\bibinfo {volume} {18}},\ \bibinfo {pages} {1089}
  (\bibinfo {year} {2022})}\BibitemShut {NoStop}%
\bibitem [{\citenamefont {Azumi}\ and\ \citenamefont
  {Matsuzaki}(1977)}]{azumi1977does}%
  \BibitemOpen
  \bibfield  {author} {\bibinfo {author} {\bibfnamefont {T.}~\bibnamefont
  {Azumi}}\ and\ \bibinfo {author} {\bibfnamefont {K.}~\bibnamefont
  {Matsuzaki}},\ }\href@noop {} {\bibfield  {journal} {\bibinfo  {journal}
  {Photochemistry and Photobiology}\ }\textbf {\bibinfo {volume} {25}},\
  \bibinfo {pages} {315} (\bibinfo {year} {1977})}\BibitemShut {NoStop}%
\bibitem [{\citenamefont {Gao}\ and\ \citenamefont
  {G{\"u}nnemann}(2022)}]{gao2022ab}%
  \BibitemOpen
  \bibfield  {author} {\bibinfo {author} {\bibfnamefont {N.}~\bibnamefont
  {Gao}}\ and\ \bibinfo {author} {\bibfnamefont {S.}~\bibnamefont
  {G{\"u}nnemann}},\ }\href@noop {} {\bibfield  {journal} {\bibinfo  {journal}
  {Tenth International Conference on Learning Representations (ICLR)}\ }
  (\bibinfo {year} {2022})}\BibitemShut {NoStop}%
\bibitem [{\citenamefont {Scherbela}\ \emph {et~al.}(2024)\citenamefont
  {Scherbela}, \citenamefont {Gerard},\ and\ \citenamefont
  {Grohs}}]{scherbela2024towards}%
  \BibitemOpen
  \bibfield  {author} {\bibinfo {author} {\bibfnamefont {M.}~\bibnamefont
  {Scherbela}}, \bibinfo {author} {\bibfnamefont {L.}~\bibnamefont {Gerard}}, \
  and\ \bibinfo {author} {\bibfnamefont {P.}~\bibnamefont {Grohs}},\
  }\href@noop {} {\bibfield  {journal} {\bibinfo  {journal} {Nature
  Communications}\ }\textbf {\bibinfo {volume} {15}},\ \bibinfo {pages} {120}
  (\bibinfo {year} {2024})}\BibitemShut {NoStop}%
\bibitem [{\citenamefont {Hunt}\ \emph {et~al.}(2018)\citenamefont {Hunt},
  \citenamefont {Szyniszewski}, \citenamefont {Prayogo}, \citenamefont
  {Maezono},\ and\ \citenamefont {Drummond}}]{hunt2018quantum}%
  \BibitemOpen
  \bibfield  {author} {\bibinfo {author} {\bibfnamefont {R.~J.}\ \bibnamefont
  {Hunt}}, \bibinfo {author} {\bibfnamefont {M.}~\bibnamefont {Szyniszewski}},
  \bibinfo {author} {\bibfnamefont {G.~I.}\ \bibnamefont {Prayogo}}, \bibinfo
  {author} {\bibfnamefont {R.}~\bibnamefont {Maezono}}, \ and\ \bibinfo
  {author} {\bibfnamefont {N.~D.}\ \bibnamefont {Drummond}},\ }\href@noop {}
  {\bibfield  {journal} {\bibinfo  {journal} {Physical Review B}\ }\textbf
  {\bibinfo {volume} {98}},\ \bibinfo {pages} {075122} (\bibinfo {year}
  {2018})}\BibitemShut {NoStop}%
\bibitem [{\citenamefont {Yang}\ and\ \citenamefont
  {Zhao}(2023)}]{yang2023deep}%
  \BibitemOpen
  \bibfield  {author} {\bibinfo {author} {\bibfnamefont {Y.}~\bibnamefont
  {Yang}}\ and\ \bibinfo {author} {\bibfnamefont {P.}~\bibnamefont {Zhao}},\
  }\href@noop {} {\bibfield  {journal} {\bibinfo  {journal} {Physical Review
  C}\ }\textbf {\bibinfo {volume} {107}},\ \bibinfo {pages} {034320} (\bibinfo
  {year} {2023})}\BibitemShut {NoStop}%
\bibitem [{\citenamefont {Li}\ \emph {et~al.}(2022{\natexlab{a}})\citenamefont
  {Li}, \citenamefont {Li},\ and\ \citenamefont {Chen}}]{li2022ab}%
  \BibitemOpen
  \bibfield  {author} {\bibinfo {author} {\bibfnamefont {X.}~\bibnamefont
  {Li}}, \bibinfo {author} {\bibfnamefont {Z.}~\bibnamefont {Li}}, \ and\
  \bibinfo {author} {\bibfnamefont {J.}~\bibnamefont {Chen}},\ }\href@noop {}
  {\bibfield  {journal} {\bibinfo  {journal} {Nature Communications}\ }\textbf
  {\bibinfo {volume} {13}},\ \bibinfo {pages} {7895} (\bibinfo {year}
  {2022}{\natexlab{a}})}\BibitemShut {NoStop}%
\bibitem [{\citenamefont {Motta}\ and\ \citenamefont
  {Zhang}(2018)}]{motta2018ab}%
  \BibitemOpen
  \bibfield  {author} {\bibinfo {author} {\bibfnamefont {M.}~\bibnamefont
  {Motta}}\ and\ \bibinfo {author} {\bibfnamefont {S.}~\bibnamefont {Zhang}},\
  }\href@noop {} {\bibfield  {journal} {\bibinfo  {journal} {Wiley
  Interdisciplinary Reviews: Computational Molecular Science}\ }\textbf
  {\bibinfo {volume} {8}},\ \bibinfo {pages} {e1364} (\bibinfo {year}
  {2018})}\BibitemShut {NoStop}%
\bibitem [{\citenamefont {Stokes}\ \emph {et~al.}(2020)\citenamefont {Stokes},
  \citenamefont {Moreno}, \citenamefont {Pnevmatikakis},\ and\ \citenamefont
  {Carleo}}]{stokes2020phases}%
  \BibitemOpen
  \bibfield  {author} {\bibinfo {author} {\bibfnamefont {J.}~\bibnamefont
  {Stokes}}, \bibinfo {author} {\bibfnamefont {J.~R.}\ \bibnamefont {Moreno}},
  \bibinfo {author} {\bibfnamefont {E.~A.}\ \bibnamefont {Pnevmatikakis}}, \
  and\ \bibinfo {author} {\bibfnamefont {G.}~\bibnamefont {Carleo}},\
  }\href@noop {} {\bibfield  {journal} {\bibinfo  {journal} {Physical Review
  B}\ }\textbf {\bibinfo {volume} {102}},\ \bibinfo {pages} {205122} (\bibinfo
  {year} {2020})}\BibitemShut {NoStop}%
\bibitem [{\citenamefont {Sharir}\ \emph {et~al.}(2020)\citenamefont {Sharir},
  \citenamefont {Levine}, \citenamefont {Wies}, \citenamefont {Carleo},\ and\
  \citenamefont {Shashua}}]{sharir2020deep}%
  \BibitemOpen
  \bibfield  {author} {\bibinfo {author} {\bibfnamefont {O.}~\bibnamefont
  {Sharir}}, \bibinfo {author} {\bibfnamefont {Y.}~\bibnamefont {Levine}},
  \bibinfo {author} {\bibfnamefont {N.}~\bibnamefont {Wies}}, \bibinfo {author}
  {\bibfnamefont {G.}~\bibnamefont {Carleo}}, \ and\ \bibinfo {author}
  {\bibfnamefont {A.}~\bibnamefont {Shashua}},\ }\href@noop {} {\bibfield
  {journal} {\bibinfo  {journal} {Physical review letters}\ }\textbf {\bibinfo
  {volume} {124}},\ \bibinfo {pages} {020503} (\bibinfo {year}
  {2020})}\BibitemShut {NoStop}%
\bibitem [{\citenamefont {Gerard}\ \emph {et~al.}(2022)\citenamefont {Gerard},
  \citenamefont {Scherbela}, \citenamefont {Marquetand},\ and\ \citenamefont
  {Grohs}}]{gerard2022gold}%
  \BibitemOpen
  \bibfield  {author} {\bibinfo {author} {\bibfnamefont {L.}~\bibnamefont
  {Gerard}}, \bibinfo {author} {\bibfnamefont {M.}~\bibnamefont {Scherbela}},
  \bibinfo {author} {\bibfnamefont {P.}~\bibnamefont {Marquetand}}, \ and\
  \bibinfo {author} {\bibfnamefont {P.}~\bibnamefont {Grohs}},\ }\href@noop {}
  {\bibfield  {journal} {\bibinfo  {journal} {Advances in Neural Information
  Processing Systems (NeurIPS)}\ }\textbf {\bibinfo {volume} {35}},\ \bibinfo
  {pages} {10282} (\bibinfo {year} {2022})}\BibitemShut {NoStop}%
\bibitem [{\citenamefont {Gao}\ and\ \citenamefont
  {G{\"u}nnemann}(2023)}]{gao2023generalizing}%
  \BibitemOpen
  \bibfield  {author} {\bibinfo {author} {\bibfnamefont {N.}~\bibnamefont
  {Gao}}\ and\ \bibinfo {author} {\bibfnamefont {S.}~\bibnamefont
  {G{\"u}nnemann}},\ }\href@noop {} {\bibfield  {journal} {\bibinfo  {journal}
  {arXiv preprint arXiv:2302.04168}\ } (\bibinfo {year} {2023})}\BibitemShut
  {NoStop}%
\bibitem [{\citenamefont {Pescia}\ \emph {et~al.}(2023)\citenamefont {Pescia},
  \citenamefont {Nys}, \citenamefont {Kim}, \citenamefont {Lovato},\ and\
  \citenamefont {Carleo}}]{pescia2023message}%
  \BibitemOpen
  \bibfield  {author} {\bibinfo {author} {\bibfnamefont {G.}~\bibnamefont
  {Pescia}}, \bibinfo {author} {\bibfnamefont {J.}~\bibnamefont {Nys}},
  \bibinfo {author} {\bibfnamefont {J.}~\bibnamefont {Kim}}, \bibinfo {author}
  {\bibfnamefont {A.}~\bibnamefont {Lovato}}, \ and\ \bibinfo {author}
  {\bibfnamefont {G.}~\bibnamefont {Carleo}},\ }\href@noop {} {\bibfield
  {journal} {\bibinfo  {journal} {arXiv preprint arXiv:2305.07240}\ } (\bibinfo
  {year} {2023})}\BibitemShut {NoStop}%
\bibitem [{\citenamefont {Li}\ \emph {et~al.}(2024)\citenamefont {Li},
  \citenamefont {Ye}, \citenamefont {Jiang}, \citenamefont {Wen}, \citenamefont
  {Wang}, \citenamefont {Li}, \citenamefont {Li}, \citenamefont {He},
  \citenamefont {Chen}, \citenamefont {Ren},\ and\ \citenamefont
  {Wang}}]{li2024forward}%
  \BibitemOpen
  \bibfield  {author} {\bibinfo {author} {\bibfnamefont {R.}~\bibnamefont
  {Li}}, \bibinfo {author} {\bibfnamefont {H.}~\bibnamefont {Ye}}, \bibinfo
  {author} {\bibfnamefont {D.}~\bibnamefont {Jiang}}, \bibinfo {author}
  {\bibfnamefont {X.}~\bibnamefont {Wen}}, \bibinfo {author} {\bibfnamefont
  {C.}~\bibnamefont {Wang}}, \bibinfo {author} {\bibfnamefont {Z.}~\bibnamefont
  {Li}}, \bibinfo {author} {\bibfnamefont {X.}~\bibnamefont {Li}}, \bibinfo
  {author} {\bibfnamefont {D.}~\bibnamefont {He}}, \bibinfo {author}
  {\bibfnamefont {J.}~\bibnamefont {Chen}}, \bibinfo {author} {\bibfnamefont
  {W.}~\bibnamefont {Ren}}, \ and\ \bibinfo {author} {\bibfnamefont
  {L.}~\bibnamefont {Wang}},\ }\href@noop {} {\bibfield  {journal} {\bibinfo
  {journal} {Nature Machine Intelligence}\ }\textbf {\bibinfo {volume} {6}},\
  \bibinfo {pages} {209} (\bibinfo {year} {2024})}\BibitemShut {NoStop}%
\bibitem [{\citenamefont {Wigner}\ and\ \citenamefont
  {Seitz}(1934)}]{wigner1934constitution}%
  \BibitemOpen
  \bibfield  {author} {\bibinfo {author} {\bibfnamefont {E.}~\bibnamefont
  {Wigner}}\ and\ \bibinfo {author} {\bibfnamefont {F.}~\bibnamefont {Seitz}},\
  }\href@noop {} {\bibfield  {journal} {\bibinfo  {journal} {Physical Review}\
  }\textbf {\bibinfo {volume} {46}},\ \bibinfo {pages} {509} (\bibinfo {year}
  {1934})}\BibitemShut {NoStop}%
\bibitem [{\citenamefont {Spencer}\ \emph {et~al.}(2020)\citenamefont
  {Spencer}, \citenamefont {Pfau}, \citenamefont {Botev},\ and\ \citenamefont
  {Foulkes}}]{spencer2020better}%
  \BibitemOpen
  \bibfield  {author} {\bibinfo {author} {\bibfnamefont {J.~S.}\ \bibnamefont
  {Spencer}}, \bibinfo {author} {\bibfnamefont {D.}~\bibnamefont {Pfau}},
  \bibinfo {author} {\bibfnamefont {A.}~\bibnamefont {Botev}}, \ and\ \bibinfo
  {author} {\bibfnamefont {W.~M.~C.}\ \bibnamefont {Foulkes}},\ }\href@noop {}
  {\bibfield  {journal} {\bibinfo  {journal} {arXiv preprint arXiv:2011.07125}\
  } (\bibinfo {year} {2020})}\BibitemShut {NoStop}%
\bibitem [{\citenamefont {Vaswani}\ \emph {et~al.}(2017)\citenamefont
  {Vaswani}, \citenamefont {Shazeer}, \citenamefont {Parmar}, \citenamefont
  {Uszkoreit}, \citenamefont {Jones}, \citenamefont {Gomez}, \citenamefont
  {Kaiser},\ and\ \citenamefont {Polosukhin}}]{vaswani2017attention}%
  \BibitemOpen
  \bibfield  {author} {\bibinfo {author} {\bibfnamefont {A.}~\bibnamefont
  {Vaswani}}, \bibinfo {author} {\bibfnamefont {N.}~\bibnamefont {Shazeer}},
  \bibinfo {author} {\bibfnamefont {N.}~\bibnamefont {Parmar}}, \bibinfo
  {author} {\bibfnamefont {J.}~\bibnamefont {Uszkoreit}}, \bibinfo {author}
  {\bibfnamefont {L.}~\bibnamefont {Jones}}, \bibinfo {author} {\bibfnamefont
  {A.~N.}\ \bibnamefont {Gomez}}, \bibinfo {author} {\bibfnamefont
  {{\L}.}~\bibnamefont {Kaiser}}, \ and\ \bibinfo {author} {\bibfnamefont
  {I.}~\bibnamefont {Polosukhin}},\ }\href@noop {} {\bibfield  {journal}
  {\bibinfo  {journal} {Advances in Neural Information Processing Systems
  (NeurIPS)}\ }\textbf {\bibinfo {volume} {30}} (\bibinfo {year}
  {2017})}\BibitemShut {NoStop}%
\bibitem [{\citenamefont {Ba}\ \emph {et~al.}(2016)\citenamefont {Ba},
  \citenamefont {Kiros},\ and\ \citenamefont {Hinton}}]{ba2016layer}%
  \BibitemOpen
  \bibfield  {author} {\bibinfo {author} {\bibfnamefont {J.~L.}\ \bibnamefont
  {Ba}}, \bibinfo {author} {\bibfnamefont {J.~R.}\ \bibnamefont {Kiros}}, \
  and\ \bibinfo {author} {\bibfnamefont {G.~E.}\ \bibnamefont {Hinton}},\
  }\href@noop {} {\bibfield  {journal} {\bibinfo  {journal} {arXiv preprint
  arXiv:1607.06450}\ } (\bibinfo {year} {2016})}\BibitemShut {NoStop}%
\bibitem [{\citenamefont {He}\ \emph {et~al.}(2016)\citenamefont {He},
  \citenamefont {Zhang}, \citenamefont {Ren},\ and\ \citenamefont
  {Sun}}]{he2016deep}%
  \BibitemOpen
  \bibfield  {author} {\bibinfo {author} {\bibfnamefont {K.}~\bibnamefont
  {He}}, \bibinfo {author} {\bibfnamefont {X.}~\bibnamefont {Zhang}}, \bibinfo
  {author} {\bibfnamefont {S.}~\bibnamefont {Ren}}, \ and\ \bibinfo {author}
  {\bibfnamefont {J.}~\bibnamefont {Sun}},\ }in\ \href@noop {} {\emph {\bibinfo
  {booktitle} {Proceedings of the IEEE Conference on Computer Vision and
  Pattern Recognition (CVPR)}}}\ (\bibinfo {year} {2016})\ pp.\ \bibinfo
  {pages} {770--778}\BibitemShut {NoStop}%
\bibitem [{\citenamefont {Weser}\ \emph {et~al.}(2022)\citenamefont {Weser},
  \citenamefont {Liebermann}, \citenamefont {Kats}, \citenamefont {Alavi},\
  and\ \citenamefont {Li~Manni}}]{weser2022spin}%
  \BibitemOpen
  \bibfield  {author} {\bibinfo {author} {\bibfnamefont {O.}~\bibnamefont
  {Weser}}, \bibinfo {author} {\bibfnamefont {N.}~\bibnamefont {Liebermann}},
  \bibinfo {author} {\bibfnamefont {D.}~\bibnamefont {Kats}}, \bibinfo {author}
  {\bibfnamefont {A.}~\bibnamefont {Alavi}}, \ and\ \bibinfo {author}
  {\bibfnamefont {G.}~\bibnamefont {Li~Manni}},\ }\href@noop {} {\bibfield
  {journal} {\bibinfo  {journal} {The Journal of Physical Chemistry A}\
  }\textbf {\bibinfo {volume} {126}},\ \bibinfo {pages} {2050} (\bibinfo {year}
  {2022})}\BibitemShut {NoStop}%
\bibitem [{\citenamefont {Sun}\ \emph {et~al.}(2018)\citenamefont {Sun},
  \citenamefont {Berkelbach}, \citenamefont {Blunt}, \citenamefont {Booth},
  \citenamefont {Guo}, \citenamefont {Li}, \citenamefont {Liu}, \citenamefont
  {McClain}, \citenamefont {Sayfutyarova}, \citenamefont {Sharma},
  \citenamefont {Wouters},\ and\ \citenamefont {Chan}}]{sun2018pyscf}%
  \BibitemOpen
  \bibfield  {author} {\bibinfo {author} {\bibfnamefont {Q.}~\bibnamefont
  {Sun}}, \bibinfo {author} {\bibfnamefont {T.~C.}\ \bibnamefont {Berkelbach}},
  \bibinfo {author} {\bibfnamefont {N.~S.}\ \bibnamefont {Blunt}}, \bibinfo
  {author} {\bibfnamefont {G.~H.}\ \bibnamefont {Booth}}, \bibinfo {author}
  {\bibfnamefont {S.}~\bibnamefont {Guo}}, \bibinfo {author} {\bibfnamefont
  {Z.}~\bibnamefont {Li}}, \bibinfo {author} {\bibfnamefont {J.}~\bibnamefont
  {Liu}}, \bibinfo {author} {\bibfnamefont {J.~D.}\ \bibnamefont {McClain}},
  \bibinfo {author} {\bibfnamefont {E.~R.}\ \bibnamefont {Sayfutyarova}},
  \bibinfo {author} {\bibfnamefont {S.}~\bibnamefont {Sharma}}, \bibinfo
  {author} {\bibfnamefont {S.}~\bibnamefont {Wouters}}, \ and\ \bibinfo
  {author} {\bibfnamefont {G.~K.-L.}\ \bibnamefont {Chan}},\ }\href@noop {}
  {\bibfield  {journal} {\bibinfo  {journal} {Wiley Interdisciplinary Reviews:
  Computational Molecular Science}\ }\textbf {\bibinfo {volume} {8}},\ \bibinfo
  {pages} {e1340} (\bibinfo {year} {2018})}\BibitemShut {NoStop}%
\bibitem [{\citenamefont {Fahy}\ \emph {et~al.}(1990)\citenamefont {Fahy},
  \citenamefont {Wang},\ and\ \citenamefont {Louie}}]{fahy1990variational}%
  \BibitemOpen
  \bibfield  {author} {\bibinfo {author} {\bibfnamefont {S.}~\bibnamefont
  {Fahy}}, \bibinfo {author} {\bibfnamefont {X.}~\bibnamefont {Wang}}, \ and\
  \bibinfo {author} {\bibfnamefont {S.~G.}\ \bibnamefont {Louie}},\ }\href@noop
  {} {\bibfield  {journal} {\bibinfo  {journal} {Physical Review B}\ }\textbf
  {\bibinfo {volume} {42}},\ \bibinfo {pages} {3503} (\bibinfo {year}
  {1990})}\BibitemShut {NoStop}%
\bibitem [{\citenamefont {Bennett}\ \emph {et~al.}(2018)\citenamefont
  {Bennett}, \citenamefont {Wang}, \citenamefont {Annaberdiyev}, \citenamefont
  {Melton}, \citenamefont {Shulenburger},\ and\ \citenamefont
  {Mitas}}]{bennett2018new}%
  \BibitemOpen
  \bibfield  {author} {\bibinfo {author} {\bibfnamefont {M.~C.}\ \bibnamefont
  {Bennett}}, \bibinfo {author} {\bibfnamefont {G.}~\bibnamefont {Wang}},
  \bibinfo {author} {\bibfnamefont {A.}~\bibnamefont {Annaberdiyev}}, \bibinfo
  {author} {\bibfnamefont {C.~A.}\ \bibnamefont {Melton}}, \bibinfo {author}
  {\bibfnamefont {L.}~\bibnamefont {Shulenburger}}, \ and\ \bibinfo {author}
  {\bibfnamefont {L.}~\bibnamefont {Mitas}},\ }\href@noop {} {\bibfield
  {journal} {\bibinfo  {journal} {The Journal of Chemical Physics}\ }\textbf
  {\bibinfo {volume} {149}} (\bibinfo {year} {2018})}\BibitemShut {NoStop}%
\bibitem [{\citenamefont {Li}\ \emph {et~al.}(2022{\natexlab{b}})\citenamefont
  {Li}, \citenamefont {Fan}, \citenamefont {Ren},\ and\ \citenamefont
  {Chen}}]{li2022fermionic}%
  \BibitemOpen
  \bibfield  {author} {\bibinfo {author} {\bibfnamefont {X.}~\bibnamefont
  {Li}}, \bibinfo {author} {\bibfnamefont {C.}~\bibnamefont {Fan}}, \bibinfo
  {author} {\bibfnamefont {W.}~\bibnamefont {Ren}}, \ and\ \bibinfo {author}
  {\bibfnamefont {J.}~\bibnamefont {Chen}},\ }\href@noop {} {\bibfield
  {journal} {\bibinfo  {journal} {Physical Review Research}\ }\textbf {\bibinfo
  {volume} {4}},\ \bibinfo {pages} {013021} (\bibinfo {year}
  {2022}{\natexlab{b}})}\BibitemShut {NoStop}%
\bibitem [{\citenamefont {Gao}\ and\ \citenamefont
  {K{\"o}hler}(2023)}]{gao2023folx}%
  \BibitemOpen
  \bibfield  {author} {\bibinfo {author} {\bibfnamefont {N.}~\bibnamefont
  {Gao}}\ and\ \bibinfo {author} {\bibfnamefont {J.}~\bibnamefont
  {K{\"o}hler}},\ }\href@noop {} {\enquote {\bibinfo {title} {{Folx -- Forward
  Laplacian for JAX}},}\ }\bibinfo {howpublished}
  {\url{https://github.com/microsoft/folx}} (\bibinfo {year}
  {2023})\BibitemShut {NoStop}%
\bibitem [{\citenamefont {L{\"o}wdin}(1955)}]{lowdin1955s2}%
  \BibitemOpen
  \bibfield  {author} {\bibinfo {author} {\bibfnamefont {P.-O.}\ \bibnamefont
  {L{\"o}wdin}},\ }\href@noop {} {\bibfield  {journal} {\bibinfo  {journal}
  {Physical Review}\ }\textbf {\bibinfo {volume} {97}},\ \bibinfo {pages}
  {1474} (\bibinfo {year} {1955})}\BibitemShut {NoStop}%
\bibitem [{\citenamefont {Wang}\ \emph {et~al.}(1995)\citenamefont {Wang},
  \citenamefont {Becke},\ and\ \citenamefont {Smith}}]{wang1995s2}%
  \BibitemOpen
  \bibfield  {author} {\bibinfo {author} {\bibfnamefont {J.}~\bibnamefont
  {Wang}}, \bibinfo {author} {\bibfnamefont {A.~D.}\ \bibnamefont {Becke}}, \
  and\ \bibinfo {author} {\bibfnamefont {V.~H.}\ \bibnamefont {Smith}},\
  }\href@noop {} {\bibfield  {journal} {\bibinfo  {journal} {Journal of
  Chemical Physics}\ }\textbf {\bibinfo {volume} {102}},\ \bibinfo {pages}
  {3477} (\bibinfo {year} {1995})}\BibitemShut {NoStop}%
\bibitem [{\citenamefont {Lewart}\ \emph {et~al.}(1988)\citenamefont {Lewart},
  \citenamefont {Pandharipande},\ and\ \citenamefont
  {Pieper}}]{lewart1988single}%
  \BibitemOpen
  \bibfield  {author} {\bibinfo {author} {\bibfnamefont {D.}~\bibnamefont
  {Lewart}}, \bibinfo {author} {\bibfnamefont {V.}~\bibnamefont
  {Pandharipande}}, \ and\ \bibinfo {author} {\bibfnamefont {S.~C.}\
  \bibnamefont {Pieper}},\ }\href@noop {} {\bibfield  {journal} {\bibinfo
  {journal} {Physical Review B}\ }\textbf {\bibinfo {volume} {37}},\ \bibinfo
  {pages} {4950} (\bibinfo {year} {1988})}\BibitemShut {NoStop}%
\end{thebibliography}%

\section*{Acknowledgments}
The authors would like to thank Matthew Foulkes, Denis Jacquemin, Michael Bearpark, Aron Cohen and Alex Gaunt for helpful discussions, Nicholas Gao for help with Folx, and James Kirkpatrick, Annette Obika, Ali Eslami, Shakir Mohamed, Danilo Rezende, Ekin Dogus Cubuk, Pushmeet Kohli and Demis Hassabis for support.

\subsection*{Funding}
DP, IvG and JSS are all supported by Google DeepMind. SA was supported by a DeepMind internship during this project. HS is supported by the Aker Scholarship. HS gratefully acknowledges the Gauss Centre for Supercomputing
e.V. (www.gauss-centre.eu) for providing
computing time through the John von Neumann Institute for Computing
(NIC) on the GCS Supercomputer JUWELS at Jülich Supercomputing
Centre (JSC); the HPC RIVR consortium and EuroHPC JU for resources on
the Vega high performance computing system at IZUM, the Institute of
Information Science in Maribor; and the UK Engineering and Physical
Sciences Research Council for resources on the Baskerville Tier 2 HPC
service. Baskerville was funded by the EPSRC and UKRI through the World
Class Labs scheme (EP/T022221/1) and the Digital Research Infrastructure
programme (EP/W032244/1) and is operated by Advanced Research Computing
at the University of Birmingham.

\subsection*{Author Contributions}
DP conceived the project, wrote the code, ran the experiments and wrote the manuscript. SA wrote and tested the code for computing density matrices and natural orbitals. HS wrote and tested the code for pseudopotentials. IvG and JSS contributed to the code and DP, IvG and JSS wrote the library on which the code was based.

\subsection*{Competing Interests}
The authors declare that they have no competing financial interests.

\subsection*{Data and Materials Availability}
Data on molecular geometries and baseline values for all calculations in the paper are available in referenced publications. Numerical data for all figures in the paper are included in tables in the Supplementary Materials. Code for natural excited states, pseudopotentials and observable operator evaluation has been added to the FermiNet GitHub repository (\url{https://github.com/google-deepmind/ferminet}) under an open source license.

\newpage

\renewcommand{\thefigure}{S\arabic{figure}}
\setcounter{figure}{0}

\renewcommand{\thetable}{S\arabic{table}}
\setcounter{table}{0}

\renewcommand{\thesection}{S\arabic{section}}
\setcounter{section}{0}

\section*{Materials and Methods}
\label{sec:method}

\section{Variational Monte Carlo}
\label{sec:vmc}

First we briefly review ground-state VMC and establish some notation. We will stick to the notation of first quantization and consider a system of $N$ particles with states $\mathbf{x} = \mathbf{x}_1, \ldots, \mathbf{x}_N$, although everything we discuss could be applied to variational Ans{\"a}tze represented in second quantization as well. We aim to find the lowest eigenfunction of a Hamiltonian operator $\hat{H}$. This can be done by reformulating the eigenfunction problem in variational form, as one of finding the minimum of the Rayleigh quotient:

\begin{equation}
    \psi^* = \arg\min_\psi \frac{\langle \psi \hat{H} \psi \rangle}{\langle \psi^2 \rangle}
    \label{eqn:rayleigh_ground_state}
\end{equation}
where the Ansatz $\psi$ is not necessarily normalized. Computing this quotient involves taking high-dimensional integrals over all possible particle states $\mathbf{x}$, and can be approximated by Monte Carlo integration. Many choices of Monte Carlo sampling distribution $p(\mathbf{x})$ are possible, but if $p(\mathbf{x}) \propto \psi^2(\mathbf{x})$, then the Rayleigh quotient take a simple form that allows for statistically unbiased empirical estimation of the energy:

\begin{equation}
    \frac{\langle \psi \hat{H} \psi \rangle}{\langle \psi^2 \rangle} = \mathbb{E}_{\mathbf{x} \sim \psi^2}\left[\psi^{-1}(\mathbf{x}) \hat{H} \psi(\mathbf{x}) \right] = \mathbb{E}_{\mathbf{x} \sim \psi^2}\left[ E_L(\mathbf{x}) \right]
    \label{eqn:expected_energy_ground_state}
\end{equation}
For this reason, $\psi^2$ is the natural choice of sampling distribution for ground state estimation. The scalar $E_L(\mathbf{x}) \triangleq \psi^{-1}(\mathbf{x}) \hat{H} \psi(\mathbf{x})$ that appears inside the expectation is the {\em local energy}, and at any eigenfunction of $\hat{H}$ it will be constant if $\hat{H}$ is a local operator.

In addition to unbiased estimation of the ground state energy, sampling from $\psi^2$ enables unbiased estimation of {\em gradients} of the energy with respect to the variational parameters:

\begin{widetext}
\begin{align}
    \nabla_\theta \frac{\langle \psi \hat{H} \psi \rangle}{\langle \psi^2 \rangle} &= 2\mathbb{E}_{\mathbf{x} \sim \psi^2}\left[ \left(E_L(\mathbf{x}) - \mathbb{E}_{\mathbf{x}' \sim \psi^2}[E_L(\mathbf{x}')]\right)\nabla_\theta \mathrm{log}|\psi(\mathbf{x})| \right] \\
    &= 2\mathbb{E}_{\mathbf{x} \sim \psi^2}\left[ E_L(\mathbf{x}) \nabla_\theta \mathrm{log}|\psi(\mathbf{x})| \right] - 2 \mathbb{E}_{\mathbf{x}' \sim \psi^2}[E_L(\mathbf{x}')]\mathbb{E}_{\mathbf{x} \sim \psi^2}\left[\nabla_\theta \mathrm{log}|\psi(\mathbf{x})| \right]
    \label{eqn:expected_energy_ground_state_gradient}
\end{align}

where $\theta$ are the variational parameters of the Ansatz $\psi$. On the face of it, the product of two expectations in Eq.~\ref{eqn:expected_energy_ground_state_gradient} could present a challenge for unbiased estimation from finite samples. It turns out this is not an issue if we combine independent walkers from the same batch to estimate the gradient:

\begin{align}
 &\mathbb{E}_{\mathbf{x}_1,\ldots,\mathbf{x}_N}\left[\frac{1}{N} \sum_i \left(E_L(\mathbf{x}_i) - \frac{1}{N} \sum_j E_L(\mathbf{x}_j)\right) \nabla_\theta \mathrm{log}|\psi(\mathbf{x}_i)| \right] = \nonumber \\
  &\mathbb{E}_{\mathbf{x}_1,\ldots,\mathbf{x}_N}\left[\frac{1}{N} \sum_i E_L(\mathbf{x}_i)  \nabla_\theta \mathrm{log}|\psi(\mathbf{x}_i)| \right] - \mathbb{E}_{\mathbf{x}_1,\ldots,\mathbf{x}_N}\left[\frac{1}{N^2}\sum_{ij} E_L(\mathbf{x}_j)\nabla_\theta \mathrm{log}|\psi(\mathbf{x}_i)|\right] = \nonumber\\
  &\mathbb{E}_{\mathbf{x}_1,\ldots,\mathbf{x}_N}\left[\frac{N-1}{N^2} \sum_i E_L(\mathbf{x}_i)  \nabla_\theta \mathrm{log}|\psi(\mathbf{x}_i)| \right] - \mathbb{E}_{\mathbf{x}_1,\ldots,\mathbf{x}_N}\left[\frac{1}{N^2}\sum_{i\ne j} E_L(\mathbf{x}_j)\nabla_\theta \mathrm{log}|\psi(\mathbf{x}_i)|\right] = \nonumber\\
  &\mathbb{E}_{\mathbf{x}_1,\ldots,\mathbf{x}_N}\left[\frac{N-1}{N^2} \sum_i E_L(\mathbf{x}_i)  \nabla_\theta \mathrm{log}|\psi(\mathbf{x}_i)| \right] - \mathbb{E}_{\mathbf{x}_1,\ldots,\mathbf{x}_N}\left[\frac{1}{N^2}\sum_{i\ne j} E_L(\mathbf{x}_j)\nabla_\theta \mathrm{log}|\psi(\mathbf{x}_i)|\right] = \nonumber\\
  &\frac{N-1}{N}\left(\mathbb{E}_{\mathbf{x}}\left[E_L(\mathbf{x})  \nabla_\theta \mathrm{log}|\psi(\mathbf{x})| \right] - \mathbb{E}_{\mathbf{x}'}\left[ E_L(\mathbf{x}')\right]\mathbb{E}_{\mathbf{x}}\left[\nabla_\theta \mathrm{log}|\psi(\mathbf{x}_i)|\right] \right) = \nonumber \\
  &\frac{N-1}{2N} \nabla_\theta  \frac{\langle \psi \hat{H} \psi \rangle}{\langle \psi^2 \rangle}
\end{align}
Where in the second-to-last line we used the fact that $\mathbb{E}[\mathbf{X}]\mathbb{E}[\mathbf{Y}] = \mathbb{E}[\mathbf{X}\mathbf{Y}]$ for two independent random variables $\mathbf{X}$ and $\mathbf{Y}$, and the fact that all walkers in the batch are sampled independently.
\end{widetext}

\section{Natural Excited States}
\label{sec:nes-vmc-supplemental}

Going from ground states to excited states, we aim to find the lowest $K$ eigenfunctions of $\hat{H}$. We refer to a single set of $N$ particle states as a {\em particle set}, and denote different particle sets with an upper index, so that $\mathbf{x}^i$ denotes a set of $N$ particles $\mathbf{x}^i_1,\ldots,\mathbf{x}^i_N$. We will use $\mathbf{x}$ to denote the complete state of all particle sets $\mathbf{x}^1,\ldots,\mathbf{x}^K$. Let $\psi_i$ denote a (possibly unnormalized) N-particle wavefunction, then we are trying to find wavefunctions $\psi_1,\ldots,\psi_K$ which approximate the lowest excited states. Let $\psimat(\mathbf{x}) \in \mathbb{R}^{K \times K}$ denote the matrix combining all electron sets with all wavefunctions:

\begin{equation}
    \psimat(\mathbf{x}) \triangleq
    \begin{pmatrix}
    \psi_1(\mathbf{x}^1) & \ldots & \psi_K(\mathbf{x}^1) \\
    \vdots & & \vdots \\
    \psi_1(\mathbf{x}^K) & \ldots & \psi_K(\mathbf{x}^K)
    \end{pmatrix}
\end{equation}
The determinant of this matrix $\Psi(\mathbf{x}) = \mathrm{det}(\psimat(\mathbf{x}))$ can be thought of as an unnormalized Slater determinant, except that instead of single-particle orbitals, it is made up of N-particle wavefunctions. We call $\Psi(\mathbf{x}) = \mathrm{det}(\psimat(\mathbf{x}))$ the {\em total Ansatz}, while the individual $\psi_i$ are the {\em single-state Ans{\"a}tze}.

Rather than optimizing the single-state Ans{\"a}tze in order from lowest to highest energy, we will only optimize the total Ansatz to minimize the total energy of all states. This is conceptually quite similar to many state-specific and state-averaged approaches in VMC, except that we will not explicitly enforce the orthogonality of the different single-state Ans{\"a}tze. Note that taking any linear combination of single-state Ans{\"a}tze $\psi{'}_i = \sum_j a_{ij} \psi_j$ only changes the total Ansatz by a constant factor. Also note that if two single-state Ans{\"a}tze are the same, the total Ansatz becomes zero. Thus, by representing the total Ansatz as a determinant of single-state Ans{\"a}tze, we can prevent the collapse of different Ans{\"a}tze onto the same state, without requiring them to be orthogonal.

For an arbitrary operator $\op$ that acts on $N$-particle wavefunctions, let $\oppsimat(\mathbf{x})$ denote the matrix of all values of this operator applied to all single-state Ans{\"a}tze and particle sets:

\begin{equation}
    \oppsimat(\mathbf{x}) \triangleq
    \begin{pmatrix}
    \op\psi_1(\mathbf{x}^1) & \ldots & \op\psi_K(\mathbf{x}^1) \\
    \vdots & & \vdots \\
    \op\psi_1(\mathbf{x}^K) & \ldots & \op\psi_K(\mathbf{x}^K)
    \end{pmatrix}
\end{equation}
Let $\overlap$ and $\opoverlap$ denote the overlap matrix between states and the matrix of expectations of states for the operator $\op$:

\begin{align}
    \overlap \triangleq
    \begin{pmatrix}
        \langle \psi_1^2 \rangle & \ldots & \langle \psi_1 \psi_K \rangle \\
        \vdots & & \vdots \\
        \langle \psi_K \psi_1 \rangle & \ldots & \langle \psi_K^2 \rangle
    \end{pmatrix} \\
    \opoverlap \triangleq
    \begin{pmatrix}
        \langle \psi_1 \op \psi_1 \rangle & \ldots & \langle \psi_1 \op \psi_K \rangle \\
        \vdots & & \vdots \\
        \langle \psi_K \op \psi_1 \rangle & \ldots & \langle \psi_K \op \psi_K \rangle
    \end{pmatrix}
\end{align}

If $\hat{H}$ is the Hamiltonian for the system under consideration, then we can define an expanded Hamiltonian that acts on the total Ansatz $\Psi$ as $\hat{\mathcal{H}} = \hat{H}_1 \oplus \ldots \oplus \hat{H}_K$, where $\hat{H}_i$ is the Hamiltonian that acts only on particle set $i$. The ground state energy of $\hat{\mathcal{H}}$ is the sum of the lowest $K$ energies of $\hat{H}$, and the ground state wavefunction $\Psit$ is a determinant of the $K$ lowest states $\psit_1,\ldots,\psit_K$ of $\hat{H}$. This can be seen by writing out the Rayleigh quotient for the total Ansatz and expanded Hamiltonian:

\begin{widetext}
\begin{equation}
    \frac{\langle \Psi \hat{\mathcal{H}} \Psi \rangle}{\langle \Psi^2 \rangle} = \frac{\langle \Psi \hat{\mathcal{H}} \Psi \rangle}{\mathrm{det}(\overlap)} = \frac{\sum_i \langle \Psi \hat{H}_i \Psi \rangle}{\mathrm{det}(\overlap)} 
    =\frac{
    \sum_i \mathrm{det}\begin{pmatrix}
        \langle \psi^2_1 \rangle & \ldots & \langle \psi_1 \hat{H} \psi_i \rangle & \ldots & \langle \psi_1 \psi_K \rangle \\
        \vdots & & & & \vdots \\ 
        \langle \psi_K \psi_1 \rangle & \ldots & \langle \psi_K \hat{H} \psi_i \rangle & \ldots & \langle \psi_K^2 \rangle
    \end{pmatrix}}{
    \mathrm{det}(\overlap)
    }
    \label{eqn:total_rayleigh}
\end{equation}
\end{widetext}
The terms in the numerator are rank-one updates to the denominator, and using the matrix determinant lemma, Eq.~\ref{eqn:total_rayleigh} can be rewritten as

\begin{equation}
    \frac{\langle \Psi \hat{\mathcal{H}} \Psi \rangle}{\langle \Psi^2 \rangle} = \mathrm{Tr}\left[\overlap^{-1} \hoverlap\right]
    \label{eqn:matrix_rayleigh}
\end{equation}

The minimum of the expression in Eq.~\ref{eqn:matrix_rayleigh} is a set of functions $\psit_1,\ldots,\psit_K$ that span the bottom $K$ eigenfunctions of $\hat{H}$. However, note that we have no way to know from this expression alone how to decompose the total energy into individual energies.

Because the objective function in Eq.~\ref{eqn:matrix_rayleigh} is just a special case of a ground state, we can minimize it by conventional energy minimization for VMC. The Rayleigh quotient can be rewritten as an expectation of the local (total) energy:

\begin{equation}
    \mathbb{E}_{\mathbf{x}\sim\Psi^2}\left[ \Psi^{-1}(\mathbf{x}) \hat{\mathcal{H}} \Psi(\mathbf{x}) \right] = \mathrm{Tr}\left[ \mathbb{E}_{\mathbf{x}\sim\Psi^2}  [\psimat^{-1}(\mathbf{x}) \hpsimat(\mathbf{x})] \right]
    \label{eqn:expected_local_energy_matrix}
\end{equation}
Here we generate samples of $K$ electron sets simultaneously (or equivalently, $NK$ particles) by sampling from a density proportional to $\Psi^2$. Unbiased gradients of this objective function can be computed as in standard VMC and plugged into any number of gradient-based optimization algorithms.

To derive Eqn.~\ref{eqn:expected_local_energy_matrix}, first we write out the effect of the total Hamiltonian on the total Ansatz using the Leibniz definition of the determinant:

\begin{align}
    \hat{\mathcal{H}} \Psi(\mathbf{x}) &= \hat{\mathcal{H}} \sum_{\sigma} (-1)^{|\sigma|} \prod_{i=1}^K \psi_{\sigma_i}(\mathbf{x}^i) \\
    &= \hat{H}_1 \oplus \ldots \oplus \hat{H}_K \sum_{\sigma} (-1)^{|\sigma|} \prod_{i=1}^K \psi_{\sigma_i}(\mathbf{x}^i) \\
    &= \sum_{j=1}^K \sum_{\sigma} (-1)^{|\sigma|} \hat{H} \psi_{\sigma_j}(\mathbf{x}^j) \prod_{i\ne j} \psi_{\sigma_i}(\mathbf{x}^i) \\
    &= \sum_{j=1}^K
    \mathrm{det}\begin{pmatrix}
    \psi_1(\mathbf{x}^1) & \ldots & \psi_K(\mathbf{x}^1) \\
    \vdots & & \vdots \\
    \hat{H} \psi_1(\mathbf{x}^j) & \ldots & \hat{H} \psi_K(\mathbf{x}^j) \\
    \vdots & & \vdots \\
    \psi_1(\mathbf{x}^K) & \ldots & \psi_K(\mathbf{x}^K) \\
    \end{pmatrix}
    \label{eqn:leibniz_hpsi}
\end{align}
where the sum over $\sigma$ is over all $K$-permutations. The matrices in the sum in Eq.~\ref{eqn:leibniz_hpsi} are rank-one updates to $\mathbf{\Psi}(\mathbf{x})$ of the form $\mathbf{e}_j (\hat{H}\mathbf{\Psi}_j - \mathbf{\Psi}_j)$ where $\mathbf{\Psi}_j$ is the $j$th row of $\mathbf{\Psi}$ and $\hat{H}\mathbf{\Psi}_j$ is the row vector with elements $\hat{H}\psi_i(\mathbf{x}^j)$. Plugging this into the matrix determinant lemma yields

\begin{align}
    \hat{\mathcal{H}}\Psi(\mathbf{x}) &= \sum_{j=1}^K\mathrm{det}\mathbf{\Psi}(\mathbf{x}) \left(1 +   (\hat{H}\mathbf{\Psi}_j - \mathbf{\Psi}_j)\mathbf{\Psi}^{-1}(\mathbf{x})\mathbf{e}_j\right) \\
    &= \sum_{j=1}^K\Psi(\mathbf{x})\left(1 +  \hat{H}\mathbf{\Psi}_j\mathbf{\Psi}^{-1}(\mathbf{x}) \mathbf{e}_j - \mathbf{\Psi}_j\mathbf{\Psi}^{-1}(\mathbf{x}) \mathbf{e}_j\right) \\
    &= \sum_{j=1}^K\Psi(\mathbf{x})\left(1 +  \hat{H}\mathbf{\Psi}_j \mathbf{\Psi}^{-1}(\mathbf{x}) \mathbf{e}_j - \mathbf{e}_j^T \mathbf{e}_j\right) \\
    &= \sum_{j=1}^K\Psi(\mathbf{x})\left(\hat{H}\mathbf{\Psi}_j \mathbf{\Psi}^{-1}(\mathbf{x}) \mathbf{e}_j \right) \label{eqn:matrix_det_lemma}
    \end{align}
And plugging this into the expression for the local total energy yields
    \begin{align}
\Psi(\mathbf{x})^{-1} \hat{\mathcal{H}} \Psi(\mathbf{x}) &= \sum_{j=1}^K \hat{H}\mathbf{\Psi}_j \mathbf{\Psi}^{-1}(\mathbf{x}) \mathbf{e}_j \\
&= \mathrm{Tr}\left[\hat{H}\mathbf{\Psi}(\mathbf{x})\mathbf{\Psi}^{-1}(\mathbf{x})\right] \\
&= \mathrm{Tr}\left[\mathbf{\Psi}^{-1}(\mathbf{x})\hat{H}\mathbf{\Psi}(\mathbf{x})\right]
\end{align}
The rest of Eq~\ref{eqn:expected_local_energy_matrix} follows from the linearity of expectations.

This is clearly a direct generalization of the VMC objective in Eq.~\ref{eqn:expected_energy_ground_state}. In the same way that $\psi^2$ is the natural choice of sampling distribution for ground state VMC, this makes it clear that $\Psi^2$ is the natural choice of sampling distribution for excited-state VMC. We define the {\em local energy matrix} to be the term inside the expectation in Eq.~\ref{eqn:expected_local_energy_matrix}:

\begin{equation}
    \mathbf{E}_L(\mathbf{x}) \triangleq \psimat^{-1}(\mathbf{x}) \hpsimat(\mathbf{x})
    \label{eqn:local_energy_matrix}
\end{equation}
When $K=1$ this reduces to the scalar local energy in Eq.~\ref{eqn:expected_energy_ground_state}. The benefits of this choice of sampling distribution go beyond the simple functional form of $\mathbf{E}_L(\mathbf{x})$, and we will see that this matrix in fact contains all the information needed to compute the full spectrum of energies.

Note that the expression in Eq.~\ref{eqn:expected_local_energy_matrix} looks quite similar to Eq.~\ref{eqn:matrix_rayleigh}. Not only are the traces of $\overlap^{-1} \hoverlap$ and $\mathbb{E}_{\Psi^2}  [\mathbf{E}_L(\mathbf{x})]$ equal, but in fact the two matrices are identical. This can be seen by following the same derivation as in Eqs.~\ref{eqn:total_rayleigh}-\ref{eqn:expected_local_energy_matrix}, but rather than with the operator $\hat{\mathcal{H}}$, with an operator $\hat{H}_{ij}$ such that $\hat{H}_{ij} \psi_k(\mathbf{x}^i) = \hat{H} \psi_k(\mathbf{x}^j)$ and is the identity for all other $\mathbf{x}^{i'}$. Plugging in this operator to the Rayleigh quotient, one can show that it gives row $i$ and column $j$ in $\overlap^{-1} \hoverlap$, while plugging it into the Monte Carlo form of the objective gives row $i$ and column $j$ of $\mathbb{E}_{\Psi^2}  [\mathbf{E}_L(\mathbf{x})]$

While we only need to use the diagonal of the matrix $\mathbb{E}_{\Psi^2} [\mathbf{E}_L(\mathbf{x})]$ for optimization, the full matrix has precisely the information which we need to separate the different excited states in proper order. When we plug in the exact eigenfunctions $\psit_1,\ldots,\psit_K$, we find that ${\psimatopt}^{-1} \hpsimatopt = {\psimatopt}^{-1} \psimatopt \mathbf{\Lambda} = \mathbf{\Lambda}$, where $\mathbf{\Lambda}$ is the diagonal matrix of energies, and even if the single-state Ans{\"a}tze are linear combinations of eigenfunctions $\psi_i = \sum_j a_{ij} \psit_j$, then ${\psimatopt}^{-1} \hpsimatopt = \mathbf{A}^{-1} \mathbf{\Lambda} \mathbf{A}$. This suggests that, if we are in the vicinity of the true ground state of the total Ansatz, then by accumulating and then diagonalizing the matrix $\mathbb{E}_{\Psi^2} [\mathbf{E}_L(\mathbf{x})] = \mathbf{U} \mathbf{\Lambda} \mathbf{U}^{-1}$, we can recover the individual energies of the states in order, rather than simply the total energy. Note that in general $\mathbf{U} = \mathbf{A}^{-1} \mathbf{\Sigma}$ where $\mathbf{\Sigma} = \left(\begin{smallmatrix}
  \sigma_1 &  & \\
  & \ddots & \\
  & & \sigma_K
\end{smallmatrix}\right)$ is an arbitrary diagonal matrix which is not identifiable.

Not only is diagonalizing $\mathbb{E}_{\Psi^2} [\mathbf{E}_L(\mathbf{x})]$ sufficient to recover the energies -- it also provides us with the necessary change of basis to evaluate other observables $\op$, even off-diagonal observables $\langle \psi_i \op \psi_j \rangle$ between states. This can be seen due to the identity $\mathbb{E}_{\Psi^2} [\psimat^{-1} \oppsimat] = \overlap^{-1} \opoverlap$, and for single-state Ans{\"a}tze which are a linear combination of eigenfunctions, $\overlap^{-1} \opoverlap = \mathbf{A}^{-1} \opoverlapopt \mathbf{A}$. So if we accumulate and diagonalize $\mathbb{E}_{\Psi^2} [\mathbf{E}_L(\mathbf{x})]$ and use the resulting eigenvectors to compute $\mathbf{U}^{-1} \mathbb{E}_{\Psi^2} [\psimat^{-1} \oppsimat] \mathbf{U}$, then in the vicinity of the true ground state of the total Ansatz the result will be approximately $\mathbf{\Sigma}^{-1} \opoverlapopt \mathbf{\Sigma}$. Along the diagonal, this gives exactly the expectations $\langle \psit_i \op \psit_i\rangle$. Off the diagonal, this yields $\frac{\sigma_i}{\sigma_j}\langle \psit_i \op \psit_j \rangle$. If we multiply the matrix elementwise by its transpose, the $\sigma_i$ terms cancel out, and we recover $\langle \psit_i \op \psit_j \rangle^2$, which gives the expectation up to a sign factor. This sign factor is not physically observable however, and in practice for computing quantities like the oscillator strength, only the expectation squared is needed.

\section{Neural Network Ans{\"a}tze}

The use of variational Monte Carlo for ground state calculations was typically used to find a trial wavefunction for more accurate projector QMC methods like diffusion Monte Carlo\cite{foulkes2001quantum} or auxiliary field Monte Carlo\cite{motta2018ab}. However, in recent years, advances in deep neural networks have led to their use as accurate Ans{\"a}tze for studying spin systems\cite{carleo2017solving}, electronic structure\cite{hermann2023ab} and nuclear systems\cite{yang2023deep}, often reaching levels of accuracy rivaling projector QMC methods. This has led to a renewed interest in VMC as a standalone method. While a variety of different neural network architectures can be used depending on the problem, such as restricted Boltzmann machines\cite{carleo2017solving}, convolutional neural networks\cite{stokes2020phases}, and autoregressive models\cite{sharir2020deep}, a number of custom architectures have been developed specifically for many-body electronic structure problems in first quantization\cite{luo2019backflow, pfau2020ab, hermann2020deep, gerard2022gold, von2023self, gao2023generalizing, pescia2023message, li2024forward}. Most of these Ans{\"a}tze start from a linear combination of Slater determinants:

\begin{equation}
    \psi(\mathbf{x}) = \sum_k \omega_k \mathrm{det}\begin{pmatrix} \phi^k_1(\mathbf{x}_1) & \ldots & \phi^k_N(\mathbf{x}_1) \\ \vdots & & \vdots \\ \phi^k_1(\mathbf{x}_N) & \ldots & \phi^k_N(\mathbf{x}_N) \end{pmatrix}
    \label{eqn:slater_det}
\end{equation}
It has long been recognized\cite{wigner1934constitution} that the single-particle orbitals in a Slater determinant can be generalized to depend on {\em all} particles, so long as they depend on all but one in a permutation-independent manner:
\begin{equation}
    \psi(\mathbf{x}) = \sum_k \omega_k \mathrm{det}\begin{pmatrix} \phi^k_1(\mathbf{x}_1; \{\mathbf{x}_{/1}\}) & \ldots & \phi^k_N(\mathbf{x}_1; \{\mathbf{x}_{/1}\}) \\ \vdots & & \vdots \\ \phi^k_1(\mathbf{x}_N; \{\mathbf{x}_{/N}\}) & \ldots & \phi^k_N(\mathbf{x}_N; \{\mathbf{x}_{/N}\}) \end{pmatrix}
    \label{eqn:generalized_slater_det}
\end{equation}
where $\{\mathbf{x}_{/i}\}$ denotes the set of all particles {\em except} $\mathbf{x}_i$. In the event that the particles are spin-assigned, the orbitals can also be expressed as $\phi^k_i(\mathbf{x}^\uparrow_j; \{\mathbf{x}^\uparrow_{/j}\}, \{\mathbf{x}^\downarrow\})$ where the function is only invariant to changing the order of particles of the same spin. Most neural network Ans{\"a}tze for electrons in real space implement this idea by using permutation-equivariant deep neural networks to represent the orbitals, sometimes with a multiplicative Jastrow factor to account for pairwise interactions\cite{pfau2020ab, hermann2020deep, von2023self}.

We give a brief summary here of the FermiNet and Psiformer architectures\cite{pfau2020ab, spencer2020better, von2023self}, the two neural network Ans{\"a}tze used in this paper. Both Ans{\"a}tze take electron positions $\mathbf{r}_1, \ldots, \mathbf{r}_N$ as inputs, and assume that the spins are fixed, so that the first $N_\alpha$ electrons are spin up, while the last $N_\beta = N-N_\alpha$ electrons are spin down. Where relevant, we will add a superscript of $\alpha$ or $\beta$ to $\mathbf{r}_i$ to indicate the spin. For both the FermiNet and Psiformer, the vector differences between atoms and ions $\mathbf{r}_i - \mathbf{R}_I$ and scalar distances $|\mathbf{r}_i - \mathbf{R}_I|$ are concatenated as input features. Additionally in the FermiNet, the differences $\mathbf{r}_i - \mathbf{r}_j$ and distances $|\mathbf{r}_i - \mathbf{r}_j|$ between pairs of electrons form the input to the two-electron stream of the network.

In both the FermiNet and Psiformer, these inputs are propagated through multiple layers of nonlinear computation. In the Psiformer, each layer consists of a standard self-attention layer\cite{vaswani2017attention} with layer normalization\cite{ba2016layer} and a residual connection\cite{he2016deep} followed by parallel linear-nonlinear layers with a hyperbolic tangent nonlinearity. In the FermiNet, the two-electron stream consists entirely of parallel linear-nonlinear layers with hyperbolic tangent nonlinearities and residual connections. The one-electron stream integrates information from the two-electron stream as well as between different activations in the one-electron stream. If $\mathbf{h}^{\ell\alpha}_i$ is the activation vector at layer $\ell$ for electron $i$ of spin $\alpha$ in the one-electron stream and $\mathbf{h}^{\ell\alpha\beta}_{ij}$ is the activation vector for electrons $i$ and $j$ of spin $\alpha$ and $\beta$ in the two electron stream, then

\begin{widetext}
\begin{align}
\mathbf{f}^{\ell \alpha}_i =& \Biggl(
    \mathbf{h}^{\ell\alpha}_i,
    \frac{1}{n^\uparrow}\sum_{j=1}^{n^\uparrow} \mathbf{h}^{\ell\uparrow}_j, \frac{1}{n^\downarrow} \sum_{j=1}^{n^\downarrow} \mathbf{h}^{\ell\downarrow}_j, \frac{1}{n^\uparrow} \sum_{j=1}^{n^\uparrow} \mathbf{h}^{\ell\alpha\uparrow}_{ij},
    \frac{1}{n^\downarrow} \sum_{j=1}^{n^\downarrow} \mathbf{h}^{\ell\alpha\downarrow}_{ij} \Biggr) \nonumber \\
\mathbf{h}^{(\ell+1)\alpha}_i =& \mathrm{tanh}\left(\mathbf{W}^\ell \mathbf{f}^{\ell \alpha}_i + \mathbf{b}^\ell\right) + \mathbf{h}^{\ell\alpha}_i
\end{align}
\end{widetext}
gives the activation at the next layer.

Following the sequence of repeated nonlinear layers, the final activations $\mathbf{h}^{L\alpha}_i$ are linearly projected to a set of orbitals, one spin up and one spin down, and multiplied by an exponentially-decaying envelope which enforces the boundary condition that the wavefunction should decay to zero at long range, described in more detail below in Sec.~\ref{sec:bottleneck_envelope}. The form of these orbitals can be written as:

\begin{align}
    \phi^{k\alpha}_i(\mathbf{r}^\alpha_j; \{\mathbf{r}^\alpha_{/j}\}; \{\mathbf{r}^{\bar{\alpha}}\}) =&
    \left(\mathbf{w}^{k\alpha}_i \cdot \mathbf{h}^{L\alpha}_j + g^{k\alpha}_i\right)e^{k\alpha}_i(\mathbf{r}^\alpha_j)
    \label{eqn:orbital}
\end{align}

Finally, for the Psiformer, a multiplicative Jastrow factor $\mathrm{exp}\left(\mathcal{J}_\theta(\mathbf{r}_1^\uparrow,\ldots,\mathbf{r}_{N}^\downarrow)\right)$ is applied after the determinant to capture the correct behavior at the electron-electron cusps, due to the lack of two-electron inputs to the neural network part of the Ansatz. A simple two-parameter Jastrow factor seems to suffice:

\begin{align}
    \mathcal{J}_\theta(\mathbf{r}_1^\uparrow,\ldots,\mathbf{r}_{N}^\downarrow)= & \sum_{i<j; \alpha = \beta} - \frac{1}{4}\frac{ \alpha_\mathrm{par}^2}{\alpha_\mathrm{par} + |\mathbf{r}^\alpha_i - \mathbf{r}^\beta_j|} + \nonumber \\
    &\sum_{i,j; \alpha \ne \beta} - \frac{1}{2}\frac{ \alpha_\mathrm{anti}^2}{\alpha_\mathrm{anti} + |\mathbf{r}^\alpha_i - \mathbf{r}^\beta_j|}
    \label{eqn:jastrow}
\end{align}

Extending these Ans{\"a}tze to represent multiple states is quite straightforward. Each state is still expressed as a sum of determinants of generalized neural network orbitals, there are simply more orbitals:
\begin{equation}
    \psi_i(\mathbf{x}) = \sum_{ik} \omega_{ik} \mathrm{det}\begin{pmatrix} \phi^{ik}_1(\mathbf{x}_1; \{\mathbf{x}_{/1}\}) & \ldots & \phi^{ik}_N(\mathbf{x}_1; \{\mathbf{x}_{/1}\}) \\ \vdots & & \vdots \\ \phi^{ik}_1(\mathbf{x}_N; \{\mathbf{x}_{/N}\}) & \ldots & \phi^{ik}_N(\mathbf{x}_N; \{\mathbf{x}_{/N}\}) \end{pmatrix}
    \label{eqn:more_generalized_slater_det}
\end{equation}
Nothing is changed about the neural network architecture itself, just the number of orbitals is increased proportionally to the number of states.

\section{Bottleneck Envelope}
\label{sec:bottleneck_envelope}

For the experiments with benzene, the memory overhead of the default FermiNet and Psiformer became prohibitive, largely due to the cost of the multiplicative envelope which is used to enforce open boundary conditions. For both the FermiNet and Psiformer, each orbital $\phi^{k\alpha}_j(\mathbf{r}^\alpha_i; \{\mathbf{r}^\alpha_{/i}\}, \{\mathbf{r}^{\bar{\alpha}}\})$ included a multiplicative envelope of the form:

\begin{equation}
    e^{k\alpha}_j(\mathbf{r}^\alpha_i) = \sum_I \pi^{k\alpha}_{jI} \mathrm{exp}\left(\sigma^{k\alpha}_{jI} |\mathbf{r}^\alpha_i - \mathbf{R}_I| \right)
    \label{eqn:envelope}
\end{equation}
where $i$ indexes electrons, $I$ indexes atomic nuclei, $\alpha\in\{\uparrow,\,\downarrow\}$ indexes spin, $j$ indexes different orbitals in a determinant and $k$ indexes different determinants. Because the number of orbitals scales linearly with the number of electrons {\em and} excited states, this becomes prohibitively large for large molecules with many excited states. Instead, we fix the number of distinct envelopes, and for each orbital we take a weighted sum of these envelopes:
\begin{equation}
    e^{k\alpha}_j(\mathbf{r}^\alpha_i) = \sum_\ell w^{k\alpha}_{j\ell} \sum_I \pi^{\alpha}_{\ell I} \mathrm{exp}\left(\sigma^{\alpha}_{\ell I} |\mathbf{r}^\alpha_i - \mathbf{R}_I| \right)
    \label{eqn:bottleneck_envelope}
\end{equation}
This reduces the memory overhead when computing the sum over $I$ significantly. In our experiments, we used $|\ell|=32$, which is far lower than $|k||j| = 336$ for benzene (21 electrons of each spin and 16 determinants). We refer to this as the {\em bottleneck} envelope, as it can be seen as projecting the envelopes into a lower dimension space and then back into the space of orbitals. Although we did not have any numerical difficulties using the bottleneck envelope for benzene, we did find on some systems that there were occasional numerical stability issues, and so we only used the bottleneck envelope for the very largest systems we investigated in this paper.

\section{Singlet Targeting}
\label{sec:singlet_targeting}

For the largest double excitations investigated -- cyclopentadienone and tetrazine -- a large number of low-lying triplet states meant that it was impractical to compute enough states to compare against other results in the literature without some modification. To exclusively target singlet states, we minimized the energy for a modified Hamiltonian which increased the energy of triplet states:

\begin{equation}
    \hat{H}'_\lambda = \hat{H} + \lambda \mathcal{\hat{S}}^2
\end{equation}
where $\hat{H}$ is the standard Born-Oppenheimer Hamiltonian and $\mathcal{\hat{S}}^2$ is the spin magnitude operator, as defined in Sec.~\ref{sec:spin_magnitude}. Since $\hat{H}$ and $\hat{\mathcal{S}}^2$ commute, the eigenfunctions of $\hat{H}'_\lambda$ are the same as for $\hat{H}$, just with the energy of states with spin $S$ shifted up by $S(S+1)\lambda$. So long as $\lambda$ is chosen to be larger than the difference between the lowest singlet and triplet (or even greater spin) states, then the lowest excitations of $\hat{H}'_\lambda$ will be exclusively singlets. A similar approach was used for FCIQMC to target singlet states of O$_2$ and [Mn$_3^{\mathrm{(IV)}}$O$_4]$ clusters \cite{weser2022spin}. For cyclopentadienone, we set $\lambda=1$, while for tetrazine we set $\lambda=0.5$, roughly double the required scale to remove triplets in both cases.

\section{Ensemble Penalty Methods}
\label{sec:penalty}

Penalty methods are the only other class of excited state method for VMC which has been used with neural networks to date \cite{choo2018symmetries, entwistle2023electronic}. Prior work on penalty methods for neural network excited state VMC attempted to use penalty terms for which unbiased gradients could be computed, but found that optimization suffered from issues with higher states collapsing onto lower ones \cite{entwistle2023electronic}. However, after the first draft of this manuscript appeared, a new analysis of ensemble penalty methods (that is, methods where all states are optimized simultaneously) showed that the state collapse issue can be alleviated by putting different weights on the energies of different states in the optimization objective \cite{wheeler2023ensemble}. If the objective function is chosen to be:

\begin{equation}
    \sum_i w_i \frac{\langle \psi_i \hat{H} \psi_i\rangle}{\langle\psi^2\rangle} + \lambda \sum_{j>i} \frac{\langle \psi_i \psi_j \rangle^2}{\langle \psi_i^2 \rangle \langle \psi_j^2\rangle}
    \label{eqn:ensemble_penalty_objective}
\end{equation}
then, so long as the penalty weight $\lambda$ satisfies

\begin{equation}
    \lambda > \mathrm{max}_{j>i} (E_j - E_i) \frac{w_i w_j}{w_i - w_j}
\end{equation}
where $E_i$ is the energy of the $i$th state of $\hat{H}$, the different states should converge to the bottom excited states of $\hat{H}$ without collapsing.

Eq.~\ref{eqn:ensemble_penalty_objective} can be written in Monte Carlo form as

\begin{widetext}
\begin{equation}
    \sum_i w_i \mathbb{E}_{\mathbf{x}_i}\left[E_L(\mathbf{x}_i)\right] + \lambda \sum_{j>i} \mathbb{E}_{\mathbf{x}_i}\left[\psi_i^{-1}(\mathbf{x}_i)\psi_j(\mathbf{x}_i)\right] \mathbb{E}_{\mathbf{x}_j}\left[\psi_j^{-1}(\mathbf{x}_j)\psi_i(\mathbf{x}_j)\right]
    \label{eqn:monte_carlo_ensemble_penalty_objective}
\end{equation}
where $\mathbf{x}_i$ is sampled from a distribution proportional to $\psi_i^2$, and the gradient of Eq.~\ref{eqn:monte_carlo_ensemble_penalty_objective} can be written as

\begin{align}
    \sum_i w_i \mathbb{E}_{\mathbf{x}_i}\left[(E_L(\mathbf{x}_i)-\mathbb{E}_{\mathbf{x}'_i}\left[E_L(\mathbf{x}'_i)\right])\nabla \mathrm{log}|\psi_i(\mathbf{x}_i)|\right] + \nonumber \\
    \lambda \sum_{j>i} \mathbb{E}_{\mathbf{x}_j}\left[O^{ji}_L(\mathbf{x}_j)\right] \mathbb{E}_{\mathbf{x}_i}\left[(O^{ij}_L(\mathbf{x}_i)-\mathbb{E}_{\mathbf{x}'_i}\left[O^{ij}_L(\mathbf{x}'_i)\right])\nabla \mathrm{log}|\psi_i(\mathbf{x}_i)|\right]
    \label{eqn:ensemble_penalty_gradient}
\end{align}
\end{widetext}
where $O^{ij}_L(\mathbf{x}) \triangleq \psi_i^{-1}(\mathbf{x})\psi_j(\mathbf{x})$ is the local overlap. Following a similar argument to that in Sec.~\ref{sec:vmc}, it can be shown that, because $\mathbf{x}_i$ and $\mathbf{x}_j$ are sampled independently, we can construct statistically unbiased estimates of the gradient in Eq.~\ref{eqn:ensemble_penalty_gradient} from empirical samples.

To compare the effectiveness of NES-VMC with existing methods, we implemented the ensemble penalty described above for neural network Ans{\"a}tze. To avoid numerical instability, we clipped the individual energies of each state as well as each element of the overlap matrix independently when computing the gradients, similarly to how the energies were clipped in \cite{pfau2020ab} -- any element more than 5 times greater than the total variation from the median was clipped. Otherwise, the sampling and optimization were nearly identical to standard training of the FermiNet and Psiformer. We ran the ensemble penalty method on nitroxyl (HNO) and fluoromethylene (HCF), and experimented with different pretraining strategies, as described below in Sec.~\ref{sec:pretraining}. We also attempted to run the ensemble penalty method on benzene, but found it became numerically unstable after a few hundred iterations. For all systems, we set the weights to $w_i = i^{-1}$, indexing the ground state as $i=1$, and set the penalty strength $\lambda = 1$. Results are shown in Fig.~\ref{fig:pretraining_and_penalty} and Table~\ref{tab:penalty}. We discuss the results in more depth, along with the effect of pretraining, in Sec.~\ref{sec:penalty_and_pretrain_results}.

\section{Pretraining}
\label{sec:pretraining}

If the different states are not linearly independent at initialization, then the total Ansatz will be identically zero and the local energy will be undefined. To avoid this pathological initialization, we pretrain the single-state Ans{\"a}tze to be linearly independent. In the original FermiNet paper \cite{pfau2020ab}, the individual orbitals were pretrained to match Hartree-Fock orbitals computed by PySCF \cite{sun2018pyscf}. For excited states, we need to construct $k$ different Slater determinants. We experimented with two different pretraining methods, which we refer to as the ``ordered" and ``random" pretraining methods. Most experiments used ordered pretraining, but we found that for a number of systems the ordered pretraining would lead to certain states being missed, especially double excitations, and that random pretraining would lead to these states being correctly identified. Additionally, this sensitivity to choice of pretraining does not seem to be unique to NES-VMC, and seems to affect penalty methods as well. A comparison on several systems is shown in Fig.~\ref{fig:pretraining_and_penalty}. Here we describe both pretraining strategies, and then provide a comparison on a number of systems where ordered pretraining failed, with both NES-VMC and ensemble penalty methods. All experiments in the main text used the ordered pretraining unless otherwise specified.

\subsection{Ordered Pretraining}

\begin{figure*}[]
    \centering
    \includegraphics[width=1.0\textwidth]{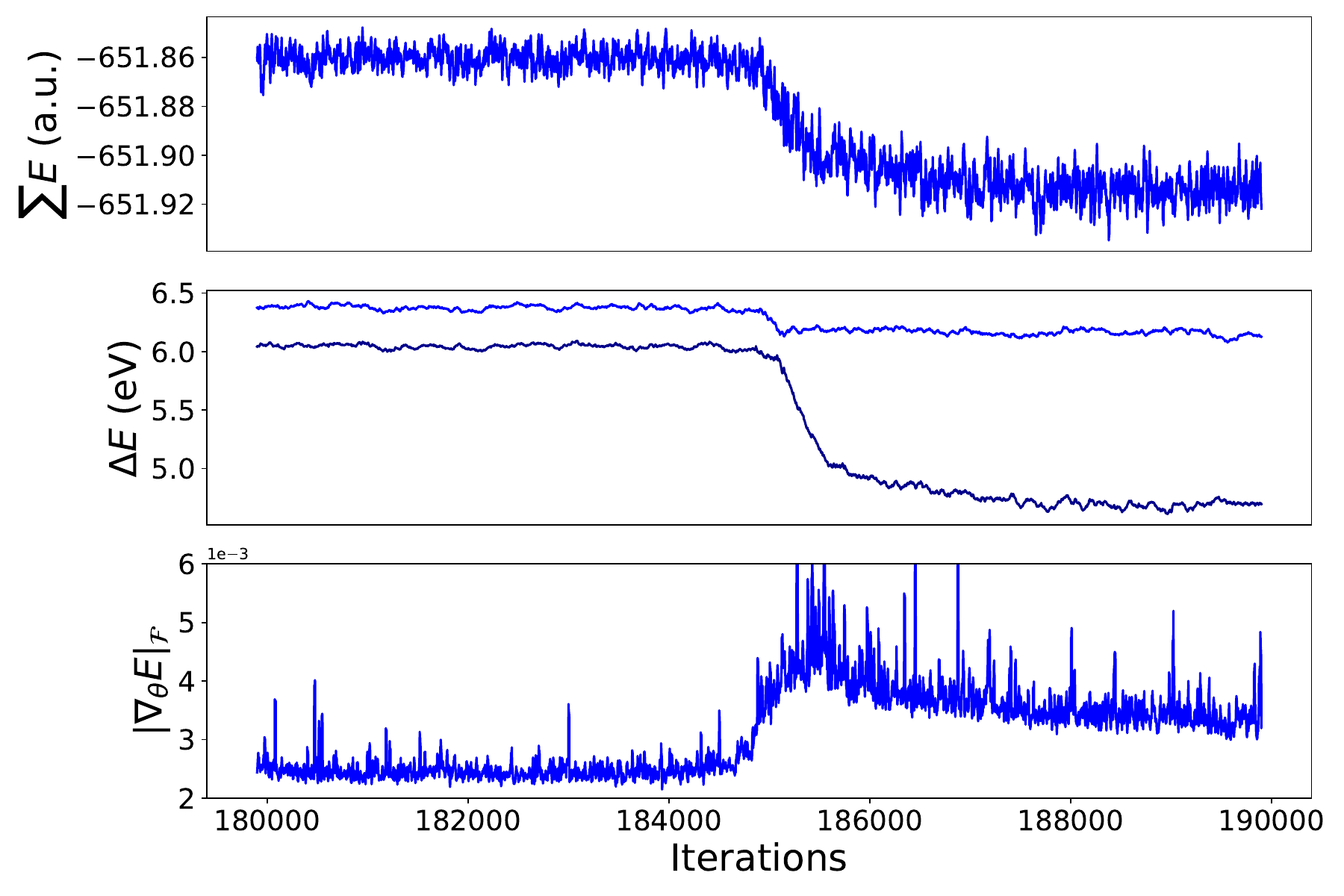} \\
    \caption{Illustration of NES-VMC escaping a saddle point in optimization on nitroxyl. Top: the total energy over states. Middle: the approximate vertical excitation energy of the top two states, estimated by taking a moving average of the local energy matrix and diagonalizing. Bottom: the norm of the gradient of the energy with respect to the metric defined by the approximate Fisher information matrix used by KFAC. The network is initially converged to the wrong state, missing the double excitation. The gradient norm increasing transiently as the objective drops is the expected behavior at a saddle point. The results are from the Psiformer with ordered pretraining.}
    \label{fig:saddle_point_escape}
\end{figure*}

In ordered pretraining, we compute all the single and double excitations of the Hartree-Fock solution and order them by energy, and pretrain the single-state Ans{\"atze} to match the Slater determinants of the $k$ lowest excitations. We also use a cc-pVDZ basis set rather than STO-6G to guarantee a sufficient number of single and double excitations. While the excitations found by Hartree-Fock do not necessarily correspond to the true lowest excitations, pretraining is primarily for finding a low-energy non-degenerate starting point, and the Ansatz would usually converge uniformly to the true lowest energy states. For some systems, the Ansatz initialized by ordered pretraining would initially converge towards states of higher energy, which are saddle points of the total energy rather than minima. Most of the time, the Ansatz would eventually escape from this saddle point. An example is shown in Fig.~\ref{fig:saddle_point_escape} of nitroxyl with the Psiformer, where the energy plateaus, followed by a sudden drop. For an even smaller number of systems, but especially those with double excitations, the optimization would remain stuck at this saddle point, and so we investigated alternative pretraining strategies to avoid this saddle point entirely.

\subsection{Random Pretraining}

In random pretraining, each single-state Ansatz is pretrained to match a Slater determinant where the orbitals are a random combination of Hartree-Fock orbitals. An entirely random set of orbitals would have extremely high energy, so we leave the core electron orbitals unchanged, but take a random linear combination of occupied valence electron orbitals plus the two lowest unoccupied orbitals from Hartree-Fock for the remaining orbitals. Weights were sampled from a normal distribution with variance given by the inverse of the number of orbitals. Although this does not lead to orthogonal initial states, it does lead to linearly independent states, which is sufficient for initialization. While this pretraining strategy leads to a higher initial total energy than the ordered pretraining, it was effective at avoiding getting stuck at saddle points for systems with double excitations like nitroxyl (HNO), nitrosomethane, butadiene and glyoxal. We used the random pretraining exclusively on systems for which we had encountered issues with ordered pretraining.

\subsection{Comparison on Small Molecules}
\label{sec:penalty_and_pretrain_results}

\begin{figure*}[]
    \centering
     \hspace*{-1.5cm}
    \includegraphics[width=0.8\textwidth]{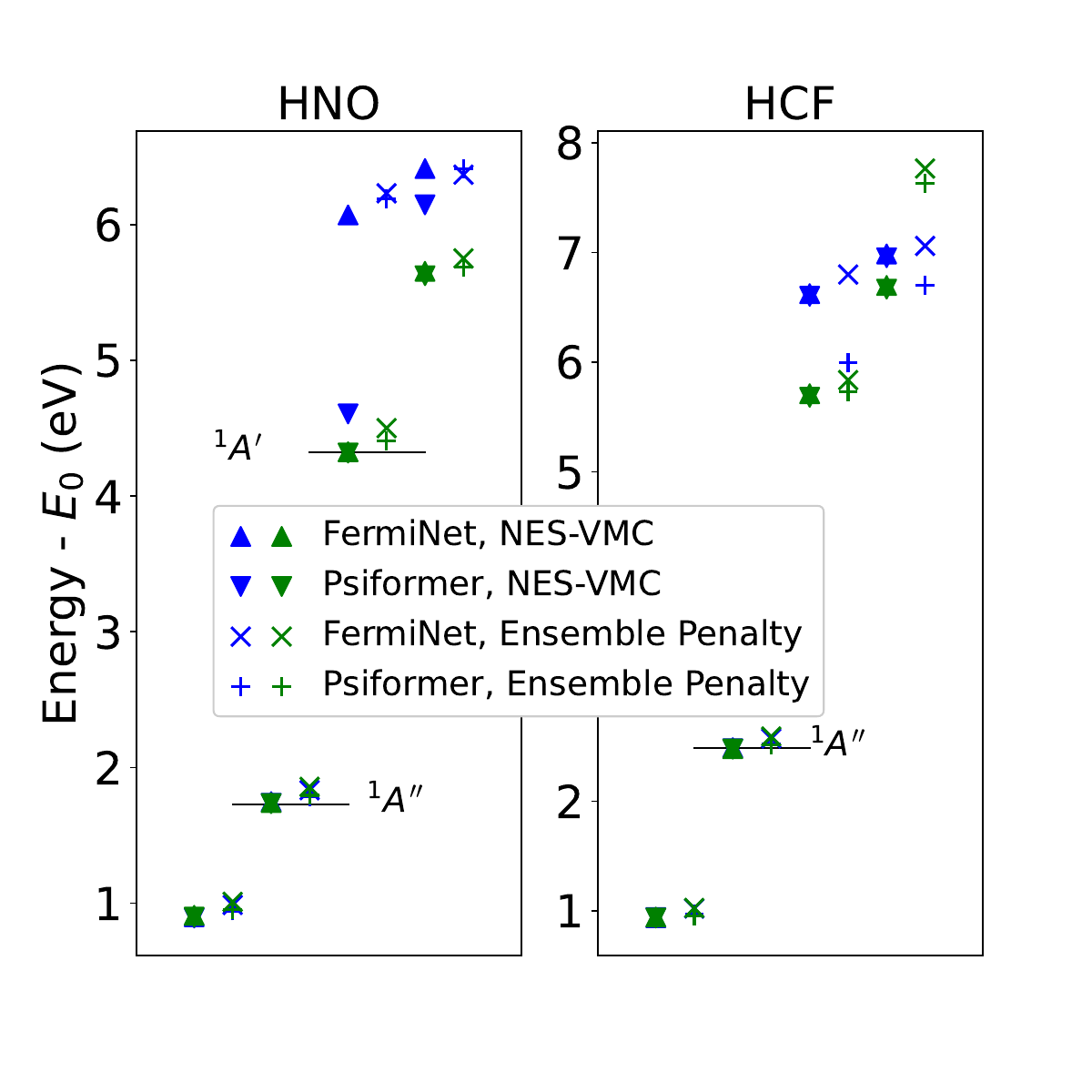} \\
    \caption{Comparison of pretraining strategies on systems from Fig.~\ref{fig:oscillator_strengths} for which ordered pretraining missed states, with both NES-VMC and the ensemble penalty method. Blue denotes ordered pretraining and green denotes random pretraining. Full numerical results for NES-VMC are included in Table~\ref{tab:oscillator_strengths} and for the ensemble penalty method in Table~\ref{tab:penalty}.}
    \label{fig:pretraining_and_penalty}
\end{figure*}

We compared the different pretraining methods on nitroxyl (HNO) and fluoromethylene (HCF) with both the FermiNet and Psiformer, using both NES-VMC and the method described in Sec.~\ref{sec:penalty}. Results are shown in Fig~\ref{fig:pretraining_and_penalty}, with ordered pretraining in blue and random pretraining in green. Numerical results for NES-VMC are included in Table~\ref{tab:oscillator_strengths}, and results for the ensemble penalty method are in Table~\ref{tab:penalty}. For both systems, ordered pretraining leads to a state being missed, while random pretraining leads to all states being accurately computed, including the $^1A'$ double excitation in HNO at 4.32 eV and the singlet state of HCF at 5.65 eV. These results are consistent across multiple runs.

Critically, the sensitivity to initialization is a general feature of how we optimize neural network Ans{\"a}tze, and {\em not} an issue specific to NES-VMC. The ensemble penalty method misses the exact same states missed by NES-VMC. In some cases the ensemble penalty method gets the ordering of the states wrong, though we show them in order for clarity. Additionally, the ensemble penalty method is less numerically accurate than NES-VMC most of the time, reaching chemical accuracy on only a single identifiable state. The mean absolute error relative to the TBE for the states identifiable by both ensemble penalty methods and NES-VMC (so excluding the $^1A'$ state of HNO with the Psiformer and ordered pretraining, which the ensemble penalty method missed) was 3.2 eV for the ensemble penalty method and 0.25 eV for NES-VMC -- an order of magnitude difference. We were not able to successfully run the ensemble method on larger systems like benzene. While further improvements to penalty methods may lead to performance comparable to NES-VMC in the future, at this moment in time it is fair to say that natural excited states is the only method capable of reaching high accuracy on larger systems with deep neural networks.

\section{Numerical Stability}

Computation of the local energy becomes numerically unstable at nodes (i.e. when the Ansatz is zero). For variational Monte Carlo calculation of ground states, because walkers are sampled from $\psi^2$, it is vanishingly unlikely that a walker will be found close enough to a node that the local energy becomes undefined. However, in natural excited states VMC, the walkers are sampled from $\Psi^2 = \mathrm{det}\left(\psimat\right)^2$, so even if $\Psi^2$ is nonzero, some of the entries $\psi_i(\mathbf{x}^j)$ in the matrix $\psimat$ might be zero up to numerical precision. Since we compute the local energy matrix by first computing local energies in the log domain, then multiplying by $\psi_i(\mathbf{x}^j)$, NaNs would appear in the local energy matrix at points where $\psi_i(\mathbf{x}^j)=0$. To avoid this, we introduce a number of heuristics to stabilize optimization. First, if any entries of $\psimat$ are zero, we run additional steps of MCMC (up to 10 times per iteration, to avoid an infinite loop). Secondly, if the local energy of any of the walkers is undefined, we remove that walker from the computation of energies and gradients in that step. Lastly, if the gradient still has undefined values, that iteration of optimization is skipped. If more than 100 iterations are skipped in a row, the optimization fails, but we did not find that this happened in any of our experiments. Out of all these heuristics, only the last should really be necessary, as it is more general than the first two.

When calculating the matrices $\psimat$ and $\oppsimat$, we always compute the values of $\psi_i(\mathbf{x}^j)$ and the local operator $O^i_L(\mathbf{x}^j) = \psi_i^{-1}(\mathbf{x}^j) \hat{\mathcal{O}} \psi_i(\mathbf{x}^j)$ in the log domain, then subtract the maximum value of $\mathrm{log} \psi_i(\mathbf{x}^j)$ over $i$ and $j$ from the values of both $\mathrm{log} \psi_i(\mathbf{x}^j)$ and $O^i_L(\mathbf{x}^j) + \mathrm{log} \psi_i(\mathbf{x}^j)$ before converting from the log domain to real domain. This avoids issues with numerical underflow or overflow that can occur when computing $\psi$ and $\hat{\mathcal{O}}\psi$ directly in the real domain.

For larger systems ($>$25 electrons: butadiene, glyoxal, tetrazine, cyclopentadienone and benzene) we found that the distribution of local energies became very heavy-tailed, such that taking the empirical mean of the local energy matrix for all walkers could lead to errors on the order of several mHa or $\sim$0.1 eV. To mitigate this, we took the median of elements of the local energy matrices over minibatches of 16 walkers, and took the empirical mean over these medians. This improved the accuracy for the majority of states of large systems. For the rest of the systems in this paper, the empirical mean over all walkers was used.

\section{Pseudopotentials}

For molecules with heavier atoms in Sec.~\ref{sec:oscillator_strengths} such as HCl, thioformaldehyde and silylidene, we replaced the core electrons of second-row atoms ($Z=11\ldots 18$) with a pseudopotential, or effective core potential. We found that this generally led to more accurate results than all-electron calculations, possibly because the energy scale of the core electrons is far greater than the valence electrons for heavier atoms. We followed the general approach for VMC of approximating the core electrons by a nonlocal one-electron potential\cite{fahy1990variational}, specifically adapting the correlation-consistent effective core potential (ccECP)\cite{bennett2018new} from PySCF. A similar approach has been demonstrated with neural network Ans{\"a}tze to be effective at modeling transition metals\cite{li2022fermionic}.

\section{Hyperparameters}

For all experiments, we stayed close to the default hyperparameters used in the original papers on the FermiNet\cite{pfau2020ab, spencer2020better} and Psiformer\cite{von2023self}. We used neural network Ans{\"atze} with 16 dense determinants, 4 layers and 256 units per layer in the one-electrons stream. For the FermiNet we used 32 units per layer in the two-electron stream, and for the Psiformer we used 4 self-attention heads with 64 units per head, and layer normalization. We follow the gradient scaling from the Psiformer paper, and hence use a learning rate of 0.05.

For some larger systems, we increased the number of pretraining steps from 10,000 to 50,000 or 100,000, in line with previous results suggesting that this improved overall accuracy\cite{von2023self}. While we did not find this made a significant difference, we include details for completeness. We also found that {\em relative} energies were almost always well converged after 100,000 iterations, so for some larger systems we stopped training then, rather than for the full 200,000 iterations recommended for optimal absolute energies.

Ordered pretraining was used by default, except for two places. Firstly, for HNO and HCF in Fig.~\ref{fig:oscillator_strengths}, random pretraining was found to converge to lower states more robustly, as discussed in Sec.~\ref{sec:penalty_and_pretrain_results}. Secondly, in all the experiments in Sec.~\ref{sec:double_excitations} we tried both random and ordered pretraining and reported the best results. We found that, similarly to HNO, random pretraining was more likely to robustly converge to the double excitations on nitrosomethane, butadiene and glyoxal. On tetrazine and cyclopentadienone, where we added an extra spin magnitude term to the Hamiltonian to remove triplet states (see Sec.~\ref{sec:singlet_targeting}), we actually found that the ordered pretraining was generally more effective. For cyclopentadienone with the Psiformer, training with ordered pretraining experienced numerical instability in the middle of training, but we found that random pretraining also worked well and included those results.

In each MCMC step, electrons were sampled in blocks rather than one at a time or all simultaneously, as described in the Psiformer paper. Each electron set was treated as at least one block, and for some larger systems, each electron set was further subdivided into multiple blocks. A summary of the different settings used for different experiments is given in Table~\ref{tab:hyperparams}.

\section{Scaling and Time Complexity}

In theory, the asymptotic scaling of the NES-VMC optimization as the number of excited states grows is dominated by the calculation of the determinant when evaluating $\Psi = \mathrm{det}(\psimat)$. This scales as $\mathcal{O}(K^3)$ for a single forward evaluation, and $\mathcal{O}(K^4)$ for evaluating the total local energy. However in practice, the computation of the determinant is relatively cheap at the scales investigated in this paper, up to $K=10$, and the scaling is dominated by the cost of computing the matrix of local energies needed for $\hpsimat$. Thus a quadratic scaling more accurately reflects the empirical behavior of most experiments in this paper.

While it does not affect the asymptotic complexity of the calculation, different approaches to computing the kinetic energy operator can speed up the calculations by a significant constant factor. In the original FermiNet work, the local kinetic energy was calculated by first computing the gradient of the Ansatz by backpropagation, then looping over computing different columns of the Hessian through a forward Jacobian-vector product. Later, it was shown that computing the local kinetic energy in a single forward pass can be efficient in both time and memory \cite{li2024forward}, and we integrated our code with the open-source package Folx to take advantage of this \cite{gao2023folx}. Many of the largest experiments in the paper use this forward Laplacian method, which in our experience leads to a 2-5x speedup depending on the system. The numerical results were identical to within expected run-to-run variability using either method.

\begin{figure*}[]
    \centering
    \includegraphics[width=0.6\textwidth]{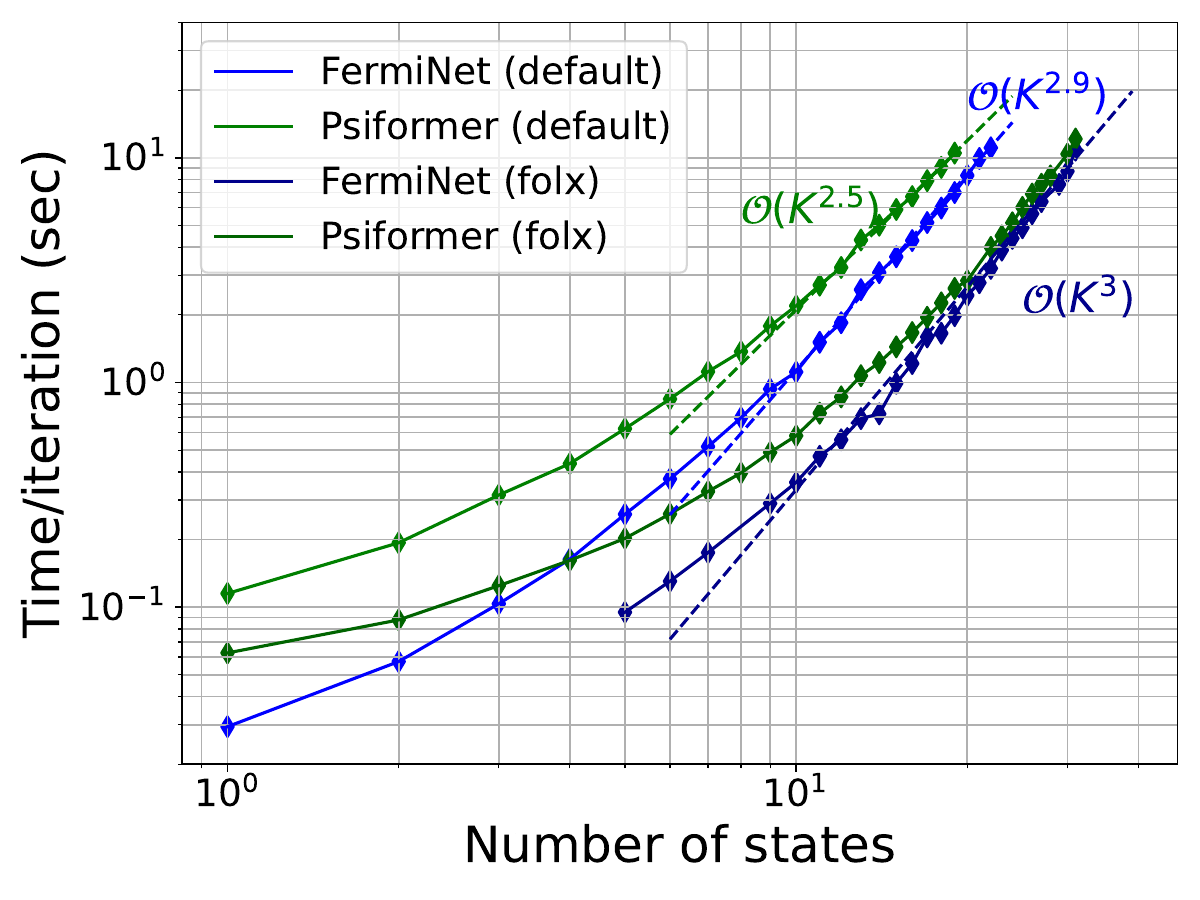} \\
    \caption{Time per iteration of NES-VMC optimization on the neon atom on one A100 GPU with a batch size of 64 as a function of the number of excited states. The dashed lines illustrate the polynomial scaling for different Ans{\"a}tze. Using Folx to calculate the kinetic energy leads to a roughly 4x speedup, and comes closer to the expected asymptotically cubic scaling.}
    \label{fig:timings}
\end{figure*}

In Fig~\ref{fig:timings}, we evaluate the empirical scaling with number of states of NES-VMC applied to the neon atom, with both the original Laplacian calculation and forward Laplacian with Folx. All calculations were done with a batch size of 64 on a single A100 GPU, which is much smaller scale than the experiments elsewhere in the paper, which allowed us to investigate scaling for larger $K$ without having to worry about accuracy. It can be seen that for both the FermiNet and Psiformer, the asymptotic time complexity seems to be slightly below cubic with the original Laplacian method, but this scaling does not dominate until $K>10$. With Folx, we were able to calculate a larger number of states, and the complexity is roughly cubic, while all calculations were roughly 4x faster than their counterpart with the original Laplacian method. While the cubic scaling may seem costly, it is worth noting that state-of-the-art eigensolvers for finite matrices such as the Lanczos method scale cubically with the number of eigenvalues as well. When $K<10$, the roughly quadratic scaling of NES-VMC is comparable to penalty methods, which still require explicit computation of the overlap matrix, and do not yet work as well as NES-VMC with neural networks.

\begin{figure*}[]
    \centering
    \hspace*{-1.5cm}
    \includegraphics[width=1.2\textwidth]{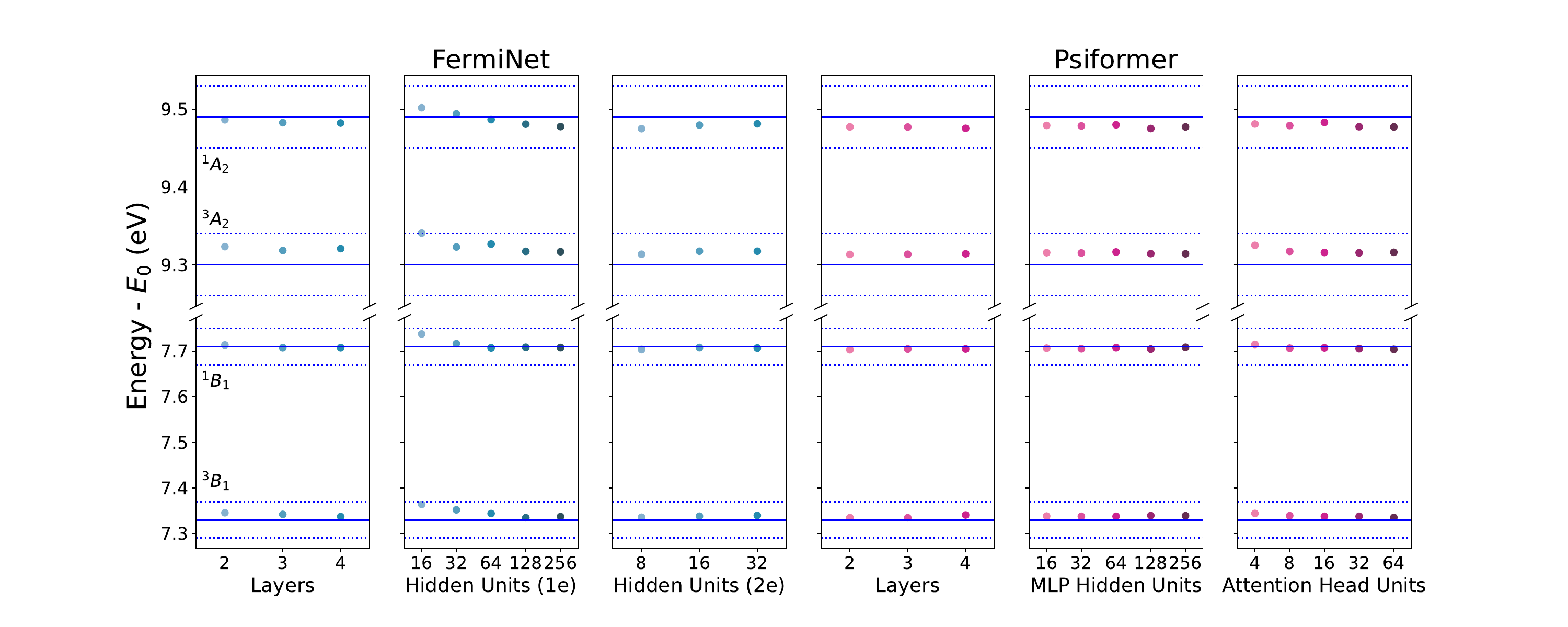} \\
    \caption{Comparison of vertical excitation energies on H$_2$O for various network sizes for the FermiNet and Psiformer. Dotted lines denote chemical accuracy. Consistent with results on ground state energies of the FermiNet\cite{pfau2020ab}, the width of the one-electron stream (1e) is the primary factor in convergence to the ground truth. The Psiformer is more robust than the FermiNet to reduction in the width of the intermediate layers.}
    \label{fig:water_ablations}
\end{figure*}

Finally, the consideration of time complexity here is only for networks of a fixed size, while networks may need to increase in size and capacity to handle either more complex systems or larger numbers of states. To evaluate the effect of network size on numerical accuracy, we looked at H$_2$O, as it is relatively small and we have ground truth data for the first 4 excited states, and varied three parameters of both the FermiNet and Psiformer relative to the default architecture used throughout the paper. For both Ans{\"a}tze we reduced the number of layers and the width of the one-electron stream. For the FermiNet, we also reduced the number of units in the two-electron stream, while for the Psiformer we reduced the number of units in each attention head, as there is no two-electron stream in the Psiformer. Previous results suggested that changing either the number of units in each head or the number of heads has a similar effect \cite{von2023self}. Results are shown in Fig.~\ref{fig:water_ablations}. Consistent with previous results on the FermiNet, we found that the network was most sensitive to changing the width of the one-electron stream, while neither the number of layers nor number of units in the two-electron stream had much impact on the results. For the Psiformer, even changing the width of the MLP layers did not have a significant effect, and changing the number of units in the attention heads only seemed to make a difference in the limit of extremely few units. Overall, the results are very robust to network ablations. This suggests that the Ansatz we used was much larger than necessary for the smallest systems in this paper, and it remains to be seen how large we can push the system sizes considered before the Ansatz needs to be made larger.

\section{Variance Estimation}

Because the energies of individual states are not computed directly, the estimation of the variance of the energy for different states must be treated with some care. In addition to computing the expected value of the local energy matrix $\mathbf{E}_L$, we compute the covariance matrix of the vectorized local energy matrix $\mathbf{\Sigma}_{\mathbf{E}_L} = \mathrm{Cov}\left[\mathrm{vec}(\mathbf{E}_L)\right]\in\mathbb{R}^{K^2\times K^2}$. Then after diagonalizing the energy matrix $\mathbb{E} [\mathbf{E}_L] = \mathbf{U} \mathbf{\Lambda} \mathbf{U}^{-1}$, we can transform the covariance to give the covariance of $\mathbf{\Lambda}$ as $\mathbf{\Sigma}_\mathbf{\Lambda} = \left(\mathbf{U}^{-1} \otimes \mathbf{U}^T\right)\mathbf{\Sigma}_{\mathbf{E}_L} \left(\mathbf{U}^{-1} \otimes \mathbf{U}^T\right)^T$, since $\mathbf{\Lambda} = \mathbf{U}^{-1} \mathbb{E} [\mathbf{E}_L] \mathbf{U}$ can be expressed in vectorized form as $\mathrm{vec}(\mathbf{\Lambda}) = \left(\mathbf{U}^{-1} \otimes \mathbf{U}^T\right)\mathrm{vec}\left(\mathbb{E} [\mathbf{E}_L]\right)$. The diagonal of $\mathbf{\Sigma}_{\mathbf{\Lambda}}$ gives the marginal variance of each element of $\mathrm{vec}(\mathbf{\Lambda})$, and reshaping the diagonal as a $K\times K$ matrix, the diagonal of {\em this} matrix gives the variance of the individual energies of each state. A similar procedure can be used to estimate the uncertainty in other observables.

\section{Observables}

As described in Sec.~\ref{sec:nes-vmc-supplemental}, observables can be evaluated by computing the matrix $\mathbb{E}_{\Psi^2}\left[\psimat^{-1} \oppsimat\right]$ and then separating the different states based on the eigenvectors of the energy matrix $\mathbb{E}_{\Psi^2}\left[\psimat^{-1} \hpsimat\right]$. This applies to off-diagonal observables $\langle \psi_i \op \psi_j \rangle$ as well, with some additional effort. The exact computational details of the observables considered in this paper -- the spin magnitude $\mathcal{S}^2$, the density matrix and the dipole moment -- require some additional elaboration, which we provide here.

\subsection{Spin Magnitude}
\label{sec:spin_magnitude}

In first quantization, the total spin magnitude operator $\hat{\mathcal{S}}^2$ is defined as \cite{lowdin1955s2, wang1995s2}: 

\begin{equation}
    \hat{\mathcal{S}}^2 = -\frac{N(N-4)}{4} + \sum_{i<j} \hat{P}_{ij}
    \label{eqn:s2}
\end{equation}
where $\hat{P}_{ij}$ is the operator which swaps the spins of particles $i$ and $j$. In all our experiments, the particles were spin-assigned, so the exact number of $\alpha$ and $\beta$ electrons is known. The expected value of $\hat{\mathcal{S}^2}$ can be evaluated by Monte Carlo for a single state as:

\begin{align}
    \braket{\hat{\mathcal{S}}^2} = &-\frac{\Na - \Nb}{4}\left(\Na - \Nb + 2\right) + \Nb \nonumber\\
    &-\mathbb{E}_{\psi^2}\left[\sum_{\substack{i = 1\ldots \Na \\j = \Na+1 \ldots N}} \frac{\psi(\ldots, \sigma_i \mathbf{r}_j, \ldots, \sigma_j \mathbf{r}_i, \ldots)}{\psi(\ldots, \sigma_i \mathbf{r}_i, \ldots, \sigma_j \mathbf{r}_j, \ldots)} \right]
    \label{eqn:s2_monte_carlo}
\end{align}
where $\sigma \mathbf{r}$ denotes a particle state with the spin $\sigma$ and position $\mathbf{r}$. Note that in Eq~\ref{eqn:s2_monte_carlo}, we swap the particle positions rather than spins, and add a minus sign to the expectation. This makes the implementation simpler for spin-assigned wavefunctions where the first $\Na$ inputs are assumed to have spin $\alpha$.

\begin{figure*}[]
    \centering
    \includegraphics[width=0.6\textwidth]{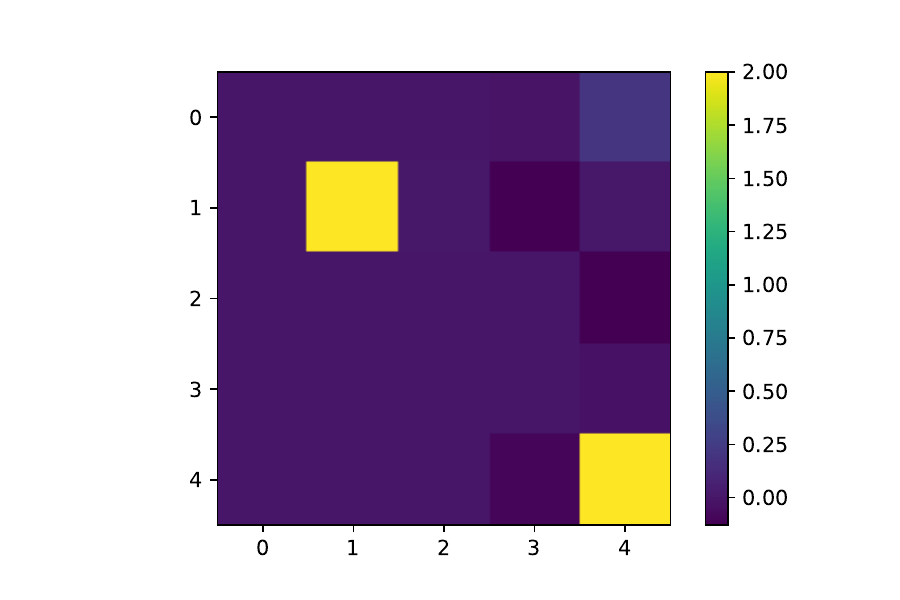} \\
    \caption{Matrix of $\langle\hat{\mathcal{S}^2}\rangle$ values for the first 5 excited states of ethylene at equilibrium geometry. Two triplet states can clearly be identified from the diagonal. The off-diagonal terms are all near zero.}
    \label{fig:ethene_s2}
\end{figure*}

To compute this for all excited states, we compute the matrices $\hat{\mathcal{S}^2}\psimat(\mathbf{x})$ and $\psimat(\mathbf{x})$, first in the log domain, then subtract off the largest value of $\mathrm{log}\psi_i(\mathbf{x}^j)$ from both matrices for numerical stability, then accumulate the expectation $\mathbb{E}_{\Psi^2}[\psimat^{-1} \hat{\mathcal{S}^2}\psimat]$ and finally demixing the states $\mathbf{U}^{-1}\mathbb{E}_{\Psi^2}[\psimat^{-1} \hat{\mathcal{S}^2}\psimat]\mathbf{U}$ where $\mathbf{U}$ are the eigenvectors of the energy matrix. Since $\hat{\mathcal{S}^2}$ and $\hat{H}$ commute, the demixed matrix should be diagonal, and we do not need to worry about correcting for the nonidentifiability of $\Sigma$. An example result for ethylene is shown in Fig.~\ref{fig:ethene_s2}.

\subsection{Density Matrices and Natural Orbitals}
\label{sec:dm_and_no}

The one-electron reduced density matrix is defined as

\begin{align}
    &\Gamma(\mathbf{x}_1; \mathbf{x}'_1) = \nonumber\\
    &\int d\mathbf{x}_2 \ldots d\mathbf{x}_N \psi(\mathbf{x}_1, \mathbf{x}_2, \ldots, \mathbf{x}_N) \psi(\mathbf{x}'_1, \mathbf{x}_2, \ldots, \mathbf{x}_N)
    \label{eqn:1e-rdm}
\end{align}
The eigenfunctions of this operator are the {\em natural orbitals}, which play a role similar to marginal distributions for probability densities, and it is common to use natural orbitals as a way to analyze the qualitative nature of different Ans{\"a}tze\cite{lewart1988single}. For a single Slater determinant wavefunction, the natural orbitals are equal to the true orbitals, up to linear combination. To work with the density matrix in a computationally tractable manner, we can project it into an orthogonal basis set $\phi_1(\mathbf{x}), \ldots \phi_M(\mathbf{x})$. Elements of this projected density matrix $\Gamma_{ij}$ are given by

\begin{equation}
    \Gamma_{ij} = \int d\mathbf{x}_1 d\mathbf{x}'_1 \phi_i(\mathbf{x}_1) \phi_j(\mathbf{x}'_1) \Gamma(\mathbf{x}_1; \mathbf{x}'_1)
    \label{eqn:1e-rdm-projected}
\end{equation}
which can be rewritten as the expectation of an observable:

\begin{align}
    \Gamma_{ij} &= \braket{\hat{\Gamma}_{ij}} \\
    \hat{\Gamma}_{ij}(\mathbf{x}; \mathbf{x}') &= \phi_i(\mathbf{x}_1) \phi_j(\mathbf{x}'_1) \prod_{k=2}^N \delta(\mathbf{x}_k - \mathbf{x}'_k)
    \label{eqn:1e-rdm-observable}
\end{align}
Evaluating this observable by the formula outlined in Sec.~\ref{sec:nes-vmc-supplemental} takes some care, as $\hat{\Gamma}_{ij}$ is a nonlocal operator. This means that Monte Carlo estimation of $\hat{\Gamma}_{ij}\psimat$ must be done by integrating over values of $\mathbf{x}'_1$ as well as $\mathbf{x}_1, \ldots, \mathbf{x}_N$. In the case of a single state, a Monte Carlo estimator for $\Gamma_{ij}$ is given by
\begin{equation}
    \braket{\hat{\Gamma}_{ij}} = \mathbb{E}_{\mathbf{x} \sim \psi^2, \mathbf{x}'_1\sim\rho}\left[ \frac{\phi_i(\mathbf{x}_1) \phi_j(\mathbf{x}'_1) \psi(\mathbf{x}'_1, \mathbf{x}_2, \ldots)}{\rho(\mathbf{x}'_1)\psi(\mathbf{x}_1, \mathbf{x}_2, \ldots)} \right]
    \label{eqn:1e-rdm-monte-carlo}
\end{equation}
where $\rho$ is an arbitrary probability distribution over single-particle states. While any $\rho$ can work in theory, a choice of $\rho$ which more closely matches the electron density will give an estimator with lower variance.

In the case of multiple excited states, we calculate not a single density matrix but an entire rank-4 tensor in $\mathbb{R}^{M \times M \times K \times K}$ over both basis set elements and excited states. We sample $K$ sets of $N$ particles $\mathbf{x}^1_1, \ldots, \mathbf{x}^K_N$ from $\Psi^2$ as with local observables, but additionally sample $K$ particles $x'^{1}_1, \ldots, x'^{K}_1 \sim \rho$ i.i.d. The matrix $\psimat$ is computed as in the local observable case, but calculating $\hat{\Gamma}_{ij}\psimat \in \mathbb{R}^{K \times K}$ is more complicated. For the excited state $k$ and particle set $\ell$, the appropriate element of $\hat{\Gamma}_{ij}\psimat$ is given by:

\begin{equation}
    \hat{\Gamma}_{ij}\Psi_{k\ell} = \phi_i(\mathbf{x}^\ell_1) \phi_j(\mathbf{x}'^\ell_1) \rho^{-1}(\mathbf{x}'^\ell_1) \psi^k(\mathbf{x}'^\ell_1, \mathbf{x}^\ell_2, \ldots, \mathbf{x}^\ell_N)
\end{equation}
Finally, as with all other observables, we must unmix the accumulated $\mathbb{E}[\psimat^{-1} \hat{\Gamma}_{ij}\psimat]$ by the eigenvectors of the energy matrix. The diagonals of the demixed matrices can then be assembled into density matrices $\Gamma^k_{ij}$ for each state $k$, and then {\em these} can be diagonalized to recover natural orbitals. 

In all experiments in the paper, we used the Hartree-Fock orbitals as computed by PySCF \cite{sun2018pyscf} in the def2-TZVPD basis set as the orthogonal basis set elements $\phi_i$, and used the Hartree-Fock electron density $\rho(\mathbf{x}_1) = \frac{1}{N} \sum_{i=1}^N \phi_i^2(\mathbf{x}_1)$ as the sampling distribution for $\mathbf{x}'_1$. In practice, we were interested in the natural orbitals for different spin electrons. This meant we would run unrestricted Hartree-Fock, and compute the density matrices twice, once with the $\alpha$-electron orbitals and once with the $\beta$-electron orbitals. In the latter case, we had to be sure to substitute $\mathbf{x}'_1$ for a $\beta$ electron rather than an $\alpha$ electron in the spin-assigned wavefunction.

\subsection{Off-Diagonal Expectations and Dipole Moments}
\label{sec:appendix_oscillator}

For $\hat{\mathcal{S}}^2$ and $\hat{\Gamma}_{ij}$, we primarily care about the expectations for single states $\braket{\psi_i \hat{\mathcal{O}} \psi_i}$. For computing oscillator strengths, we must work out the transition dipole moments, which are defined as expectations between different states $\braket{\psi_i \hat{\mathcal{O}} \psi_j}$. The oscillator strength between states $i$ and $j$ is a dimensionless quantity defined as

\begin{equation}
    f_{ij} = \frac{2}{3}\frac{m}{\hbar^2} (E_i - E_j) \sum_{k \in x, y, z} \braket{\psi_i q \hat{r}_k \psi_j}^2
\end{equation}
where $m$ is the particle mass and $q$ is the particle charge, both 1 for electrons in atomic units (as is $\hbar$) and $\hat{r}_x$, $\hat{r}_y$ and $\hat{r}_z$ are the operators for the sum of the x, y and z positions of all particles in the system. The transition dipole moment $\mathbf{d}^{ij}$ is the 3-vector made up of these off-diagonal expectations $d^{ij}_k = \braket{\psi_i q \hat{r}_k \psi_j}$. The Monte Carlo estimator for each element of the transition dipole moment vector is quite simple: $\hat{r}_k\psimat = \mathbf{R}_k \psimat$ where $\mathbf{R}_k$ is a diagonal matrix with $\sum_i r_{ik}^\ell$ in the $\ell$th diagonal -- that is, the sum of the position along coordinate $k$ taken over all electrons in particle set $\ell$. Note that we have dropped the charge, as we are in atomic units. The only additional wrinkle in comparison to other observables we have discussed is that, after demixing the accumulated matrix by the eigenvectors of the energy matrix $\mathbf{\Delta}_k = \mathbf{U}^{-1} \mathbb{E}_{\Psi^2}[\psimat^{-1} \mathbf{R}_k \psimat] \mathbf{U}$, we must multiply this matrix elementwise by its transpose $\mathbf{\Delta}_k^2 = \mathbf{\Delta}_k \circ \mathbf{\Delta}_k^T$ to cancel out a nonidentifiable scaling term. This gives the off-diagonal expectations squared $\braket{\psi_i \hat{r}_k \psi_j}^2$, which is precisely the term needed for the oscillator strength.

\newpage\clearpage

\begin{table*}[t]
    \centering
    \begin{tabular}{ccccc}\hline\hline
         System & Ansatz & Pretraining & Blocks/set & Training \\\hline
         Atoms (Fig.~\ref{fig:atomic_spectra}) & Both & 10k & 1 & 200k \\
         Oscillator Strengths (Fig.~\ref{fig:oscillator_strengths}) & Both & 10k & 1 & 200k \\
         Carbon Dimer (Fig.~\ref{fig:carbon-dimer}) & Psiformer & 100k & 1 & 100k \\
         Ethylene (Fig.~\ref{fig:ethene}) & Both & 10k & 1 & 100k \\
         Double Excitations (Fig.~\ref{fig:double_excitations}) & Both & 50k & 2 & 200k \\
         Benzene (Fig.~\ref{fig:benzene}) & FermiNet & 100k & 4 & 200k \\
         Benzene (Fig.~\ref{fig:benzene}) & Psiformer & 100k & 4 & 100k \\
         \hline\hline
    \end{tabular}
    \caption{Settings used for different experiments}
    \label{tab:hyperparams}
\end{table*}

\begin{table*}[t]
    \vspace*{-1cm}
    \hspace*{-1cm}
    \tiny
    \centering
    \bgroup
    \def\arraystretch{0.5}%
    \begin{tabular}{ccccccccccccc}\hline\hline
          & \multicolumn{3}{c}{NES-VMC Energy (Ha)} & \multicolumn{3}{c}{NES-VMC $\Delta E$ (Ha)} & & & & \multicolumn{3}{c}{Error (mHa)} \\
        \cmidrule(lr){2-4} \cmidrule(lr){5-7} \cmidrule(lr){11-13} & \multicolumn{2}{c}{FermiNet} & Psiformer & \multicolumn{2}{c}{FermiNet} & Psiformer & & & & \multicolumn{2}{c}{FermiNet} & Psiformer \\
        
        \cmidrule(lr){2-3} \cmidrule(lr){4-4} \cmidrule(lr){5-6} \cmidrule(lr){7-7} \cmidrule(lr){11-12} \cmidrule(lr){13-13} System & $k=5$ & $k=10$ & $k=10$ & $k=5$ & $k=10$ & $k=10$ & Config. & Term & Expt. & $k=5$ & $k=10$ & $k=10$ \\  \hline
Li & -7.478059(3) & -7.478059(3) & -7.478065(3) &  -- & -- & -- & $2s$ & $^2S$ & -- & -- & -- & -- \\ 
 & -7.410157(5) & -7.410154343(6) & -7.410156(3) &  0.067902(6) & 0.067904(3) & 0.067908(5) & $2p$ & $^2P^\circ$ & 0.067907 & -0.005(6) & -0.002(3) & 0.001(5) \\ 
 & -7.4101481(2) & -7.410154343(6) & -7.410147(4) &  0.067911(3) & 0.067904(3) & 0.067917(5) & '' & '' & 0.067907 & 0.004(3) & -0.002(3) & 0.010(5) \\ 
 & -7.4101481(2) & -7.410145(3) & -7.410144(5) &  0.067911(3) & 0.067914(4) & 0.067921(6) & '' & '' & 0.067907 & 0.004(3) & 0.007(4) & 0.014(6) \\ 
 & -7.35379(1) & -7.354037(5) & -7.354095(5) &  0.12427(1) & 0.124022(6) & 0.123970(6) & $3s$ & $^2S$ & 0.123960 & 0.31(1) & 0.062(6) & 0.010(6) \\ 
 &  & -7.337074(9) & -7.337147(5) &  & 0.14098(1) & 0.140918(6) &  $3p$ & $^2P^\circ$ & 0.140907 & & 0.08(1) & 0.010(6) \\ 
 &  & -7.3370438(7) & -7.337136(5) &  & 0.141015(3) & 0.140928(6) &  '' & '' & 0.140907 & & 0.108(3) & 0.021(6) \\ 
 &  & -7.3370438(7) & -7.337130(4) &  & 0.141015(3) & 0.140935(5) &  '' & '' & 0.140907 & & 0.108(3) & 0.028(5) \\ 
 &  & -7.335469(6) & -7.335513(4) &  & 0.142590(7) & 0.142552(5) &  $3d$ & $^2D$ & 0.142536 & & 0.054(7) & 0.016(5) \\ 
 &  & -7.335454(7) & -7.335500(4) &  & 0.142605(8) & 0.142565(5) &  '' & '' & 0.142536 & & 0.069(8) & 0.029(5) \\  \hline 
Be & -14.667340(8) & -14.667315(7) & -14.667329(6) &  -- & -- & -- & $2s^2$ & $^1S$ & -- & -- & -- & -- \\ 
 & -14.56723(1) & -14.567224(8) & -14.56723(1) &  0.10011(1) & 0.10009(1) & 0.10010(1) & $2s2p$ & $^3P^\circ$ & 0.100149 & -0.04(1) & -0.06(1) & -0.05(1) \\ 
 & -14.5672150(1) & -14.567214(1) & -14.567219(1) &  0.100125(8) & 0.100101(7) & 0.100110(6) & '' & '' & 0.100149 & -0.024(8) & -0.048(7) & -0.039(6) \\ 
 & -14.5672150(1) & -14.567214(1) & -14.567219(1) &  0.100125(8) & 0.100101(7) & 0.100110(6) & '' & '' & 0.100149 & -0.024(8) & -0.048(7) & -0.039(6) \\ 
 & -14.473414(9) & -14.4734154(7) & -14.47344(1) &  0.19393(1) & 0.193899(7) & 0.19389(1) & $2s2p$ & $^1P^\circ$ & 0.193942 & -0.02(1) & -0.043(7) & -0.05(1) \\ 
 &  & -14.4734154(7) & -14.4734211(8) &  & 0.193899(7) & 0.193908(6) &  '' & '' & 0.193942 & & -0.043(7) & -0.034(6) \\ 
 &  & -14.473405(8) & -14.4734211(8) &  & 0.19391(1) & 0.193908(6) &  '' & '' & 0.193942 & & -0.03(1) & -0.034(6) \\ 
 &  & -14.42992(2) & -14.430005(9) &  & 0.23739(2) & 0.23732(1) &  $2s3s$ & $^3S$ & 0.237298 & & 0.10(2) & 0.03(1) \\ 
 &  & -14.41773(2) & -14.41810(1) &  & 0.24958(2) & 0.24923(1) &  $2s3s$ & $^1S$ & 0.249128 & & 0.45(2) & 0.10(1) \\ 
 &  & -14.40802(1) & -14.40815(1) &  & 0.25929(1) & 0.25917(1) &  $2p^2$ & $^1D$ & 0.259175 & & 0.12(1) & -0.00(1) \\  \hline 
B & -24.65383(1) & -24.6537704(5) & -24.65383(1) &  -- & -- & -- & $2s^22p$ & $^2P^\circ$ & -- & -- & -- & -- \\ 
 & -24.653805(1) & -24.6537704(5) & -24.653811(1) &  0.00002(1) & 0.0000000(7) & 0.00002(1) & '' & '' & 0.000000 & 0.02(1) & 0.0000(7) & 0.02(1) \\ 
 & -24.653805(1) & -24.65374(1) & -24.653811(1) &  0.00002(1) & 0.00003(1) & 0.00002(1) & '' & '' & 0.000000 & 0.02(1) & 0.03(1) & 0.02(1) \\ 
 & -24.52198(1) & -24.5219383(1) & -24.521969(7) &  0.13185(1) & 0.1318321(5) & 0.13186(1) & $2s2p^2$ & $^4P$ & 0.131528 & 0.32(1) & 0.3038(5) & 0.33(1) \\ 
 & -24.52196(1) & -24.5219383(1) & -24.521969(7) &  0.13187(1) & 0.1318321(5) & 0.13186(1) & '' & '' & 0.131528 & 0.34(1) & 0.3038(5) & 0.33(1) \\ 
 &  & -24.52190(1) & -24.52196(3) &  & 0.13187(1) & 0.13187(3) &  '' & '' & 0.131528 & & 0.35(1) & 0.34(3) \\ 
 &  & -24.47114(2) & -24.47134(1) &  & 0.18263(2) & 0.18249(1) &  $2s^23s$ & $^2S$ & 0.182388 & & 0.24(2) & 0.10(1) \\ 
 &  & -24.435720(2) & -24.435757(3) &  & 0.218050(2) & 0.21807(1) &  $2s2p^2$ & $^2D$ & 0.218006 & & 0.044(2) & 0.07(1) \\ 
 &  & -24.435720(2) & -24.435757(3) &  & 0.218050(2) & 0.21807(1) &  '' & '' & 0.218006 & & 0.044(2) & 0.07(1) \\ 
 &  & -24.43547(3) & -- &  & 0.21830(3) & -- &  '' & '' & 0.218006 & & 0.29(3) & -- \\ 
 &  & -- & -24.43219(1) &  & -- & 0.22164(2) &  $2s^23p$ & $^2P^\circ$ & 0.221497 & & -- & 0.14(2) \\ \hline 
C & -37.844746(3) & -37.84478(2) & -37.84478(1) &  -- & -- & -- & $2s^22p^2$ & $^3P$ & -- & -- & -- & -- \\ 
 & -37.844746(3) & -37.84472(2) & -37.84473(2) &  0.000000(5) & 0.00007(3) & 0.00004(2) & '' & '' & 0.000000 & 0.000(5) & 0.07(3) & 0.04(2) \\ 
 & -37.84464(2) & -37.84468(2) & -37.84471(1) &  0.00011(2) & 0.00011(3) & 0.00006(2) & '' & '' & 0.000000 & 0.11(2) & 0.11(3) & 0.06(2) \\ 
 & -37.79839(2) & -37.79843(2) & -37.798439(1) &  0.04635(2) & 0.04636(3) & 0.04634(1) & $2s^2p^2$ & $^1D$ & 0.046306 & 0.05(2) & 0.05(3) & 0.03(1) \\ 
 & -37.79833(2) & -37.79840(5) & -37.798439(1) &  0.04641(2) & 0.04638(5) & 0.04634(1) & '' & '' & 0.046306 & 0.11(2) & 0.08(5) & 0.03(1) \\ 
 &  & -37.79839(5) & -37.798431(3) &  & 0.04639(5) & 0.04634(1) &  '' & '' & 0.046306 & & 0.09(5) & 0.04(1) \\ 
 &  & -37.79834(2) & -37.798431(3) &  & 0.04644(2) & 0.04634(1) &  '' & '' & 0.046306 & & 0.14(2) & 0.04(1) \\ 
 &  & -37.79827(2) & -37.79836(1) &  & 0.04651(2) & 0.04642(2) &  '' & '' & 0.046306 & & 0.21(2) & 0.11(2) \\ 
 &  & -37.74617(1) & -37.74623(1) &  & 0.09861(2) & 0.09855(2) &  $2s^2p^2$ & $^1S$ & 0.098501 & & 0.11(2) & 0.04(2) \\ 
 &  & -37.69158(2) & -37.69160(1) &  & 0.15321(2) & 0.15318(2) &  $2s2p^3$ & $^5S^\circ$ & 0.153574 & & -0.37(2) & -0.40(2) \\  \hline 
N & -54.58890(2) & -54.58876(2) & -54.58884(5) &  -- & -- & -- & $2s^22p^3$ & $^4S^\circ$ & -- & -- & -- & -- \\ 
 & -54.50109(2) & -54.50118(3) & -54.50123(3) &  0.08780(4) & 0.08758(4) & 0.08761(6) & $2s^22p^3$ & $^2D^\circ$ & 0.087609 & 0.19(4) & -0.03(4) & 0.00(6) \\ 
 & -54.50108(3) & -54.501102(3) & -54.50120(5) &  0.08781(4) & 0.08766(2) & 0.08764(7) & '' & '' & 0.087609 & 0.21(4) & 0.05(2) & 0.03(7) \\ 
 & -54.501034(1) & -54.501102(3) & -54.50119(2) &  0.08786(2) & 0.08766(2) & 0.08766(5) & '' & '' & 0.087609 & 0.25(2) & 0.05(2) & 0.05(5) \\ 
 & -54.501034(1) & -54.50105(4) & -54.50119(2) &  0.08786(2) & 0.08772(5) & 0.08766(5) & '' & '' & 0.087609 & 0.25(2) & 0.11(5) & 0.05(5) \\ 
 &  & -54.50092(3) & -54.50115(3) &  & 0.08785(4) & 0.08769(6) &  '' & '' & 0.087609 & & 0.24(4) & 0.08(6) \\ 
 &  & -54.45736313(2) & -54.45745(3) &  & 0.13140(2) & 0.13140(6) &  $2s^22p^3$ & $^2P^\circ$ & 0.131401 & & 0.00(2) & -0.00(6) \\ 
 &  & -54.45736313(2) & -54.45742(1) &  & 0.13140(2) & 0.13142(5) &  '' & '' & 0.131401 & & 0.00(2) & 0.02(5) \\ 
 &  & -54.45721(3) & -54.45742(1) &  & 0.13156(4) & 0.13142(5) &  '' & '' & 0.131401 & & 0.16(4) & 0.02(5) \\ 
 &  & -54.20689(6) & -54.20798(3) &  & 0.38187(7) & 0.38086(6) &  $2s^22p^2(^3P)3s$ & $^4P$ & 0.379705 & & 2.17(7) & 1.16(6) \\  \hline 
O & -75.0667572(6) & -75.06672(3) & -75.066849(9) &  -- & -- & -- & $2s^22p^4$ & $^3P$ & -- & -- & -- & -- \\ 
 & -75.0667572(6) & -75.066658(1) & -75.066849(9) &  0.0000000(9) & 0.00006(3) & 0.00000(1) & '' & '' & 0.000000 & 0.0000(9) & 0.06(3) & 0.00(1) \\ 
 & -75.06628(3) & -75.066658(1) & -75.06679(3) &  0.00047(3) & 0.00006(3) & 0.00006(3) & '' & '' & 0.000000 & 0.47(3) & 0.06(3) & 0.06(3) \\ 
 & -74.99477(3) & -74.994713(3) & -74.99488(3) &  0.07198(3) & 0.07201(3) & 0.07197(3) & $2s^22p^4$ & $^1D$ & 0.071944 & 0.04(3) & 0.06(3) & 0.02(3) \\ 
 & -74.99466(3) & -74.994713(3) & -74.9947662(8) &  0.07210(3) & 0.07201(3) & 0.072083(9) & '' & '' & 0.071944 & 0.15(3) & 0.06(3) & 0.138(9) \\ 
 &  & -74.99466(5) & -74.99477(6) &  & 0.07206(6) & 0.07208(6) &  '' & '' & 0.071944 & & 0.12(6) & 0.14(6) \\ 
 &  & -74.99461(5) & -74.9947662(8) &  & 0.07211(6) & 0.072083(9) &  '' & '' & 0.071944 & & 0.17(6) & 0.138(9) \\ 
 &  & -74.99451(4) & -74.99471(6) &  & 0.07220(5) & 0.07214(6) &  '' & '' & 0.071944 & & 0.26(5) & 0.20(6) \\ 
 &  & -74.91306(3) & -74.91316(3) &  & 0.15366(5) & 0.15369(3) &  $2s^22p^4$ & $^1S$ & 0.153615 & & 0.05(5) & 0.07(3) \\ 
 &  & -74.72779(8) & -74.72948(4) &  & 0.33893(9) & 0.33737(4) &  $2s^22p^3(^4S^\circ)3s$ & $^5S^\circ$ & 0.335757 & & 3.17(9) & 1.61(4) \\  \hline 
F & -99.73317(4) & -99.73293(7) & -99.73338(3) &  -- & -- & -- & $2s^22p^5$ & $^2P^\circ$ & -- & -- & -- & -- \\ 
 & -99.73306(8) & -99.732819(2) & -99.73322(5) &  0.00011(9) & 0.00011(7) & 0.00016(6) & '' & '' & 0.000000 & 0.11(9) & 0.11(7) & 0.16(6) \\ 
 & -99.73302(7) & -99.732819(2) & -99.73318(5) &  0.00014(9) & 0.00011(7) & 0.00021(6) & '' & '' & 0.000000 & 0.14(9) & 0.11(7) & 0.21(6) \\ 
 & -99.26238(8) & -99.26406(7) & -99.26543(9) &  0.47079(9) & 0.4689(1) & 0.4679(1) & $2s^22p^4(^3P)3s$ & $^4P$ & 0.466728 & 4.06(9) & 2.1(1) & 1.2(1) \\ 
 & -99.26186(9) & -99.26347(8) & -99.26539(9) &  0.4713(1) & 0.4695(1) & 0.4680(1) & '' & '' & 0.466728 & 4.6(1) & 2.7(1) & 1.3(1) \\ 
 &  & -99.26334(8) & -99.26524(4) &  & 0.4696(1) & 0.46815(5) &  '' & '' & 0.466728 & & 2.9(1) & 1.42(5) \\ 
 &  & -99.2537(1) & -99.255272(1) &  & 0.4792(1) & 0.47811(3) &  $2s^22p^4(^3P)3s$ & $^2P$ & 0.477070 & & 2.1(1) & 1.04(3) \\ 
 &  & -99.25364(8) & -99.255272(1) &  & 0.4793(1) & 0.47811(3) &  '' & '' & 0.477070 & & 2.2(1) & 1.04(3) \\ 
 &  & -99.25335(8) & -99.25526(4) &  & 0.4796(1) & 0.47813(5) &  '' & '' & 0.477070 & & 2.5(1) & 1.06(5) \\ 
 &  & -99.1943(1) & -99.19580(6) &  & 0.5387(1) & 0.53758(7) &  $2s^22p^4(^3P)3p$ & $^4P^\circ$ & 0.527905 & & 10.8(1) & 9.68(7) \\  \hline 
Ne & -128.93712(5) & -128.93677(7) & -128.93707(4) &  -- & -- & -- & $2p^6$ & $^1S$ & -- & -- & -- & -- \\ 
 & -128.32430(6) & -128.3239(2) & -128.3243(3) &  0.61282(8) & 0.6129(2) & 0.6128(3) & $2p^5(^2P^\circ_{3/2})3s$ & $^2[3/2]^\circ$ & 0.611453 & 1.37(8) & 1.5(2) & 1.4(3) \\ 
 & -128.32405(6) & -128.3238(3) & -128.3243(3) &  0.61307(8) & 0.6130(3) & 0.6128(3) & '' & '' & 0.611453 & 1.61(8) & 1.5(3) & 1.4(3) \\ 
 & -128.31737(6) & -128.3236(1) & -128.32422(6) &  0.61975(8) & 0.6131(1) & 0.61285(7) & '' & '' & 0.611453 & 8.30(8) & 1.7(1) & 1.40(7) \\ 
 & -128.31712(7) & -128.31714(7) & -128.317581(1) &  0.62000(9) & 0.6196(1) & 0.61949(4) & $2p^5(^2P^\circ_{1/2})3s$ & $^2[1/2]^\circ$ & 0.617936 & 2.06(9) & 1.7(1) & 1.56(4) \\ 
 &  & -128.31700(8) & -128.317581(1) &  & 0.6198(1) & 0.61949(4) &  '' & '' & 0.617936 & & 1.8(1) & 1.56(4) \\ 
 &  & -128.31677(8) & -128.31743(5) &  & 0.6200(1) & 0.61964(6) &  '' & '' & 0.617936 & & 2.1(1) & 1.71(6) \\ 
 &  & -128.25201(8) & -128.25639(5) &  & 0.6848(1) & 0.68068(7) &  $2p^5(^2P^\circ_{3/2})3p$ & $^2[1/2]$ & 0.678542 & & 6.2(1) & 2.14(7) \\ 
 &  & -128.24875(9) & -128.25092(5) &  & 0.6880(1) & 0.68616(6) &  $2p^5(^2P^\circ_{3/2})3p$ & $^2[5/2]$ & 0.682205 & & 5.8(1) & 3.95(6) \\ 
 &  & -128.24777(8) & -128.25009(5) &  & 0.6890(1) & 0.68698(7) &  $2p^5(^2P^\circ_{1/2})3p$ & $^2[3/2]$ & 0.684558 & & 4.4(1) & 2.42(7) \\  \hline\hline
    \end{tabular}
    \egroup
    \caption{Numerical results for atomic spectra, compared against experimental ground truth.}
    \label{tab:atomic_spectra}
\end{table*}

\begin{table*}[t]
    \tiny
    \centering
    \bgroup
    \def\arraystretch{0.5}%
    \hspace*{-.2cm}\begin{tabular}{cccccccccccc}\hline\hline
          & & \multicolumn{5}{c}{FermiNet} & \multicolumn{5}{c}{Psiformer} \\
        \cmidrule(lr){3-7} \cmidrule(lr){8-12} System & Term & Energy (Ha) & $\Delta\Delta E$ (mHa) & $\langle \hat{\mathcal{S}}^2 \rangle $ & $f$ &  $\Delta f$ & Energy (Ha) & $\Delta\Delta E$ (mHa) & $\langle \hat{\mathcal{S}}^2 \rangle $ & $f$ & $\Delta f$ \\
\hline
 BH & $^1\Sigma^+$ & -25.28921(1) & -- & 0.000 & -- & -- & -25.289249(9) & -- & 0.000 & -- & -- \\
 & - & -25.24045(2) & -- & 2.000 & 2.99e-07 & -- & -25.24047470(5) & -- & 2.000 & 1.13e-07 & -- \\
 & - & -25.24043(1) & -- & 2.000 & -2.21e-07 & -- & -25.24047470(5) & -- & 2.000 & 1.13e-07 & -- \\
 & $^1\Pi$ & -25.18376(3) & -0.38(3) & 0.000 & 0.0236 & -0.0004 & -25.18386(1) & -0.44(1) & 0.000 & 0.0235 & -0.000513 \\
 & $^1\Pi$ & -25.18374(2) & -0.36(3) & 0.000 & 0.0239 & -0.000129 & -25.18383(1) & -0.42(1) & 0.000 & 0.0239 & -9.5e-05 \\
\hline
 HCl & $^1\Sigma^+$ & -15.59729(1) & -- & 0.000 & -- & -- & -15.597583(9) & -- & 0.000 & -- & -- \\
 & - & -15.32593(1) & -- & 2.000 & -7.97e-08 & -- & -15.32633(1) & -- & 2.000 & 1.05e-06 & -- \\
 & - & -15.32591(2) & -- & 2.000 & -3.13e-08 & -- & -15.32632(1) & -- & 2.000 & -9.45e-08 & -- \\
 & $^1\Pi$ & -15.30737(1) & -0.76(2) & 0.000 & 0.0223 & -0.000723 & -15.3077421(5) & -0.846(9) & 0.000 & -9.78e-05 & -0.0231 \\
 & $^1\Pi$ & -15.30734(1) & -0.73(2) & 0.000 & 0.0223 & -0.000654 & -15.3077421(5) & -0.846(9) & 0.000 & -9.78e-05 & -0.0231 \\
\hline
 H$_2$O & $^1A_1$ & -76.43779(4) & -- & 0.000 & -- & -- & -76.43809(2) & -- & 0.000 & -- & -- \\
 & $^3B_1$ & -76.16794(4) & 0.48(6) & 2.000 & 2.73e-06 & -- & -76.16844(2) & 0.27(3) & 2.000 & 5.9e-06 & -- \\
 & $^1B_1$ & -76.15445(4) & 0.00(6) & 0.000 & 0.052 & -3.28e-05 & -76.15486(2) & -0.11(3) & 0.000 & 0.0509 & -0.00109 \\
 & $^3A_2$ & -76.09538(4) & 0.65(6) & 2.000 & 8.1e-06 & -- & -76.09583(3) & 0.49(4) & 2.000 & -7.62e-07 & -- \\
 & $^1A_2$ & -76.08935(4) & -0.31(6) & 0.000 & 1.31e-05 & -- & -76.08974(3) & -0.40(4) & 0.000 & -7.14e-07 & -- \\
\hline
 H$_2$S & $^1A_1$ & -11.38901(1) & -- & 0.000 & -- & -- & -11.389308(8) & -- & 0.000 & -- & -- \\
 & $^3A_2$ & -11.17732(1) & 0.74(1) & 2.000 & 9.44e-07 & -- & -11.177684(9) & 0.68(1) & 2.000 & 3.09e-07 & -- \\
 & $^3B_1$ & -11.17126(1) & -0.54(2) & 2.000 & 7.27e-08 & -- & -11.171657(9) & -0.64(1) & 2.000 & 8.94e-08 & -- \\
 & $^1A_2$ & -11.16411(1) & 0.73(1) & 0.000 & 1.71e-06 & -- & -11.164486(9) & 0.65(1) & 0.000 & 1.4e-07 & -- \\
 & $^1B_1$ & -11.15745(1) & -1.06(2) & 0.000 & 0.059 & -0.000967 & -11.157900(9) & -1.21(1) & 0.000 & 0.0596 & -0.000401 \\
\hline
 BF & $^1\Sigma^+$ & -124.67619(8) & -- & 0.000 & -- & -- & -124.67772(3) & -- & 0.000 & -- & -- \\
 & - & -124.54200(1) & -- & 2.000 & -2.9e-06 & -- & -124.5435944(6) & -- & 2.000 & -1.12e-06 & -- \\
 & - & -124.54200(1) & -- & 2.000 & -2.9e-06 & -- & -124.5435944(6) & -- & 2.000 & -1.12e-06 & -- \\
 & $^1\Pi$ & -124.44155(8) & -0.2(1) & 0.001 & 0.231 & -0.00295 & -124.44351(3) & -0.62(4) & 0.001 & 0.232 & -0.00109 \\
 & $^1\Pi$ & -124.4413(2) & 0.0(2) & 0.001 & 0.234 & 0.000532 & -124.44339(3) & -0.50(4) & 0.001 & 0.234 & 0.000644 \\
\hline
 CO & $^1\Sigma^+$ & -113.32326(7) & -- & 0.001 & -- & -- & -113.32477(3) & -- & 0.001 & -- & -- \\
 & - & -113.092094(2) & 0.38(7) & 2.000 & 5.61e-06 & -- & -113.094002(1) & -0.02(3) & 2.000 & -1e-06 & -- \\
 & - & -113.092094(2) & 0.38(7) & 2.000 & 5.61e-06 & -- & -113.094002(1) & -0.02(3) & 2.000 & -1e-06 & -- \\
 & $^1\Pi$ & -113.01196(8) & 0.4(1) & 0.003 & 0.0837 & 0.00123 & -113.0139(1) & -0.1(1) & 0.001 & 0.0827 & 0.000196 \\
 & $^1\Pi$ & -113.01183(8) & 0.5(1) & 0.002 & 0.0822 & -0.000262 & -113.0139(1) & 0.0(1) & 0.001 & 0.082 & -0.000476 \\
\hline
 C$_2$H$_4$ & $^1B_{3u}$ & -78.58542(6) & -- & 0.001 & -- & -- & -78.58730(2) & -- & 0.001 & -- & -- \\
 & $^3B_{1u}$ & -78.41786(6) & 0.71(8) & 2.000 & 1.65e-06 & -- & -78.41999(2) & 0.46(4) & 2.000 & 4.53e-06 & -- \\
 & $^3B_{3u}$ & -78.31636(6) & 1.52(8) & 1.997 & 7.95e-05 & -- & -78.31870(2) & 1.07(4) & 2.000 & 2.54e-05 & -- \\
 & $^1B_{3u}$ & -78.31239(6) & 0.35(8) & 0.005 & 0.0797 & 0.00368 & -78.31493(2) & -0.32(4) & 0.001 & 0.0792 & 0.00322 \\
 & $^1B_{1u}$ & -78.29347(6) & 1.63(8) & 0.003 & 0.341 & 0.0028 & -78.29640(3) & 0.57(4) & 0.002 & 0.343 & 0.00453 \\
\hline
 CH$_2$O & $^1A_1$ & -114.50546(8) & -- & 0.001 & -- & -- & -114.50756(3) & -- & 0.001 & -- & -- \\
 & $^3A_2$ & -114.37336(7) & 0.5(1) & 2.001 & 1.57e-06 & -- & -114.37572(3) & 0.27(5) & 2.000 & 1.4e-06 & -- \\
 & $^1A_2$ & -114.35941(7) & -0.6(1) & 0.002 & 7.16e-07 & -- & -114.36179(3) & -0.86(5) & 0.001 & 4.66e-07 & -- \\
 & $^3B_2$ & -114.24138(8) & 1.7(1) & 1.996 & 4.37e-05 & -- & -114.24412(3) & 1.05(5) & 2.000 & 4.26e-06 & -- \\
 & $^1B_2$ & -114.23587(8) & -0.1(1) & 0.007 & 0.0272 & 0.00721 & -114.23879(3) & -0.97(5) & 0.002 & 0.0228 & 0.00275 \\
\hline
 CH$_2$S & $^1A_1$ & -49.45811(5) & -- & 0.002 & -- & -- & -49.46048(2) & -- & 0.001 & -- & -- \\
 & $^3A_2$ & -49.38719(6) & -0.38(7) & 2.000 & 1.1e-07 & -- & -49.38977(2) & -0.58(2) & 2.000 & 7.14e-08 & -- \\
 & $^1A_2$ & -49.37754(6) & -1.02(8) & 0.002 & 3.9e-06 & -- & -49.37980(2) & -0.90(2) & 0.001 & 1.47e-06 & -- \\
 & $^3A_1$ & -49.33257(5) & -0.88(7) & 2.000 & 6.61e-07 & -- & -49.33547(2) & -1.41(3) & 2.000 & 6.96e-06 & -- \\
 & $^3B_2$ & -49.23624(5) & 10.19(7) & 1.943 & 0.00387 & -- & -49.24069(2) & 8.12(3) & 1.999 & 0.000106 & -- \\
\hline
 HNO & $^1A'$ & -130.47995(7) & -- & 0.002 & -- & -- & -130.48189(4) & -- & 0.001 & -- & -- \\
 (Ordered) & - & -130.44684(7) & -- & 2.000 & 7.95e-07 & -- & -130.44914(4) & -- & 2.000 & 3.24e-07 & -- \\
 & $^1A''$ & -130.41568(7) & 0.7(1) & 0.003 & 0.000379 & -- & -130.41808(5) & 0.24(6) & 0.002 & 0.000323 & -- \\
 & $^1A'$ & -- & -- & -- & -- & -- & -130.31268(7) & 10.45(8) & 0.010 & 6.55e-05 & -- \\
 & - & -130.25687(7) & -- & 1.999 & 4.93e-05 & -- & -130.25607(5) & -- & 1.729 & 0.00523 & -- \\
 & - & -130.24429(8) & -- & 0.006 & 0.043 & -- & -- & -- & -- & -- & -- \\
 \hline
 HNO & $^1A'$ & -130.47868(9) & -- & 0.002 & -- & -- & -130.48250(4) & -- & 0.001 & -- & -- \\
 (Random) & - & -130.44521(9) & -- & 2.000 & 8.14e-08 & -- & -130.44952(4) & -- & 2.000 & 9.99e-09 & -- \\
 & $^1A''$ & -130.41469(9) & 0.4(1) & 0.003 & 0.000331 & -- & -130.41878(4) & 0.14(6) & 0.001 & 0.000317 & -- \\
 & $^1A'$ & -130.31972(9) & 0.2(1) & 0.002 & 1.97e-05 & -- & -130.32387(6) & -0.13(8) & 0.001 & 3.58e-05 & -- \\
 & - & -130.27084(9) & -- & 2.000 & 1.24e-05 & -- & -130.27586(4) & -- & 2.000 & 1.01e-05 & -- \\
\hline
 HCF & $^1A'$ & -138.41429(9) & -- & 0.001 & -- & -- & -138.41669(4) & -- & 0.001 & -- & -- \\
 (Ordered) & - & -138.37982(9) & -- & 2.000 & -1.64e-09 & -- & -138.38220(4) & -- & 2.000 & 1.34e-06 & -- \\
 & $^1A''$ & -138.32289(9) & 0.3(1) & 0.002 & 0.00629 & 0.000286 & -138.32567(4) & -0.12(6) & 0.001 & 0.00631 & 0.000307 \\
 & - & -138.17085(9) & -- & 2.000 & -4.98e-06 & -- & -138.17415(4) & -- & 1.999 & 6.26e-05 & -- \\
 & - & -138.15756(9) & -- & 0.003 & 0.065 & -- & -138.16124(4) & -- & 0.002 & 0.0676 & -- \\
 \hline
 HCF & $^1A'$ & -138.4133(1) & -- & 0.001 & -- & -- & -138.41618(4) & -- & 0.001 & -- & -- \\
 (Random) & - & -138.37862(9) & -- & 2.000 & 4.31e-06 & -- & -138.38184(4) & -- & 2.000 & 5.71e-08 & -- \\
 & $^1A''$ & -138.3223(1) & -0.1(1) & 0.002 & 0.00556 & -0.000438 & -138.32527(4) & -0.23(6) & 0.001 & 0.00616 & 0.000157 \\
 & - & -138.2034(1) & -- & 0.002 & 0.00327 & -- & -138.20746(4) & -- & 0.001 & 0.00381 & -- \\
 & - & -138.1671(1) & -- & 1.914 & 0.00293 & -- & -138.17123(5) & -- & 1.996 & 2.79e-05 & -- \\
\hline
 H$_2$CSi & $^1A_1$ & -43.11033(4) & -- & 0.001 & -- & -- & -43.11183(1) & -- & 0.000 & -- & -- \\
 & - & -43.03848(4) & -- & 2.001 & 2.43e-06 & -- & -43.03986(1) & -- & 2.000 & -4.87e-07 & -- \\
 & $^1A_2$ & -43.03253(4) & -1.21(6) & 0.001 & 1.04e-06 & -- & -43.03401(1) & -1.19(2) & 0.001 & 1.35e-06 & -- \\
 & - & -43.01749(4) & -- & 2.000 & 1.13e-05 & -- & -43.01966(1) & -- & 2.000 & 3.08e-06 & -- \\
 & - & -43.00178(4) & -- & 2.000 & 5.48e-05 & -- & -43.00376(1) & -- & 2.000 & 6.31e-06 & -- \\\hline\hline
    \end{tabular}
    \egroup
    \caption{Energies, spin magnitudes, and oscillator strengths from the ground state for systems in Fig.~\ref{fig:oscillator_strengths}. The error in the vertical excitation energies ($\Delta\Delta E$) and oscillator strengths ($\Delta f$) relative to the theoretical best estimates\cite{chrayteh2020mountaineering, veril2021questdb} are given as well. All systems were computed with ordered pretraining unless otherwise specified. A comparison of ordered and random pretraining for systems where ordered pretraining failed to find certain states is given in Fig.~\ref{fig:pretraining_and_penalty}.}
    \label{tab:oscillator_strengths}
\end{table*}

\begin{table*}[t]
    \tiny
    \centering
    \bgroup
    \def\arraystretch{0.5}%
    \begin{tabular}{cccccccccc}\hline\hline
          & & \multicolumn{4}{c}{FermiNet} & \multicolumn{4}{c}{Psiformer} \\
        \cmidrule(lr){3-6} \cmidrule(lr){7-10} &  & \multicolumn{2}{c}{Ordered Pretraining} & \multicolumn{2}{c}{Random Pretraining} & \multicolumn{2}{c}{Ordered Pretraining} & \multicolumn{2}{c}{Random Pretraining} \\
        \cmidrule(lr){3-4} \cmidrule(lr){5-6} \cmidrule(lr){7-8} \cmidrule(lr){9-10}
        System & Term & Energy (Ha) & $\Delta\Delta E$ (mHa) & Energy (Ha) & $\Delta\Delta E$ (mHa) & Energy (Ha) & $\Delta\Delta E$ (mHa) & Energy (Ha) & $\Delta\Delta E$ (mHa) \\
\hline
HNO & $^1A'$  & -130.4788(1) & -- & -130.4789(1) & -- & -130.48193(5) & -- & -130.48189(9) & -- \\
 & -  & -130.4425(1) & -- & -130.4418(1) & -- & -130.44720(6) & -- & -130.44706(6) & -- \\
 & $^1A''$  & -130.4115(1) & -3.7(1) & -130.4107(1) & -4.6(1) & -130.41659(6) & -1.76(8) & -130.41584(6) & -2.5(1) \\
 & $^1A'$  & -- & -- & -130.3137(1) & -6.4(1) & -- & -- & -130.32021(6) & -2.9(1) \\
 & -  & -- & -- & -130.2676(1) & -- & -- & -- & -130.27310(6) & -- \\
 & -  & -130.2498(1) & -- & -- & -- & -130.25481(6) & -- & -- & -- \\
 & -  & -130.2449(1) & -- & -- & -- & -130.24639(6) & -- & -- & -- \\
\hline
HCF & $^1A'$  & -138.4132(1) & -- & -138.4121(1) & -- & -138.41616(5) & -- & -138.41600(5) & -- \\
 & -  & -138.3757(1) & -- & -138.3743(1) & -- & -138.38041(6) & -- & -138.38092(6) & -- \\
 & $^1A''$  & -138.3186(1) & -3.5(1) & -138.3170(1) & -3.9(1) & -138.32333(6) & -1.69(8) & -138.32381(7) & -1.05(9) \\
 & -  & -- & -- & -138.1976(1) & -- & -138.19593(8) & -- & -138.20535(6) & -- \\
 & -  & -138.1634(1) & -- & -- & -- & -138.17018(7) & -- & -- & -- \\
 & -  & -138.1537(1) & -- & -- & -- & -- & -- & -- & -- \\
 & -  & -- & -- & -138.1269(1) & -- & -- & -- & -138.13548(7) & -- \\
    \hline \hline
    \end{tabular}
    \egroup
    \caption{Absolute energies and error in the vertical excitation energies ($\Delta\Delta E$) for systems and states computed by the ensemble penalty method in Fig~\ref{fig:pretraining_and_penalty}. Some states were re-ordered to be in order of ascending energy, and solutions were matched to terms of known ground truth states where possible.}
    \label{tab:penalty}
\end{table*}

\begin{table*}
\footnotesize
\centering
\hspace*{-.8cm}\begin{tabular}{ccccccccccc}\hline\hline
    & & & \multicolumn{8}{c}{Oscillator strength $f$} \\
        \cmidrule(lr){4-11}
     Term & Energy (Ha) & $\langle \hat{\mathcal{S}}^2 \rangle $ & $X^1\Sigma^+_g$ & $a^3\Pi^-_u$ & $a^3\Pi^+_u$ & $c^3\Sigma^+_u$ & $A^1\Pi^-_u$ & $A^1\Pi^+_u$ & $b^3\Sigma^-_g$ & $B^1\Delta_g$ \\\hline
     $X^1\Sigma^+_g$ & -75.92258(2) & 0.001 & -- & -0.00000 & 0.00000 & 0.00000 & 0.00373 & 0.00366 & 0.00000 & 0.00000 \\
     $a^3\Pi^-_u$ & -75.91345(2) & 2.000 & -- & -- & 0.00000 & 0.00000 & 0.00000 & -0.00000 & 0.00300 & 0.00000 \\
     $a^3\Pi^+_u$ & -75.91325(3) & 2.000 & -- & -- & -- & 0.00000 & 0.00000 & 0.00000 & 0.00308 & 0.00000 \\
     $c^3\Sigma^+_u$ & -75.87881(3) & 2.000 & -- & -- & -- & -- & 0.00000 & 0.00000 & 0.00000 & -0.00000 \\
     $A^1\Pi^-_u$ & -75.87752(2) & 0.002 & -- & -- & -- & -- & -- & 0.00000 & 0.00000 & 0.00250 \\
     $A^1\Pi^+_u$ & -75.87733(3) & 0.002 & -- & -- & -- & -- & -- & -- & 0.00000 & 0.00251 \\
     $b^3\Sigma^-_g$ & -75.87527(3) & 2.000 & -- & -- & -- & -- & -- & -- & -- & 0.00000 \\
     $B^1\Delta_g$ & -75.84563(3) & 0.002 & -- & -- & -- & -- & -- & -- & -- & -- \\\hline\hline
\end{tabular}
\caption{Energies, spin magnitudes and oscillator strengths for the carbon dimer at equilibrium.}
\label{tab:carbon_dimer}
\end{table*}

\begin{table*}[t]
\scriptsize
\centering
\begin{tabular}{cccccccc}\hline\hline
     $a/a_{eq}$ & $X^1\Sigma^+_g$ & $a^3\Pi_u$ & $b^3\Sigma^-_g$ & $c^3\Sigma^+_u$ & $A^1\Pi_u$ & $B^1\Delta_g$ & $d^3\Pi_g$ \\\hline
0.8 & -75.77419(4) & -75.69933(2) &  --  & -75.75841(4) & -75.66008(2) &  -- & -75.65383(2) \\
0.9 & -75.89507(3) & -75.8582840(1) & -75.79866(3) & -75.86532(3) & -75.82059(1) &  -- & -75.79166(3) \\
0.95 & -75.91682(3) & -75.89481(1) & -75.84625(3) & -75.87997(3) & -75.8580627(1) &  -- & -75.81998(3) \\
1.0 & -75.92258(2) & -75.91335(1) & -75.87527(3) & -75.87881(3) & -75.87743(1) & -75.84563(3) &  -- \\
1.05 & -75.91770(2) & -75.91940(1) & -75.88916(3) & -75.86724(3) & -75.8842239(4) & -75.86131(3) &  -- \\
1.1 & -75.90569(3) & -75.9170(5) & -75.89303(3) & -75.84916(3) & -75.8817(1) & -75.86711(3) &  -- \\
1.2 & -75.87094(3) & -75.8941956(8) & -75.88248(3) &  --  & -75.861694(1) & -75.85953(3) &  -- \\
1.3 & -75.83653(3) & -75.86264(1) & -75.85935(3) &  --  & -75.83181(2) & -75.83955(1) &  -- \\
1.4 & -75.80875(3) & -75.8296(3) & -75.83243(2) &  --  & -75.80106(1) & -75.8155136(2) &  -- \\
1.5 & -75.78448(4) & -75.79841(1) & -75.80516(3) &  --  & -75.77237(1) & -75.79097(2) &  -- \\\hline\hline
\end{tabular}
\caption{Potential energy curves for different states of the carbon dimer. For degenerate states the average energy is given. Bond lengths are reported as multiples of the equilibrium length of 1.244\AA. Energies are in atomic units.}
\label{tab:carbon_dimer_pec}
\end{table*}

\begin{table*}[t]
\small
\centering
\begin{tabular}{cccccc}\hline\hline
& \multicolumn{2}{c}{Bond Length (\AA)}  & \multicolumn{3}{c}{Adiabatic Excitation Energy (eV)} \\
 \cmidrule(lr){2-3} \cmidrule(lr){4-6}
State & Expt.\cite{martin1992c2} & NES-VMC & Expt. & SHCI \cite{holmes2017excited} & NES-VMC \\\hline
$X^1\Sigma^+_g$ & 1.24253 & 1.241 & -- & -- & -- \\
$a^3\Pi_u$ & 1.312 & 1.321 & 0.0891 & 0.07 & 0.0837 \\
$b^3\Sigma^-_g$ & 1.369 & 1.368 & 0.7978 & 0.78 & 0.8041 \\
$A^1\Pi_u$ & 1.318 & 1.320 & 1.0404 & 1.03 & 1.0407 \\
$c^3\Sigma^+_u$ & 1.208 & 1.206 & 1.1312 & 1.16 & 1.1299 \\
$B^1\Delta_g$ & 1.385 & 1.382 & 1.4980 & 1.49 & 1.5031 \\\hline\hline
\end{tabular}
\caption{Bond lengths and adiabatic energies of different states of the carbon dimer. SHCI results use experimental bond lengths, while NES-VMC results are from the minimum of the interpolated curve. Energies are in atomic units.}
\label{tab:carbon_dimer_adiabatic}
\end{table*}

\begin{table*}[t]
    \tiny
    \centering
    \bgroup
    \def\arraystretch{0.5}%
    \hspace*{-1cm}\begin{tabular}{ccccc}\hline\hline
          & \multicolumn{2}{c}{FermiNet} & \multicolumn{2}{c}{Psiformer} \\
        \cmidrule(lr){2-3} \cmidrule(lr){4-5} & Energy (Ha) & $\langle \hat{\mathcal{S}}^2 \rangle $ & Energy (Ha) & $\langle \hat{\mathcal{S}}^2 \rangle $ \\ \hline
$\tau$=0, k=5 & -78.58389(7) & 0.002 & -78.58629(3) & 0.001 \\
& -78.42187(7) & 2.000 & -78.42446(3) & 2.000 \\
& -78.31728(6) & 1.996 & -78.32053(3) & 2.000 \\
& -78.31335(6) & 0.007 & -78.31663(3) & 0.001 \\
& -78.29620(7) & 0.005 & -78.29949(3) & 0.002 \\
$\tau$=0, k=3 & -78.58465(6) & 0.001 & -78.58626(3) & 0.001 \\
& -78.42296(6) & 2.000 & -78.42424(3) & 2.000 \\
& -78.31716(6) & 1.413 & -78.31874(3) & 0.001 \\
$\tau$=15 & -78.58083(6) & 0.001 & -78.58161(3) & 0.001 \\
& -78.42683(6) & 2.000 & -78.42763(3) & 2.000 \\
& -78.31128(6) & 0.002 & -78.31227(4) & 0.004 \\
$\tau$=30 & -78.56795(6) & 0.001 & -78.56930(3) & 0.001 \\
& -78.43636(6) & 2.000 & -78.43774(3) & 2.000 \\
& -78.33266(6) & 0.002 & -78.33452(3) & 0.002 \\
$\tau$=45 & -78.54724(5) & 0.001 & -78.54852(3) & 0.001 \\
& -78.44801(6) & 2.000 & -78.44929(3) & 2.000 \\
& -78.35192(6) & 0.002 & -78.35327(3) & 0.002 \\
$\tau$=60 & -78.51944(6) & 0.001 & -78.52094(3) & 0.001 \\
& -78.45856(5) & 2.000 & -78.45999(3) & 2.000 \\
& -78.36711(6) & 0.002 & -78.36861(3) & 0.002 \\
$\tau$=70 & -78.49858(6) & 0.001 & -78.50006(3) & 0.001 \\
& -78.46394(6) & 2.000 & -78.46536(3) & 2.000 \\
& -78.37368(6) & 0.002 & -78.37585(3) & 0.001 \\
$\tau$=80 & -78.47919(5) & 0.001 & -78.48064(3) & 0.001 \\
& -78.46745(5) & 2.000 & -78.46873(3) & 2.000 \\
& -78.37756(5) & 0.002 & -78.37959(3) & 0.002 \\
$\tau$=85 & -78.47247(6) & 0.001 & -78.47399(3) & 0.001 \\
& -78.46819(6) & 2.000 & -78.46973(3) & 2.000 \\
& -78.38004(6) & 0.001 & -78.38159(3) & 0.001 \\
$\tau$=90 & -78.46991(6) & 0.003 & -78.47133(3) & 0.003 \\
& -78.46851(6) & 1.999 & -78.47001(3) & 1.998 \\
& -78.38188(6) & 0.001 & -78.38337(3) & 0.001 \\
$\phi$=0.0 & -78.47283(5) & 0.009 & -78.47409(2) & 0.003 \\
& -78.47116(5) & 1.993 & -78.47243(2) & 1.998 \\
& -78.38395(5) & 0.001 & -78.38519(3) & 0.001 \\
$\phi$=20.0 & -78.47176(5) & 0.001 & -78.47289(2) & 0.001 \\
& -78.47033(5) & 2.000 & -78.47151(2) & 2.000 \\
& -78.38669(5) & 0.001 & -78.38824(3) & 0.001 \\
$\phi$=40.0 & -78.46626(5) & 0.042 & -78.46735(3) & 0.006 \\
& -78.46555(5) & 1.960 & -78.46678(3) & 1.995 \\
& -78.39383(5) & 0.001 & -78.39510(3) & 0.001 \\
$\phi$=60.0 & -78.45258(5) & 1.923 & -78.45374(3) & 2.034 \\
& -78.45231(5) & 0.078 & -78.45340(3) & -0.033 \\
& -78.40003(5) & 0.001 & -78.40135(3) & 0.001 \\
$\phi$=70.0 & -78.44149(5) & 1.993 & -78.44278(3) & 1.987 \\
& -78.44051(5) & 0.009 & -78.44196(3) & 0.014 \\
& -78.40085(5) & 0.001 & -78.40234(3) & 0.001 \\
$\phi$=80.0 & -78.42701(5) & 2.000 & -78.42815(3) & 1.999 \\
& -78.42552(5) & 0.002 & -78.42664(3) & 0.002 \\
& -78.39926(5) & 0.001 & -78.40073(3) & 0.001 \\
$\phi$=90.0 & -78.40856(6) & 2.002 & -78.40991(3) & 2.000 \\
& -78.40621(6) & -0.000 & -78.40779(3) & 0.001 \\
& -78.39454(6) & 0.001 & -78.3959(1) & 0.001 \\
$\phi$=95.0 & -78.39794(5) & 2.000 & -78.39952(3) & 1.999 \\
& -78.39555(5) & 0.001 & -78.39676(3) & 0.002 \\
& -78.39079(5) & 0.001 & -78.39205(3) & 0.001 \\
$\phi$=97.5 & -78.39275(5) & 2.000 & -78.39400(3) & 2.000 \\
& -78.38973(5) & 0.001 & -78.39106(3) & 0.001 \\
& -78.38829(5) & 0.002 & -78.38996(3) & 0.001 \\
$\phi$=100.0 & -78.38690(5) & 2.002 & -78.38826(3) & 2.005 \\
& -78.38587(5) & -0.000 & -78.38728(3) & -0.003 \\
& -78.38350(5) & 0.002 & -78.38503(3) & 0.001 \\
$\phi$=102.5 & -78.38331(5) & -0.001 & -78.38445(3) & 0.003 \\
& -78.38113(5) & 2.002 & -78.38248(3) & 1.998 \\
& -78.37771(5) & 0.002 & -78.37894(3) & 0.001 \\
$\phi$=105.0 & -78.38000(5) & 0.001 & -78.38142(3) & 0.001 \\
& -78.37507(6) & 2.000 & -78.37652(3) & 2.000 \\
& -78.37125(6) & 0.002 & -78.37267(3) & 0.001 \\
$\phi$=110.0 & -78.37315(5) & 0.001 & -78.37458(3) & 0.001 \\
& -78.36317(5) & 1.999 & -78.36440(3) & 2.000 \\
& -78.35850(5) & 0.003 & -78.35975(3) & 0.001 \\
$\phi$=120.0 & -78.35644(6) & 0.001 & -78.35776(3) & 0.001 \\
& -78.33870(6) & 2.000 & -78.34010(3) & 1.999 \\
& -78.33257(6) & 0.001 & -78.33368(3) & 0.002 \\
         \hline\hline
    \end{tabular}
    \egroup
    \caption{Energies and spin magnitudes for the lowest states of twisted ethylene using the FermiNet and Psiformer Ans{\"a}tze. Note that for the FermiNet at $\tau=0$, the singlet and triplet states are mixed together when $k$=3, so we compute $k$=5 at that geometry only.}
    \label{tab:ethene}
\end{table*}

\begin{table*}[t]
    \tiny
    \centering
    \begin{tabular}{ccccccccccc}\hline\hline
          & & \multicolumn{3}{c}{FermiNet} & \multicolumn{3}{c}{Psiformer} & QUEST\cite{loos2019reference, veril2021questdb} & DMC\cite{shepard2022double} & CASPT3\cite{kossoski2024reference} \\
        \cmidrule(lr){3-5} \cmidrule(lr){6-8}  \cmidrule(lr){9-9} \cmidrule(lr){10-10} \cmidrule(lr){11-11} System & State & Energy (Ha) & $\Delta E$ & $\langle \hat{\mathcal{S}}^2 \rangle $ & Energy (Ha) & $\Delta E$ & $\langle \hat{\mathcal{S}}^2 \rangle $ & $\Delta E$ & $\Delta E$ & $\Delta E$ \\
\hline
Nitrosomethane & $1^1A'$ & -169.7910(1) &  --  & 0.004 & -169.80035(5) &  --  & 0.002 &  --  &  --  &  --  \\ 
 & $1^3A''$ & -169.7486(1) & 42.4(1) & 2.000 & -169.75793(5) & 42.42(7) & 2.000 & 43 &  --  &  --  \\ 
 & $1^1A''$ & -169.7191(1) & 71.9(1) & 0.005 & -169.72847(5) & 71.88(7) & 0.003 & 72 &  --  &  --  \\ 
 & $2^1A'$ & -169.6167(1) & 174.2(2) & 0.005 & -169.62700(5) & 173.36(7) & 0.002 & 174 &  --  &  --  \\ 
 & $1^3A'$ & -169.5863(1) & 204.7(1) & 2.001 & -169.59610(5) & 204.25(7) & 2.000 & 206 &  --  &  --  \\ 
 & -- & -169.5667(1) & 224.2(2) & 1.932 & -169.57913(5) & 221.22(7) & 1.989 &  --  &  --  &  --  \\ 
\hline
Butadiene & $1^1A_g$ & -155.9733(1) &  --  & 0.010 & -155.98532(6) &  --  & 0.003 &  --  &  --  &  --  \\ 
 & $1^3B_u$ & -155.8468(1) & 126.5(2) & 2.000 & -155.86006(5) & 125.26(8) & 2.001 & 124 &  --  &  --  \\ 
 & $1^3A_g$ & -155.7864(1) & 186.8(1) & 2.005 & -155.79607(5) & 189.25(8) & 2.001 & 191 &  --  &  --  \\ 
 & $1^1B_u$ & -155.7419(1) & 231.4(2) & 0.022 & -155.75511(7) & 230.21(9) & 0.006 & 228 &  --  &  --  \\ 
 & $2^1A_g$ & -155.7319(1) & 241.3(1) & 0.068 & -155.74599(6) & 239.33(8) & 0.011 & 239 &  --  &  --  \\ 
 & -- & -155.6939(1) & 279.4(1) & 2.001 & -155.70535(6) & 279.97(8) & 2.001 &  --  &  --  &  --  \\ 
 & -- & -155.6865(1) & 286.8(1) & 0.017 & -155.69765(6) & 287.67(8) & 0.006 &  --  &  --  &  --  \\ 
\hline
Glyoxal & $1^1A_{g}$ & -227.8308(1) &  --  & 0.017 & -227.82916(6) &  --  & 0.003 &  --  &  --  &  --  \\ 
 & $1^3A_{u}$ & -227.7427(1) & 88.1(1) & 1.010 & -227.73721(6) & 91.95(8) & 2.001 & 92 &  --  &  --  \\ 
 & $1^1A_{u}$ & -227.7311(1) & 99.7(1) & 0.735 & -227.72411(6) & 105.06(8) & 0.003 & 106 &  --  &  --  \\ 
 & $1^3B_{g}$ & -227.6923(1) & 138.5(1) & 0.735 & -227.68610(6) & 143.07(8) & 2.000 & 144 &  --  &  --  \\ 
 & $1^1B_{g}$ & -227.6815(1) & 149.2(1) & 1.600 & -227.67401(6) & 155.15(8) & 0.006 & 156 &  --  &  --  \\ 
 & $1^3B_{u}$ & -227.6531(1) & 177.7(1) & 1.949 & -227.63847(6) & 190.69(8) & 2.001 & 190 &  --  &  --  \\ 
 & $2^1A_{g}$ & -227.6268(1) & 204.0(1) & 0.013 & -227.62488(6) & 204.28(8) & 0.006 & \textcolor{gray}{206} & 207 &  --  \\ 
\hline
Tetrazine & $1^1A_g$ & -296.2875(2) &  --  & 0.006 & -296.3080(1) &  --  & 0.002 &  --  &  --  &  --  \\ 
 & $1^1B_{3u}$ & -296.1987(2) & 88.8(3) & 0.008 & -296.2167(1) & 91.3(1) & 0.004 & 90 &  --  &  --  \\ 
 & $1^1A_{u}$ & -296.1492(2) & 138.3(3) & 0.010 & -296.1716(1) & 136.4(1) & 0.005 & 136 &  --  &  --  \\ 
 & $1^1B_{1g}$ & -296.1076(2) & 180.0(3) & 0.008 & -296.1293(1) & 178.7(1) & 0.004 & 181 &  --  &  --  \\ 
 & $2^1A_{g}$ & -296.1034(2) & 184.1(3) & 0.008 & -296.1221(1) & 185.9(1) & 0.005 & \textcolor{gray}{169} & 183 & 182 \\ 
\hline
Cyclopentadienone & $1^1A_1$ & -268.0744(2) &  --  & 0.004 & -268.0865(1) &  --  & 0.002 &  --  &  --  &  --  \\ 
 & $1^1A_2$ & -267.9663(2) & 108.1(3) & 0.006 & -267.9799(1) & 106.7(1) & 0.004 & 108 &  --  &  --  \\ 
 & $1^1B_2$ & -267.9501(2) & 124.3(3) & 0.005 & -267.9604(1) & 126.2(1) & 0.004 & 132 &  --  &  --  \\ 
 & $1^1B_1$ & -267.8908(2) & 183.6(3) & 0.007 & -267.9020(1) & 184.5(1) & 0.005 & \textcolor{gray}{184} &  --  & 184 \\ 
 & $2^1A_1$ & -267.8565(2) & 217.9(3) & 0.005 & -267.8694(1) & 217.1(1) & 0.003 & \textcolor{gray}{220} & 217 & 213 \\ 
 & $3^1A_1$ & -267.8236(2) & 250.8(3) & 0.009 & -267.8392(1) & 247.4(1) & 0.004 & \textcolor{gray}{224} & 253 & 247 \\ 
         \hline\hline
    \end{tabular}
    \caption{Energies and spin magnitudes for the excited states of larger systems with double excitations using the FermiNet and Psiformer Ans{\"a}tze, with comparison against theoretical best estimates from the original QUEST database \cite{loos2019reference, veril2021questdb} (with values marked ``unsafe" in gray), DMC \cite{shepard2022double} and corrections to the original QUEST values with CASPT3 \cite{kossoski2024reference}. All relative energies from the ground state ($\Delta E$) are given in milli-Hartrees.}
    \label{tab:double_excitations}
\end{table*}

\begin{table*}[t]
    \tiny
    \centering
    \hspace*{-.8cm}
    \begin{tabular}{ccccccccccccc}\hline\hline
          & \multicolumn{3}{c}{FermiNet} & \multicolumn{3}{c}{Psiformer} & \multicolumn{2}{c}{Entwistle {\em et al.}\cite{entwistle2023electronic}} & DMC\cite{pathak2021excited} & CASPT2\cite{roos1992towards} & TD-DFT\cite{adamo1999accurate} & TBE\cite{loos2020mountaineering} \\
        \cmidrule(lr){2-4} \cmidrule(lr){5-7} \cmidrule(lr){8-9} \cmidrule(lr){10-10} \cmidrule(lr){11-11} \cmidrule(lr){12-12} \cmidrule(lr){13-13} State & Energy (Ha) & $\Delta E$ & $\langle \hat{\mathcal{S}}^2 \rangle $ & Energy (Ha) & $\Delta E$ & $\langle \hat{\mathcal{S}}^2 \rangle $ & Raw & $\sigma^2$ Match & $\Delta E$ & $\Delta E$ & $\Delta E$ & $\Delta E$ \\ \hline
$^1A_{1g}$ & -232.2120(4) &  --  & 0.018 & -232.2323(1) &  --  & 0.008 & -232.0675(11) & -232.0265(13) & -- & -- & -- & -- \\
$^3B_{1u}$ & -232.0541(4) & 157.9(6) & 2.007 & -232.0758(1) & 156.5(2) & 2.003 & -231.8628(9) & -231.8628(9) & 152(1) & 145 & 140 & 153 \\
$^3E_{1u}$ & -232.0287(4) & 183.3(6) & 1.991 & -232.0516(1) & 180.7(2) & 2.002 & -- & -- & 180(1) & 165 & 173 & 179 \\
$^3E_{1u}$ & -232.0268(4) & 185.2(6) & 1.986 & -232.0512(1) & 181.1(2) & 2.002 & -- & -- & 182(1) & 165 & 173 & 179 \\
$^1B_{2u}$ & -232.0204(4) & 191.6(6) & 0.063 & -232.0438(1) & 188.5(2) & 0.015 & -- & -- & 189(1) & 173 & 198 & 186 \\
$^3B_{2u}$ & -231.9818(4) & 230.2(6) & 1.126 & -232.0159(1) & 216.3(2) & 1.993 & -- & -- & 216(1) & 200 & 185 & 213.5 \\
         \hline\hline
    \end{tabular}
    \caption{Energies and spin magnitudes for the excited states of benzene using the FermiNet and Psiformer Ans{\"a}tze, with comparison against QMC penalty methods\cite{entwistle2023electronic, pathak2021excited}, CASPT2\cite{roos1992towards} and TD-DFT\cite{adamo1999accurate} and theoretical best estimates from coupled cluster\cite{loos2020mountaineering}. All relative energies from the ground state ($\Delta E$) are given in milli-Hartrees. For the highest excited state of the FermiNet, inspection of the spin magnitude reveals contamination by the singlet $^1B_{1u}$ state, which would explain the anomalously high energy. The Psiformer does not suffer from this contamination.}
    \label{tab:benzene}
\end{table*}

\end{document}